\documentclass[12pt,oneside]{article}%
\usepackage{amsmath}
\usepackage{amssymb}
\usepackage{amsfonts}
\usepackage{graphicx}
\usepackage{cite}%
\allowdisplaybreaks
\newcommand{\eqa}{\begin{eqnarray}}
\newcommand{\ena}{\end{eqnarray}}

\def\one{\mbox{1 \kern-.59em {\rm l}}}
\begin{document}

\author{Alexander Schmidt\thanks{e-mail: schmidt@theorie.physik.uni-muenchen.de},
Hartmut Wachter\thanks{e-mail: Hartmut.Wachter@physik.uni-muenchen.de}%
\vspace*{0.2in}\\Max-Planck-Institute\\for Mathematics in the Sciences \\Inselstr. 22, D-04103 Leipzig, Germany\\\hspace{0.2in}\\Arnold-Sommerfeld-Center \\Ludwig-Maximilians-Universit\"{a}t\\Theresienstr. 37, D-80333 M\"{u}nchen, Germany}
\title{Spinor calculus for q-deformed quantum spaces II}
\date{}
\maketitle

\begin{abstract}
\noindent This is the second part of an article about q-deformed analogs of
spinor calculus. The considerations refer to quantum spaces of physical
interest, i.e. $q$-deformed Euclidean space in three or four dimensions as
well as $q$-deformed Minkowski space. The Clifford algebras corresponding to
these quantum spaces are treated. Especially, their commutation relations and
their Hopf structures are written down. Bases of the four-dimensional Clifford
algebras are constructed and their properties are discussed. Matrix
representations of the Clifford algebras lead to $q$-deformed $\gamma
$-matrices for the four-dimensional quantum spaces. Moreover, $q$-analogs of
the four-dimensional spin matrices are presented. A very complete set of trace
relations and rearrangement formulae concerning spin and $\gamma$-matrices is
given. Dirac spinors together with their bilinear covariants are defined.
Their behavior under $q$-deformed Lorentz transformation is discussed in
detail.\newpage

\end{abstract}
\tableofcontents

\section{Introduction}

Spinors play a very important role in physics. As is well-known, they are
indispensable for describing particles with spin 1/2. Moreover, spinors are an
essential ingredient of supersymmetry. In part I of this article (see Ref. \cite{qspinor1}) we started
studying a $q$-deformed version of spinor calculus. The motivation for this
undertaking stems from the observation that $q$-deformation \cite{Ku83, Wor87,
Dri85, Jim85, Drin86, RFT90, Tak90, Man88, CSSW90, PW90, SWZ91, Maj91, LWW97,
OSWZ92} should provide a very promising approach for discretizing space-time
\cite{FLW96, CW98, MajReg, GKP96, Oec99} (for other deformations of space-time
see Refs. \cite{Lu92, Cas93, Dob94, DFR95, ChDe95, ChKu04, Koch04}). Remember
that formulating quantum field theories on a lattice-like space-time structure
is a very appealing idea, since it should lead to an effective method for
regularizing quantum field theories \cite{Heis}. Due to its attractiveness,
one can find a number of attempts to attack the problem of discretizing
space-time in the former literature (see for example Refs. \cite{Fli48,
Hill55, Das60, Gol63, Sny47, Yan47}).

The construction of a $q$-deformed quantum field theory requires a very deep
and thoughtful understanding of the subject. Thus, we gave in part I of the
article a review of the foundations of the quantum group $SU_{q}(2)$
\cite{Hopf, Maj95, ChDe96, Klimyk}. Additionally, we considered their
two-dimensional co-representations , which are known as symmetrized and
antisymmetrized Manin plane (or quantum plane). The coordinates of the Manin
plane can be viewed as $q$-deformed spinors \cite{Schlieck1, Schlieck2,
Schlieck3, Mey96, WSSW90}. Furthermore, we described the construction of
higher dimensional quantum spaces out of quantum planes. In doing so, we
restricted attention to $q$-deformed Minkowski space and $q$-deformed
Euclidean space with three- or four-dimensions. These results enabled us to
introduce $q$-analogs of the well-known Pauli matrices. Although our
reasonings were in remarkable analogy to the undeformed case we had to take
care of a very delicate structure. In the $q$-deformed setting there are
different types of Pauli matrices to distinguish, so we have carefully to
decide which of these Pauli matrices applies in a certain situation. For this
reason, part I of the article provides the reader with a detailed and rather
complete discussion of the properties of these Pauli matrices. In addition to
this, we introduced $q$-analogs of spin matrices and gave a comprehensive list
of calculational rules and rearrangement formulae that could prove useful in
physical applications. Especially, we presented $q$-analogs of the very
important Weyl Spinor Fierz\textit{ }identities.

In part II we now continue our reasonings about $q$-deformed spinor calculus.
However, stress is taken on a presentation that reveals the foundations of a
$q$-deformed Dirac formalism. Thus, we use our results from part I to define
$q$-analogs of Dirac matrices and spin matrices. In this article the Dirac
matrices ($\gamma$-matrices) establish representations of the Clifford
algebras to four-dimensional quantum spaces, i.e. $q$-deformed Euclidean space
in four dimensions and $q$-deformed Minkowski space. In Sec.\thinspace
\ref{Cliffalg} we write down the commutation relations and Hopf structures of
each Clifford algebra under consideration, where for the sake of completeness
we also deal with the Clifford algebra of three-dimensional $q$-deformed
Euclidean space.

In analogy to the undeformed case a basis of a $q$-deformed Clifford algebra
can be built up from antisymmetrized products of Clifford generators. We also
define a quantum trace which together with the $\gamma$-matrices as
representations of the Clifford generators enables us to calculate a left- as
well as a right-dual basis. This task will be done in Sec.\thinspace
\ref{gammaeu4kapSec1} and Sec.\thinspace\ref{KapGammaDef}. In the undeformed
case the left-dual basis is identical to the right-dual one, but in the
$q$-deformed case it is not. The reason for this lies in the fact that for the
quantum spaces under consideration the metric is not symmetrical. Finally, we
will see that each dual basis can be used to write down some sort of
completeness relation.

The main part of the paper [cf.$\ $Sec.\thinspace\ref{RelConEuc}, Sec.
\thinspace\ref{SpinMatKapEuc}, Sec.\ \ref{RelConMin}, and
Sec.\ \ref{SpinMatKapMin}] is devoted to trace relations and rearrangement
formulae concerning $q$-de\-form\-ed $\gamma$-ma\-tri\-ces and spin matrices.
For the undeformed counterparts of these identities we refer the reader to
Refs. \cite{bailin, Core, Peskin, Wessbagger, Landau, Grimm}, for example. Our
reasonings apply to both $q$-deformed Minkowski space and four-dimensional
$q$-deformed Euclidean space. Nevertheless, there are some remarkable
differences between these spaces. Due to the various types of Pauli matrices
of $q$-deformed Minkowski space the definitions of $\gamma$-matrices and spin
matrices are not unique for that space. Thus, the discussion of trace
relations and rearrangement formulae is case sensitive in $q$-deformed
Minkowski space.

Let us also note that in complete analogy to part I we derived many identities
by first proposing reasonable ansaetze. Then we inserted the expressions for
$\gamma$-matrices and spin matrices into each ansatz and obtained a system of
equations for its unknown coefficients. The corresponding solutions were found
by means of a computer algebra system like \textit{Mathematica }\cite{Wol}.

Section \ref{CovKap} covers the subject of $q$-deformed Dirac spinors and
their bilinear covariants. Working in a kind of Weyl representation Dirac
spinors together with their Hermitian and Dirac conjugates are introduced in
very much the same way as is done for the undeformed case, i.e. we simply
write a 4-spi\-nor as two 2-spi\-nors. The corresponding bilinear covariants
are of the same form as in the undeformed case. We discuss the conjugation
properties of these covariants. Furthermore, we carefully examine the behavior
of the different types of 4-spinors under $q$-deformed Lorentz
transformations. From these results we find that our bilinear covariants again
transform as scalar, pseudoscalar, vector, pseudovector, and antisymmetric
tensor. In summary, our considerations reveal a very remarkable similarity to
the undeformed situation (see, for example, Ref. \cite{Good55}).

We end our reasonings with a short conclusion in Sec.\thinspace
\ref{Conclusion}. In App.\thinspace\ref{AppA} and App.\thinspace\ref{AppB} we
collected some expressions referring to four-dimensional $q$-deformed
Euclidean space and $q$-deformed Minkowski space, respectively. Especially, we
listed matrix representations of the various Clifford bases and expressions
for spin matrices.

Last but not least let us make some notational remarks\textit{.} Throughout
the article we use the shortcuts $\lambda\equiv q-q^{-1}$ and $\lambda
_{+}\equiv q+q^{-1}$. Sometimes an $n$-dimensional identity matrix is written
as $\mbox{1 \kern-.59em {\rm l}}$. The antisymmetric $q$-numbers are defined
by $[[n]]_{q^{a}}\equiv\frac{1-q^{an}}{1-q^{a}}$, where we assume $q>1$ and
$a\in\mathbb{R}$.

\section{$q$-Deformed Clifford algebras\label{Cliffalg}}

This section is devoted to the Clifford algebras that correspond to quantum
spaces we are interested for physical reasons, i.e. three- and
four-dimensional $q$-deformed Euclidean space as well as $q$-deformed
Min\-kow\-ski space. For each Clifford algebra we give a short review of its
defining relations and discuss its Hopf structures. In the subsequent sections
$q$-analogs of $\gamma$-matrices are introduced as four-dimensional
representations of Clifford generators.

\subsection{Clifford algebra of three-dimensional $q$-defor\-med
Eu\-cli\-de\-an space}

The Clifford algebra of three-dimensional $q$-defor\-med Eu\-cli\-de\-an space
is spanned by three generators $\xi^{k},$ $k\in\{+,3,-\}$. Their commutation
relations are determined by two conditions \cite{RFT90}. The first one is
\begin{equation}
(P_{S})^{ij}{}_{kl}\,\xi^{k}\xi^{l}=0,
\end{equation}
where $P_{S}$ denotes the symmetrizer of the three-dimensional $q$-deformed
Euclidean space. As a second claim we have
\begin{equation}
(P_{0})^{ij}{}_{kl}\,\xi^{k}\xi^{l}=c\;g^{ij}\mbox{1 \kern-.59em {\rm l}},
\end{equation}
where $P_{0}$ stands for the projector proportional to the quantum metric
$g^{ij}$ of three-dimensional $q$-deformed Euclidean space. Notice that the
constant $c\in\mathbb{R}$ is not fixed yet. If not stated otherwise, there is
always summation over repeated indices understood.

Using the projector decomposition of the $\hat{R}$-matrix of three-dimensional
$q$-deformed Euclidean space \cite{LWW97, MSW04}, i.e.%
\begin{equation}
\hat{R}=P_{+}-q^{-4}P_{-}+q^{-6}P_{0},
\end{equation}
\ the Clifford algebra can be written in the following compact form:%
\begin{equation}
\xi^{i}\xi^{j}=-q^{4}\hat{R}^{ij}{}_{kl}\,\xi^{k}\xi^{l}+c\;q^{-1}\lambda
_{+}g^{ij}\mbox{1 \kern-.59em {\rm l}}. \label{Cliffeu3gen}%
\end{equation}
However, starting from the projector decomposition of $\hat{R}^{-1}$ would
yield%
\begin{equation}
\xi^{i}\xi^{j}=-q^{-4}(\hat{R}^{-1})^{ij}{}_{kl}\,\xi^{k}\xi^{l}%
+c\;q\lambda_{+}g^{ij}\mbox{1 \kern-.59em {\rm l}}. \label{Sub}%
\end{equation}
More explicitly, the Clifford algebra in (\ref{Cliffeu3gen}) reads as
\begin{align}
\xi^{+}\xi^{+}  &  =\xi^{-}\xi^{-}=0,\nonumber\\
\xi^{3}\xi^{\pm}  &  =-q^{\mp2}\xi^{\pm}\xi^{3},\nonumber\\
\xi^{-}\xi^{+}  &  =-\xi^{+}\xi^{-}-c\lambda_{+}%
\mbox{1 \kern-.59em {\rm l}},\nonumber\\
\xi^{3}\xi^{3}  &  =\lambda\xi^{+}\xi^{-}+cq^{2}\mbox{1 \kern-.59em {\rm l}}.
\label{Cliffordeu3}%
\end{align}
Exploiting (\ref{Sub}) would lead us to the same commutation relations.

In part I of the paper we showed that the Pauli matrices $\sigma^{i}$,
$i\in{\{}+,3,-{\}}${,} for three-dimensional $q$-deformed Euclidean space also
fulfill a Clifford algebra. It becomes identical to (\ref{Cliffordeu3}) if $c$
takes on the value $q^{2}\lambda_{+}^{-1}$.

It is possible to endow the Clifford algebra (\ref{Cliffordeu3}) with two Hopf
structures. Their explicit form on the generators of the Clifford algebra can
easily be found from the Hopf structures for the quantum space coordinates
$X^{i}$ (as they are given in Refs. \cite{BW01, MSW04}, for example) by
substituting the Clifford generators $\xi^{i}$ for the quantum space
coordinates. Thus, the corresponding coproducts, antipodes, and counits for
the Clifford generators become%
\begin{align}
\Delta(\xi^{i})  &  =\xi^{i}\otimes1+\mathcal{L}^{i}{}_{j}\otimes\xi
^{j}\nonumber\\
\bar{\Delta}(\xi^{i})  &  =\xi^{i}\otimes1+\mathcal{\bar{L}}^{i}{}_{j}%
\otimes\xi^{j},\nonumber\\[0.08in]
S(\xi^{i})  &  =-S(\mathcal{L})_{j}^{i}\,\xi^{j},\nonumber\\
\bar{S}(\xi^{i})  &  =-S(\mathcal{\bar{L}})_{j}^{i}\,\xi^{j}%
,\nonumber\\[0.08in]
\epsilon(\xi^{i})  &  =\bar{\epsilon}(\xi^{i})=0, \label{HopfCliff}%
\end{align}
where $\mathcal{L}$ and $\mathcal{\bar{L}}$ stand for the $L$-matrix and its
conjugate, respectively. Let us recall that the $L$-matrix and its conjugate
can be viewed as realizations of the $SU_{q}(2)$-quantum group generators
within the quantum algebra $U_{q}(su(2))$.

However, there is one subtlety we have to take care of, since we cannot assign
the usual group-like coproduct to $\mbox{1 \kern-.59em {\rm l}}$. Instead we
have
\begin{equation}
\Delta(\mbox{1 \kern-.59em {\rm l}})=\mbox{1 \kern-.59em {\rm l}}\otimes
1+\Lambda^{-1}\otimes\mbox{1 \kern-.59em {\rm l}}-q^{3}c^{-1}\lambda
g_{ij}\,\xi^{i}\mathcal{L}^{j}{}_{k}\otimes\xi^{k},
\end{equation}
and
\begin{equation}
\bar{\Delta}(\mbox{1 \kern-.59em {\rm l}})=\mbox{1 \kern-.59em {\rm l}}\otimes
1+\Lambda\otimes\mbox{1 \kern-.59em {\rm l}}-q^{-3}c^{-1}\lambda g_{ij}%
\,\xi^{i}\mathcal{\bar{L}}^{j}{}_{k}\otimes\xi^{k}.
\end{equation}
Inserting the explicit form for the entries of the $L$-matrices (see, for
example, Ref. \cite{MSW04}) we get%
\begin{align}
-q^{3}c^{-1}\lambda g_{ij}\,\xi^{i}\mathcal{L}^{j}{}_{k}\otimes\xi^{k}=  &
-c^{-1}\lambda\Lambda^{-1/2}\;\big[q^{2}\xi^{+}(\tau^{3})^{-1/2}\otimes\xi
^{-}\nonumber\\
&  -\,q\xi^{3}\otimes\xi^{3}-q\lambda\lambda_{+}\,\xi^{3}L^{+}\otimes\xi
^{-}\nonumber\\
&  +\,\xi^{-}(\tau^{3})^{1/2}\otimes\xi^{+}+q\lambda\lambda_{+}\,\xi^{-}%
(\tau^{3})^{1/2}L^{+}\otimes\xi^{3}\nonumber\\
&  +\,q^{2}\lambda^{2}\lambda_{+}\,\xi^{-}(\tau^{3})^{1/2}(L^{+})^{2}%
\otimes\xi^{-}\big], \label{deltaeinseu3expl}%
\end{align}
and
\begin{align}
-q^{-3}c^{-1}\lambda g_{ij}\,\xi^{i}\mathcal{\bar{L}}^{j}{}_{k}\otimes\xi
^{k}=  &  \;c^{-1}\lambda\Lambda^{1/2}\;\big[q^{-2}\,\xi^{-}(\tau^{3}%
)^{-1/2}\otimes\xi^{+}\nonumber\\
&  -\,q^{-1}\xi^{3}\otimes\xi^{3}-q^{-1}\lambda\lambda_{+}\,\xi^{3}%
L^{-}\otimes\xi^{+}\nonumber\\
&  +\,\xi^{+}(\tau^{3})^{1/2}\otimes\xi^{-}+q^{-1}\lambda\lambda_{+}\,\xi
^{+}(\tau^{3})^{1/2}L^{-}\otimes\xi^{3}\nonumber\\
&  +\,q^{-2}\lambda^{2}\lambda_{+}\,\xi^{+}(\tau^{3})^{1/2}(L^{-})^{2}%
\otimes\xi^{+}\big], \label{deltaeinseu3expldach}%
\end{align}
where $L^{+}$, $L^{-}$, and $\tau^{3}$ are generators of $U_{q}(su(2))$, while
$\Lambda$ stands for a unitary scaling operator subject to%
\begin{equation}
\Lambda\mbox{1 \kern-.59em {\rm l}}=q^{-8}%
\,\mbox{1 \kern-.59em {\rm l}}\Lambda,\qquad\Lambda^{1/2}\xi^{k}=-q^{-2}\,\xi
^{k}\Lambda^{1/2},\quad k\in\{+,3,-\}.
\end{equation}
Notice that these scaling properties are consistent with the requirement that
the scaling operator respects the relations in (\ref{Cliffeu3gen}) and
(\ref{Sub}). Finally, it should be mentioned that the above expressions for
the coproducts $\Delta(\mbox{1 \kern-.59em {\rm l}})$ and $\bar{\Delta
}(\mbox{1 \kern-.59em {\rm l}})$ are consistent with
\begin{align}
S(\mbox{1 \kern-.59em {\rm l}}) =-q^6\Lambda
\mbox{1 \kern-.59em {\rm l}}, & \qquad
\bar{S}(\mbox{1 \kern-.59em {\rm l}}) =-q^{-6}\Lambda^{-1}%
\mbox{1 \kern-.59em {\rm l}},\nonumber\\[0.08in]
\epsilon(\mbox{1 \kern-.59em {\rm l}})  &  =\bar{\epsilon}%
(\mbox{1 \kern-.59em {\rm l}})=0.
\end{align}

\subsection{Clifford algebra of four-dimensional $q$-defor\-med
Eu\-cli\-de\-an space}

The Clifford algebra for the four-dimensional $q$-deformed Euclidean space is
constructed in very much the same way as was done for the three-dimensional
$q$-deformed Euclidean space. Again, we start from the requirements
\begin{equation}
(P_{S})^{ij}{}_{kl}\,\xi^{k}\xi^{l}=0,\qquad(P_{0})^{ij}{}_{kl}\,\xi^{k}%
\xi^{l}=c\,g^{ij}\mbox{1 \kern-.59em {\rm l}}, \label{ReqClif4}%
\end{equation}
but now we have to use the projectors of the four-dimensional $q$-deformed
Euclidean space \cite{Klimyk}. From (\ref{ReqClif4}) together with the
projector decomposition of the $\hat{R}$-matrix of the quantum group
$SO_{q}(4)$ (see, for example, Ref. \cite{MSW04}) we can derive as defining
relations of the Clifford algebra%
\begin{equation}
\xi^{i}\xi^{j}=-q\hat{R}_{kl}^{ij}\,\xi^{k}\xi^{l}-q^{-1}c\lambda_{+}%
g^{ij}\mbox{1 \kern-.59em {\rm l}}. \label{Cliffeu41}%
\end{equation}
Alternatively, we get
\begin{equation}
\xi^{i}\xi^{j}=-q^{-1}(\hat{R}^{-1})_{kl}^{ij}\,\xi^{k}\xi^{l}-qc\lambda
_{+}g^{ij}\mbox{1 \kern-.59em {\rm l}}, \label{Cliffeu42}%
\end{equation}
if we instead work with the projector decomposition of $\hat{R}^{-1}$. Both
(\ref{Cliffeu41}) and (\ref{Cliffeu42}) imply the following commutation
relations for the Clifford generators $\xi^{k}$, $k=1,\ldots,4$:%
\begin{align}
\xi^{k}\xi^{k}  &  =0,\qquad k=1,\ldots,4,\nonumber\\
\xi^{j}\xi^{1}  &  =-q\,\xi^{1}\xi^{j},\nonumber\\
\xi^{4}\xi^{j}  &  =-q\,\xi^{j}\xi^{4},\qquad j=2,3,\nonumber\\
\xi^{4}\xi^{1}  &  =-\xi^{1}\xi^{4}-c\lambda_{+}%
\mbox{1 \kern-.59em {\rm l}},\nonumber\\
\xi^{3}\xi^{2}  &  =-\xi^{2}\xi^{3}+\lambda\,\xi^{1}\xi^{4}-q^{-1}c\lambda
_{+}\mbox{1 \kern-.59em {\rm l}}. \label{Cliffeu4}%
\end{align}

In Sec. \ref{gammaeu4kap} we will calculate $\gamma$-matrices for the
four-dimensional $q$-de\-formed Euclidean space. These $\gamma$-matrices will
obey the relations in (\ref{Cliffeu4}) if we choose\ $c=-\lambda_{+}^{-1}$.

In analogy to the three-dimensional case there exist two Hopf structures for
the Clifford algebra of four-dimensional $q$-deformed Euclidean space. Once
again, these Hopf structures are of the same form as those for the coordinates
of four-dimensional $q$-deformed Euclidean space [cf. the expressions in
(\ref{HopfCliff}) with $\mathcal{L}$ and $\mathcal{\bar{L}}$ now standing for
the $L$-matrices of $U_{q}(so(4))$].

Next, we come to the coproducts of $\mbox{1 \kern-.59em {\rm l}}$ for which we
have to set
\begin{align}
\Delta(\mbox{1 \kern-.59em {\rm l}})  &  =\mbox{1 \kern-.59em {\rm l}}\otimes
1+\Lambda^{-1}\otimes\mbox{1 \kern-.59em {\rm l}}+qc^{-1}\lambda g_{ij}%
\,\xi^{i}\mathcal{L}^{j}{}_{k}\otimes\xi^{k},\nonumber\\
\bar{\Delta}(\mbox{1 \kern-.59em {\rm l}})  &
=\mbox{1 \kern-.59em {\rm l}}\otimes1+\Lambda\otimes
\mbox{1 \kern-.59em {\rm l}}+q^{-1}c^{-1}\lambda g_{ij}\,\xi^{i}%
\mathcal{\bar{L}}^{j}{}_{k}\otimes\xi^{k}. \label{Copr1Eu}%
\end{align}
Written out explicitly the non-classical contribution to the coproduct
$\Delta(\mbox{1 \kern-.59em {\rm l}})$ reads as
\begin{align}
qc^{-1}\lambda g_{ij}\,\xi^{i}\mathcal{L}^{j}{}_{k}\otimes\xi^{k}=  &
-\,q^{-1}c^{-1}\lambda\Lambda^{-1/2}\;\big[-q^{2}\lambda^{2}\xi^{1}K_{1}%
^{1/2}K_{2}^{1/2}L_{1}^{+}L_{2}^{+}\otimes\xi^{1}\nonumber\\
&  -\,q\lambda\xi^{1}K_{1}^{-1/2}K_{2}^{1/2}L_{2}^{+}\otimes\xi^{2}%
-q\lambda\xi^{1}K_{1}^{1/2}K_{2}^{-1/2}L_{1}^{+}\otimes\xi^{3}\nonumber\\
&  +\,\xi^{1}K_{1}^{-1/2}K_{2}^{-1/2}L_{1}^{+}L_{2}^{+}\otimes\xi^{4}%
+q^{2}\lambda\xi^{2}K_{1}^{1/2}K_{2}^{1/2}L_{2}^{+}\otimes\xi^{1}\nonumber\\
&  -\,q\xi^{2}K_{1}^{1/2}K_{2}^{-1/2}L_{1}^{+}L_{2}^{+}\otimes\xi^{3}%
-q^{2}\lambda\xi^{3}K_{1}^{1/2}K_{2}^{1/2}L_{1}^{+}\otimes\xi^{1}\nonumber\\
&  -\,q\xi^{3}K_{1}^{-1/2}K_{2}^{1/2}\otimes\xi^{2}+q^{2}\xi^{4}K_{1}%
^{1/2}K_{2}^{1/2}\otimes\xi^{1}\big]. \label{deltaeinseu41}%
\end{align}
The corresponding expression for the conjugated coproduct can be obtained from
(\ref{deltaeinseu41}) by applying the substitutions
\begin{equation}
\xi^{1}\leftrightarrow\xi^{4},\quad\xi^{2}\leftrightarrow\xi^{3},\quad
L_{1}^{+}\rightarrow L_{1}^{-},\quad L_{2}^{+}\rightarrow L_{2}^{-},\quad
q\rightarrow q^{-1}.
\end{equation}
Notice that the elements $L_{i}^{\pm}$, $K_{i}$, $i=1$,$2$, generate the
quantum algebra $U_{q}(so(4))$ and the scaling operator $\Lambda$ obeys the
commutation relations
\begin{equation}
\Lambda\mbox{1 \kern-.59em {\rm l}}=q^{-4}%
\,\mbox{1 \kern-.59em {\rm l}}\Lambda,\qquad\Lambda^{1/2}\xi^{k}=-q^{-1}\,\xi
^{k}\Lambda^{1/2},\quad k=1,\ldots,4.
\end{equation}
For the antipodes and counits of $\mbox{1 \kern-.59em {\rm l}}$ we have
\begin{align}
S(\mbox{1 \kern-.59em {\rm l}})    =-q^4\Lambda
\mbox{1 \kern-.59em {\rm l}},\qquad &
\bar{S}(\mbox{1 \kern-.59em {\rm l}})  =-q^{-4}\Lambda^{-1}%
\mbox{1 \kern-.59em {\rm l}},\nonumber\\[0.08in]
\epsilon(\mbox{1 \kern-.59em {\rm l}})  &  =\bar{\epsilon}%
(\mbox{1 \kern-.59em {\rm l}})=0.
\end{align}

\subsection{Clifford algebra of $q$-defor\-med Min\-kow\-ski space}

The Clifford algebra to $q$-deformed Min\-kow\-ski space is again determined
by the requirements
\begin{equation}
(P_{S})^{\mu\nu}{}_{\rho\sigma}\,\xi^{\rho}\xi^{\sigma}=0,\qquad(P_{0}%
)^{\mu\nu}{}_{\rho\sigma}\,\xi^{\rho}\xi^{\sigma}=c\eta^{\mu\nu}%
\mbox{1 \kern-.59em {\rm l}}, \label{ReqClifMin}%
\end{equation}
where the projectors as well as the quantum metric $\eta^{\mu\nu}$ now refer
to $q$-deformed Min\-kow\-ski space. Using the projector decomposition of the
$\hat{R}$-matrix of $q$-deformed Min\-kow\-ski space (see Ref. \cite{LWW97},
for example) the above conditions can be combined to give the following
equivalent\ relations:%
\begin{align}
\xi^{\mu}\xi^{\nu}  &  =-\hat{R}^{\mu\nu}{}_{\rho\sigma}\,\xi^{\rho}%
\xi^{\sigma}+cq\lambda_{+}\,\eta^{\mu\nu}%
\mbox{1 \kern-.59em {\rm l}},\label{Cliffmink1}\\
\xi^{\mu}\xi^{\nu}  &  =-(\hat{R}^{-1})^{\mu\nu}{}_{\rho\sigma}\,\xi^{\rho}%
\xi^{\sigma}+cq^{-1}\lambda_{+}\,\eta^{\mu\nu}\mbox{1 \kern-.59em {\rm l}}.
\label{Cliffmink2}%
\end{align}
From (\ref{Cliffmink1}) as well as (\ref{Cliffmink2}) we found as independent
commutation relations%
\begin{align}
\xi^{A}\xi^{A}  &  =0,\qquad A\in\{+,-\},\nonumber\\
\xi^{3}\xi^{\pm}  &  =-q^{\mp2}\xi^{\pm}\xi^{3},\nonumber\\
\xi^{3}\xi^{3}  &  =\lambda\xi^{+}\xi^{-}+cq^{2}%
\mbox{1 \kern-.59em {\rm l}},\nonumber\\
\xi^{0}\xi^{\pm}  &  =-\xi^{\pm}\xi^{0}\pm q^{\mp1}\lambda\xi^{\pm}\xi
^{3},\nonumber\\
\xi^{0}\xi^{3}  &  =-\xi^{3}\xi^{0}+\lambda\xi^{+}\xi^{-}+qc\lambda
\mbox{1 \kern-.59em {\rm l}},\nonumber\\
\xi^{0}\xi^{0}  &  =-c\mbox{1 \kern-.59em {\rm l}},\nonumber\\
\xi^{-}\xi^{+}  &  =-\xi^{+}\xi^{-}-c\lambda_{+}\mbox{1 \kern-.59em {\rm l}}.
\label{Cliffmink123}%
\end{align}
It should also be mentioned that (\ref{Cliffmink123}) contains the relations
of the Clifford algebra to three-dimensional $q$-deformed Euclidean space.
Indeed, dropping all relations with $\xi^{0}$ leaves us with the relations in
(\ref{Cliffordeu3}).

In Sec.\thinspace\ref{gammaminkkap} we are going to introduce $\gamma
$-matrices for $q$-deformed Min\-kow\-ski space. The relations obtained from
(\ref{Cliffmink123}) for the choice $c=-\lambda_{+}^{-1}$ are fulfilled by
these $\gamma$-matrices.

In dealing with monomials of the $\xi^{\mu}$ it is sometimes convenient to
bring them into a normal ordered form. In this respect, the following formulae
derived from the relations in (\ref{Cliffmink123}) could prove useful:
\begin{align}
(\xi^{0})^{2n}  &  =(-1)^{n}c^{n}\mbox{1 \kern-.59em {\rm l}},\quad(\xi
^{0})^{2n-1}=(-1)^{n-1}c^{n-1}\xi^{0},\nonumber\\
(\xi^{3})^{2n}  &  =q^{2n}c^{n}\mbox{1 \kern-.59em {\rm l}}+q^{-2(n-1)}%
c^{n-1}\lambda\lbrack\lbrack n]]_{q^{4}}\xi^{+}\xi^{-},\nonumber\\
(\xi^{3})^{2n-1}  &  =q^{2n}c^{n}\xi^{3}-q^{-2n}c^{n-1}\lambda\lbrack\lbrack
n]]_{q^{4}}\xi^{+}\xi^{3}\xi^{-},\nonumber\\
(\xi^{0})^{2n-1}(\xi^{3})^{2n-1}=  &  \;(-1)^{n}(cq)^{2n}\big[-q\lambda
\mbox{1 \kern-.59em {\rm l}}+c^{-1}\xi^{3}\xi^{0}\nonumber\\
&  -c^{-1}\lambda\xi^{+}\xi^{-}+c^{-2}q^{-4n}\lambda\lbrack\lbrack n]]_{q^{4}%
}\xi^{+}\xi^{3}\xi^{0}\xi^{-}\big]. \label{Cliffmonmink}%
\end{align}

To make the Clifford algebra of $q$-deformed Minkowski space into a Hopf
algebra we proceed\ in very much the same way as was done for the Euclidean
spaces. Again, coproduct, antipode, and counit for\ the generators $\xi^{\mu}$
are of the same form as for the quantum space coordinates $X^{\mu}$. For
$\mbox{1 \kern-.59em {\rm l}}$ we have to take
\begin{align}
\Delta(\mbox{1 \kern-.59em {\rm l}})  &  =\mbox{1 \kern-.59em {\rm l}}\otimes
\mbox{1 \kern-.59em {\rm l}}+\Lambda^{-1}\otimes
\mbox{1 \kern-.59em {\rm l}}+q^{-2}c^{-1}\lambda\lambda_{+}^{-1}\eta_{\mu\nu
}\,\xi^{\mu}{\mathcal{L}}^{\nu}{}_{\kappa}\otimes\xi^{\kappa},\nonumber\\
\bar{\Delta}(\mbox{1 \kern-.59em {\rm l}})  &
=\mbox{1 \kern-.59em {\rm l}}\otimes\mbox{1 \kern-.59em {\rm l}}+\Lambda
\otimes\mbox{1 \kern-.59em {\rm l}}-q^{2}c^{-1}\lambda\lambda_{+}^{-1}%
\eta_{\mu\nu}\,\xi^{\mu}{\mathcal{\bar{L}}}^{\nu}{}_{\kappa}\otimes\xi
^{\kappa}, \label{CopSymMin}%
\end{align}
and 
\begin{align}
S(\mbox{1 \kern-.59em {\rm l}}) =-q^{-4}\Lambda
\mbox{1 \kern-.59em {\rm l}}, &\qquad  
\bar{S}(\mbox{1 \kern-.59em {\rm l}}) =-q^4\Lambda^{-1}%
\mbox{1 \kern-.59em {\rm l}}.
\end{align}
The non-classical contributions in (\ref{CopSymMin}) can be expressed in terms
of the generators of the $q$-deformed Lorentz algebra \cite{SWZ91, OSWZ92},
i.e.
\begin{align}
&  q^{-2}c^{-1}\lambda\lambda_{+}^{-1}\eta_{\mu\nu}\,\xi^{\mu}{\mathcal{L}%
}^{\nu}{}_{\kappa}\otimes\xi^{\kappa}=\nonumber\\
&  =-q^{-5/2}c^{-1}\lambda\lambda_{+}^{-1}\Lambda^{-1/2}\;\big[\,q^{5/2}%
\lambda^{2}\xi^{+}(\tau^{3})^{-1/2}T^{-}S^{1}\otimes\xi^{+}\\
&  \hspace{0.19in}-\lambda\lambda_{+}^{-1/2}\xi^{+}T^{-}\tau^{1}\otimes\xi
^{3}+q^{3}\lambda\lambda_{+}^{-1/2}\xi^{+}S^{1}\otimes\xi^{3}\nonumber\\
&  \hspace{0.19in}+\lambda\lambda_{+}^{-1/2}\xi^{+}T^{-}\tau^{1}\otimes\xi
^{0}+q\lambda\lambda_{+}^{-1/2}\xi^{+}S^{1}\otimes\xi^{0}\nonumber\\
&  \hspace{0.19in}-q^{5/2}\xi^{+}(\tau^{3})^{1/2}\tau^{1}\otimes\xi
^{-}-q\lambda\lambda_{+}^{-1/2}\xi^{3}(\tau^{3})^{-1/2}S^{1}\otimes\xi
^{+}\nonumber\\
&  \hspace{0.19in}+q^{2}\lambda\lambda_{+}^{-1/2}\xi^{3}(\tau^{3})^{-1/2}%
T^{-}\sigma^{2}\otimes\xi^{+}+q^{1/2}\lambda_{+}^{-1}\xi^{3}\tau^{1}\otimes
\xi^{3}\nonumber\\
&  \hspace{0.19in}+q^{5/2}\lambda_{+}^{-1}\xi^{3}\sigma^{2}\otimes\xi
^{3}-q^{3/2}\lambda^{2}\lambda_{+}^{-1}\xi^{3}T^{-}T^{2}\otimes\xi
^{3}\nonumber\\
&  \hspace{0.19in}-q^{1/2}\lambda_{+}^{-1}\xi^{3}\tau^{1}\otimes\xi
^{0}+q^{1/2}\lambda_{+}^{-1}\xi^{3}\sigma^{2}\otimes\xi^{0}\nonumber\\
&  \hspace{0.19in}+q^{3/2}\lambda^{2}\lambda_{+}^{-1}\xi^{3}T^{-}T^{2}%
\otimes\xi^{0}-q^{4}\lambda\lambda_{+}^{-1/2}\xi^{3}(\tau^{3})^{1/2}%
T^{2}\otimes\xi^{-}\nonumber\\
&  \hspace{0.19in}-q^{2}\lambda\lambda_{+}^{-1/2}\xi^{0}(\tau^{3})^{-1/2}%
T^{-}\sigma^{2}\otimes\xi^{+}-q^{3}\lambda_{+}^{-1/2}\xi^{0}(\tau^{3}%
)^{-1/2}S^{1}\otimes\xi^{+}\nonumber\\
&  \hspace{0.19in}-q^{5/2}\lambda_{+}^{-1}\xi^{0}\sigma^{2}\otimes\xi
^{3}+q^{5/2}\lambda_{+}^{-1}\xi^{0}\tau^{1}\otimes\xi^{3}\nonumber\\
&  \hspace{0.19in}+q^{3/2}\lambda^{2}\lambda_{+}^{-1}\xi^{0}T^{-}T^{2}%
\otimes\xi^{3}-q^{1/2}\lambda_{+}^{-1}\xi^{0}\sigma^{2}\otimes\xi
^{0}\nonumber\\
&  \hspace{0.19in}-q^{5/2}\lambda_{+}^{-1}\xi^{0}\tau^{1}\otimes\xi
^{0}-q^{3/2}\lambda^{2}\lambda_{+}^{-1}\xi^{0}T^{-}T^{2}\otimes\xi
^{0}\nonumber\\
&  \hspace{0.19in}+q^{4}\lambda^{\text{2}}\lambda_{+}^{-1/2}\xi
^{0}(\tau^{3})^{1/2}T^{2}\otimes\xi^{-}-q^{1/2}\xi^{-}(\tau^{3})^{-1/2}%
\sigma^{2}\otimes\xi^{+}\nonumber\\
&  \hspace{0.19in}+q^{2}\lambda\lambda_{+}^{-1/2}\xi^{-}T^{2}\otimes\xi
^{3}-q^{2}\lambda\lambda_{+}^{-1/2}\xi^{-}T^{2}\otimes\xi^{0}\big],
\end{align}
and, similarly,
\begin{align}
&  -q^{2}c^{-1}\lambda\lambda_{+}^{-1}\eta_{\mu\nu}\,\xi^{\mu}{\mathcal{\bar
{L}}}^{\nu}{}_{\kappa}\otimes\xi^{\kappa}=\nonumber\\
&  =c^{-1}q^{5/2}\lambda\lambda_{+}^{-1}\Lambda^{1/2}\;\big[\,\lambda
\lambda_{+}^{-1/2}\xi^{+}(\tau^{3})^{-1/2}S^{1}\otimes\xi^{3}\nonumber\\
&  \hspace{0.19in}-\lambda\lambda_{+}^{-1/2}\xi^{+}(\tau^{3})^{-1/2}%
S^{1}\otimes\xi^{0}-q^{-1/2}\xi^{+}\tau^{1}\otimes\xi^{-}\nonumber\\
&  \hspace{0.19in}-q^{-2}\lambda\lambda_{+}^{-1/2}\xi^{3}S^{1}\otimes\xi
^{+}+q^{-5/2}\lambda_{+}^{-1}\xi^{3}(\tau^{3})^{1/2}\tau^{1}\otimes\xi
^{3}\nonumber\\
&  \hspace{0.19in}+q^{-1/2}\lambda_{+}^{-1}\xi^{3}(\tau^{3})^{-1/2}\sigma
^{2}\otimes\xi^{3}+q^{-3/2}\lambda^{2}\lambda_{+}^{-1}\xi^{3}(\tau^{3}%
)^{-1/2}T^{+}S^{1}\otimes\xi^{3}\nonumber\\
&  \hspace{0.19in}-q^{-1/2}\lambda_{+}^{-1}\xi^{3}(\tau^{3})^{-1/2}\sigma
^{2}\otimes\xi^{0}+q^{-1/2}\lambda_{+}^{-1}\xi^{3}(\tau^{3})^{1/2}\tau
^{1}\otimes\xi^{0}\nonumber\\
&  \hspace{0.19in}-q^{-3/2}\lambda^{2}\lambda_{+}^{-1}\xi^{3}(\tau^{3}%
)^{-1/2}T^{+}S^{1}\otimes\xi^{0}-\lambda\lambda_{+}^{-1/2}\xi^{3}T^{+}\tau
^{1}\otimes\xi^{-}\nonumber\\
&  \hspace{0.19in}-q\lambda\lambda_{+}^{-1/2}\xi^{3}T^{2}\otimes\xi^{-}%
+q^{-2}\lambda\lambda_{+}^{-1/2}\xi^{0}S^{1}\otimes\xi^{+}\nonumber\\
&  \hspace{0.19in}-q^{-5/2}\lambda_{+}^{-1}\xi^{0}(\tau^{3})^{1/2}\tau
^{1}\otimes\xi^{3}+q^{-5/2}\lambda_{+}^{-1}\xi^{0}(\tau^{3})^{-1/2}\sigma
^{2}\otimes\xi^{3}\nonumber\\
&  \hspace{0.19in}-q^{-3/2}\lambda^{2}\lambda_{+}^{-1}\xi^{0}(\tau^{3}%
)^{-1/2}T^{+}S^{1}\otimes\xi^{3}-q^{-5/2}\lambda_{+}^{-1}\xi^{0}(\tau
^{3})^{-1/2}\sigma^{2}\otimes\xi^{0}\nonumber\\
&  \hspace{0.19in}-q^{-1/2}\lambda_{+}^{-1}\xi^{0}(\tau^{3})^{1/2}\tau
^{1}\otimes\xi^{0}+q^{-3/2}\lambda^{2}\lambda_{+}^{-1}\xi^{0}(\tau^{3}%
)^{-1/2}T^{+}S^{1}\otimes\xi^{0}\nonumber\\
&  \hspace{0.19in}+\lambda\lambda_{+}^{-1/2}\xi^{0}T^{+}\tau^{1}\otimes\xi
^{-}-q^{-1}\lambda\lambda_{+}^{-1/2}\xi^{0}T^{2}\otimes\xi^{-}\nonumber\\
&  \hspace{0.19in}-q^{-5/2}\xi^{-}\sigma^{2}\otimes\xi^{+}+q^{-2}%
\lambda\lambda_{+}^{-1/2}\xi^{-}(\tau^{3})^{-1/2}T^{+}\sigma^{2}\otimes\xi
^{3}\nonumber\\
&  \hspace{0.19in}+q^{-1}\lambda\lambda_{+}^{-1/2}\xi^{-}(\tau^{3})^{1/2}%
T^{2}\otimes\xi^{3}-q^{-2}\lambda\lambda_{+}^{-1/2}\xi^{-}(\tau^{3}%
)^{-1/2}T^{+}\sigma^{2}\otimes\xi^{0}\nonumber\\
&  \hspace{0.19in}+q\lambda\lambda_{+}^{-1/2}\xi^{-}(\tau^{3})^{1/2}%
T^{2}\otimes\xi^{0}-q^{3/2}\lambda^{2}\xi^{-}T^{+}T^{2}\otimes\xi^{-}\big].
\end{align}
Notice that the scaling operator $\Lambda$ now has to satisfy the relations
\begin{equation}
\Lambda\mbox{1 \kern-.59em {\rm l}}=q^{4}%
\,\mbox{1 \kern-.59em {\rm l}}\Lambda,\qquad\Lambda^{1/2}\xi^{\mu}=-q\,\xi^{\mu
}\Lambda^{1/2},\quad\mu\in\{+,3,0,-\}.
\end{equation}

For later purpose, we wish to formulate a Clifford algebra for generators with
lower indices. This Clifford algebra shall be determined by
\begin{equation}
(P_{S})^{\mu\nu}{}_{\rho l}\,\tilde{\xi}_{\mu}\tilde{\xi}_{\nu}=0,\qquad
(P_{0})^{\mu\nu}{}_{\rho l}\,\tilde{\xi}_{\mu}\tilde{\xi}_{\nu}=\tilde{c}%
\eta_{\mu\nu}\mbox{1 \kern-.59em {\rm l}}.
\end{equation}
The above conditions are equivalent to
\begin{align}
\tilde{\xi}_{\mu}\tilde{\xi}_{\nu}  &  =-\hat{R}^{\rho l}{}_{\mu\nu}%
\,\tilde{\xi}_{\rho}\tilde{\xi}_{l}+\tilde{c}q\lambda_{+}\eta_{\mu\nu
}\mbox{1 \kern-.59em {\rm l}},\label{cliffminkinv1}\\
\tilde{\xi}_{\mu}\tilde{\xi}_{\nu}  &  =-(\hat{R}^{-1})^{\rho l}{}_{\mu\nu
}\,\tilde{\xi}_{\rho}\tilde{\xi}_{l}+\tilde{c}q^{-1}\lambda_{+}\eta_{\mu\nu
}\mbox{1 \kern-.59em {\rm l}}. \label{cliffminkinv2}%
\end{align}
More explicitly, we have%
\begin{align}
\tilde{\xi}_{A}\tilde{\xi}_{A}  &  =0,\qquad A\in\{+,-\},\nonumber\\
\tilde{\xi}_{3}\tilde{\xi}_{\pm}  &  =-q^{\mp2}\tilde{\xi}_{\pm}\tilde{\xi
}_{3},\nonumber\\
\tilde{\xi}_{3}\tilde{\xi}_{3}  &  =\lambda\tilde{\xi}_{+}\tilde{\xi}%
_{-}+q^{2}\tilde{c}\mbox{1 \kern-.59em {\rm l}},\nonumber\\
\tilde{\xi}_{0}\tilde{\xi}_{\pm}  &  =-\tilde{\xi}_{\pm}\tilde{\xi}_{0}\mp
q^{\mp1}\lambda\tilde{\xi}_{\pm}\tilde{\xi}_{3},\nonumber\\
\tilde{\xi}_{0}\tilde{\xi}_{3}  &  =-\tilde{\xi}_{3}\tilde{\xi}_{0}%
-\lambda\tilde{\xi}_{+}\tilde{\xi}_{-}-q\tilde{c}\lambda
\mbox{1 \kern-.59em {\rm l}},\nonumber\\
\tilde{\xi}_{0}\tilde{\xi}_{0}  &  =-\tilde{c}%
\mbox{1 \kern-.59em {\rm l}},\nonumber\\
\tilde{\xi}_{-}\tilde{\xi}_{+}  &  =-\tilde{\xi}_{+}\tilde{\xi}_{-}-\tilde
{c}\lambda_{+}\mbox{1 \kern-.59em {\rm l}}. \label{CliffCov}%
\end{align}

In Sec.\thinspace\ref{gammaminkkap} we are going to deal with so-called
'\textit{inverse}' $\gamma$-matrices. Taking for $\tilde{c}$ the same value as
for $c$ in (\ref{Cliffmink123}), the 'inverse' $\gamma$-matrices will then
provide a representation of the Clifford algebra (\ref{CliffCov}), as can be
checked by inserting the expressions for the 'inverse' $\gamma$-matrices into
the relations in (\ref{CliffCov}).

Finally, it should be mentioned that the relations in (\ref{CliffCov}) are
obtained by applying the replacements $\xi^{\mu}\rightarrow\tilde{\xi}_{\mu}$
to the formulae in (\ref{Cliffmink123}). However, the reader should have in
mind that (\ref{CliffCov}) describes an algebra being different from
(\ref{Cliffmink123}). Especially, we do\textit{ not }have $\tilde{\xi}_{\mu
}=\xi_{\mu}(\equiv\eta_{\mu\nu}\xi^{\nu})$. It should also be obvious that the
identities in (\ref{Cliffmonmink}) remain valid under the substitutions
$\xi^{\mu}\rightarrow\tilde{\xi}_{\mu}$.

\section{$q$-Deformed Dirac and Spin matrices}

In this section we introduce $q$-analogs of $\gamma$-matrices and
four-dimensional spin matrices ($\Sigma$-matrices). Moreover, we discuss some
of their features and present useful formulae including these matrices. Our
reasonings refer to four-dimensional $q$-deformed Euclidean space and
$q$-deformed Min\-kow\-ski space.

\subsection{$q$-Deformed Eu\-cli\-de\-an space in four dimensions
\label{gammaeu4kap}}

\subsubsection{The $q$-deformed Dirac matrices\label{gammaeu4kapSec1}}

In part I of the article we treated Pauli matrices for $q$-deformed Euclidean
space in four dimensions. These Pauli matrices can be combined to give the
$\gamma$-matrices of four-dimensional $q$-deformed Euclidean space (see also
Refs. \cite{Blo03, AKR96, WSSW90, Song92, Pod97}):%
\begin{equation}
(\gamma^{\mu})_{a}{}^{b}\equiv\left(
\begin{array}
[c]{cc}%
0 & (\sigma^{\mu})_{\alpha}{}^{\dot{\beta}}\\
(\bar{\sigma}^{\mu})_{\dot{\alpha}}{}^{{\beta}} & 0
\end{array}
\right)  ,\quad\mu=1,\ldots,4. \label{defgammaeu4N}%
\end{equation}
In addition to this, we define%
\begin{equation}
(\gamma^{\mu})_{ab}\equiv\left(
\begin{array}
[c]{cc}%
0 & (\sigma^{\mu})_{\alpha\dot{\beta}}\\
(\tilde{\sigma}^{\mu})_{\dot{\alpha}\beta} & 0
\end{array}
\right)  ,\quad(\gamma^{\mu})^{ab}\equiv\left(
\begin{array}
[c]{cc}%
0 & (\sigma^{\mu})^{\alpha\dot{\beta}}\\
(\tilde{\sigma}^{\mu})^{\dot{\alpha}\beta} & 0
\end{array}
\right)  . \label{defgammaeu42}%
\end{equation}
Recalling the index properties of the Pauli matrices $\sigma^{\mu}$ and
$\tilde{\sigma}^{\mu}$ [cf. part I], the above convention enables us to
formulate the following rules for raising and lowering bispinor indices of
$\gamma$-matrices:%
\begin{equation}
(\gamma^{\mu})^{ab}=(D_{L}^{\text{T}})^{aa^{\prime}}(\gamma^{\mu})_{a^{\prime
}}{}^{b},\quad(\gamma^{\mu})_{ab}=(\gamma^{\mu})_{a}{}^{b^{\prime}%
}(D_{R})_{b^{\prime}b},
\end{equation}
where
\begin{equation}
(D_{L})^{aa^{\prime}}=\left(
\begin{array}
[c]{cc}%
\varepsilon^{\alpha\alpha^{\prime}} & 0\\
0 & \varepsilon^{\dot{\alpha}\dot{\alpha}^{\prime}}%
\end{array}
\right)  ,\quad(D_{R})_{b^{\prime}b}=\left(
\begin{array}
[c]{cc}%
\varepsilon_{\beta^{\prime}\beta} & 0\\
0 & \varepsilon_{\dot{\beta}^{\prime}\dot{\beta}}%
\end{array}
\right)  . \label{DMat}%
\end{equation}
Notice that $\varepsilon$ denotes the $q$-deformed spinor metric [see also
App.\thinspace\ref{AppA3}].

Inserting the expressions for the Pauli matrices $\sigma^{\mu}$ and
$\tilde{\sigma}^{\mu}$ the $\gamma$-matrices in (\ref{defgammaeu4N}) become%
\begin{align}
(\gamma^{1})_{\alpha}{}^{\beta} &  =-q^{-1/2}\left(
\begin{array}
[c]{cccc}%
0 & 0 & 0 & 1\\
0 & 0 & 0 & 0\\
0 & 1 & 0 & 0\\
0 & 0 & 0 & 0
\end{array}
\right)  , & (\gamma^{2})_{\alpha}{}^{\beta} &  =\left(
\begin{array}
[c]{cccc}%
0 & 0 & 0 & 0\\
0 & 0 & 0 & -q^{-1/2}\\
q^{1/2} & 0 & 0 & 0\\
0 & 0 & 0 & 0
\end{array}
\right)  ,\nonumber\\[0.04in]
(\gamma^{3})_{\alpha}{}^{\beta} &  =\left(
\begin{array}
[c]{cccc}%
0 & 0 & q^{1/2} & 0\\
0 & 0 & 0 & 0\\
0 & 0 & 0 & 0\\
0 & -q^{-1/2} & 0 & 0
\end{array}
\right)  , & (\gamma^{4})_{\alpha}{}^{\beta} &  =-q^{1/2}\left(
\begin{array}
[c]{cccc}%
0 & 0 & 0 & 0\\
0 & 0 & 1 & 0\\
0 & 0 & 0 & 0\\
1 & 0 & 0 & 0
\end{array}
\right)  .
\end{align}
As can be verified by inserting, these $\gamma$-matrices fulfill\ the
relations in (\ref{Cliffeu4}), i.e. they establish a representation of the
Clifford algebra to four-dimensional $q$-deformed Euclidean space.
Interestingly, the matrices
\begin{equation}
(\gamma^{\mu})_{a}{}^{b}=\left(
\begin{array}
[c]{cc}%
0 & d(\sigma^{\mu})_{\alpha}{}^{\dot{\beta}}\\
d^{-1}(\tilde{\sigma}^{\mu})_{\dot{\alpha}}{}^{{\beta}} & 0
\end{array}
\right)  ,\quad d\in\mathbb{C},
\end{equation}
give equally good representations of the Clifford generators. The reason for
this lies in the fact that $d$ cancels out against $d^{-1}$ if we multiply two
$\gamma$-matrices.

The Pauli matrices which went into the calculation of the above $\gamma
$-matrices refer to symmetrized spinors. As we know from part I there are also
Pauli matrices for antisymmetrized spinors. However, a little thought can show
that the following results do not depend on the choice one makes in
(\ref{defgammaeu4N}) for the\ Pauli matrices .

We can also find a $q$-analog of the matrix $\gamma^{5}$. With the help of the
totally antisymmetric tensor of four-dimensional $q$-deformed Euclidean space
(see App. \ref{AppA3}) we find:%
\begin{equation}
\gamma^{5}\equiv\frac{q}{[[2]]_{q^{2}}[[3]]_{q^{2}}}\,\varepsilon_{\mu\nu
\rho\sigma}\,\gamma^{\sigma}\gamma^{\rho}\gamma^{\nu}\gamma^{\mu}=\left(
\begin{array}
[c]{cccc}%
-1 & 0 & 0 & 0\\
0 & -1 & 0 & 0\\
0 & 0 & 1 & 0\\
0 & 0 & 0 & 1
\end{array}
\right)  . \label{Gamma5Exp}%
\end{equation}
Notice that the numerical factor was chosen in such a way that the result
agrees with its undeformed counterpart. One readily checks that $\gamma^{5}$
anticommutes with all $\gamma$-matrices:%
\begin{equation}
{\{}\gamma^{5},\gamma^{\mu}{\}}=0,\quad\mu=1,\ldots,4.
\end{equation}
A short look at (\ref{Gamma5Exp}) tells us that $\gamma^{5}$ has unit square,
i.e. $(\gamma^{5})^{2}=1$.

In analogy to the undeformed case there are several possibilities to build a
basis of the Clifford algebra out of the $\gamma$-matrices (from now on the
$\gamma$-matrices play the role of the Clifford generators). One basis is made
up\ of antisymmetrized products of $\gamma$-matrices:
\begin{align}
&  \mbox{1 \kern-.59em {\rm l}},\;\gamma^{1},\;\gamma^{2},\;\gamma
^{3},\;\gamma^{4},\;\gamma^{\lbrack1}\gamma^{2]_{q}},\;\gamma^{\lbrack1}%
\gamma^{3]_{q}},\;\gamma^{\lbrack1}\gamma^{4]_{q}},\;\gamma^{\lbrack2}%
\gamma^{3]_{q}},\;\gamma^{\lbrack2}\gamma^{4]_{q}},\nonumber\\
&  \gamma^{\lbrack3}\gamma^{4]_{q}},\;\gamma^{\lbrack1}\gamma^{2}%
\gamma^{3]_{q}},\;\gamma^{\lbrack1}\gamma^{2}\gamma^{4]_{q}},\;\gamma
^{\lbrack1}\gamma^{3}\gamma^{4]_{q}},\;\gamma^{\lbrack2}\gamma^{3}%
\gamma^{4]_{q}},\;\gamma^{\lbrack1}\gamma^{2}\gamma^{3}\gamma^{4]_{q}}.
\label{gammabasis1eu4}%
\end{align}
In what follows we will denote these elements in the order of their appearance
by $\Gamma^{\text{e}}\equiv\mbox{1 \kern-.59em {\rm l}},$ $\Gamma^{1},$
$\Gamma^{2},...,$ $\Gamma^{1234}$. Notice that the square brackets in
(\ref{gammabasis1eu4}) indicate antisymmetrization of vector indices in the
sense of $q$-deformation. In this manner, we have
\begin{align}
\gamma^{\lbrack\mu}\gamma^{\nu]_{q}}  &  \equiv(P_{A})^{\mu\nu}{}_{\mu
^{\prime}\nu^{\prime}}\gamma^{\mu^{\prime}}\gamma^{\nu},\nonumber\\
\gamma^{\lbrack\mu}\gamma^{\nu}\gamma^{\rho]_{q}}  &  \equiv-\frac{1}%
{2q^{2}[[3]]_{q^{2}}}\,\varepsilon^{\mu\nu\rho\sigma}\varepsilon_{\mu^{\prime
}\nu^{\prime}\rho^{\prime}\sigma}\,\gamma^{\rho^{\prime}}\gamma^{\nu^{\prime}%
}\gamma^{\mu^{\prime}},\nonumber\\
\gamma^{\lbrack1}\gamma^{2}\gamma^{3}\gamma^{4]_{q}}  &  \equiv2[[2]]_{q^{2}%
}^{2}[[3]]_{q^{2}}\,\varepsilon^{1234}\,\varepsilon_{\mu\nu\rho\sigma}%
\,\gamma^{\sigma}\gamma^{\rho}\gamma^{\nu}\gamma^{\mu}\nonumber\\
&  =2q^{-1}[[2]]_{q^{2}}^{3}[[3]]_{q^{2}}^{2}\;\gamma^{5}, \label{prgammaeu4}%
\end{align}
where $P_{A}$ stands for the antisymmetrizer of the four-dimensional
$q$-deformed Euclidean space.

Let us recall that in part I of the article we introduced a quantum trace for
Pauli matrices. This notion can be extended to products of $\gamma$-matrices
as follows:%
\begin{equation}
\mbox{Tr}_{q}\;\gamma^{\mu}\gamma^{\nu}\equiv\mbox{Tr}\;D_{L}^{\text{T}%
}\,\gamma^{\mu}\gamma^{\nu}D_{R}. \label{trgammaeu4}%
\end{equation}
where usual matrix multiplication is understood on the right-hand side of
(\ref{trgammaeu4}).

The quantum trace can be viewed as a sesquilinear form. Thus, it seems
reasonable to seek matrices $\Gamma_{l,A},$ $A\in\{$e, $1$, $2$, \ldots,
$1234\}$, with the property
\begin{equation}
\mbox{Tr}_{q}\,(\Gamma_{l,A}\,\Gamma^{B})=\delta^{B}{}_{A}.
\end{equation}
This requirement defines\ a left-dual basis of the Clifford algebra with basis
elements $\Gamma^{B}$. For the elements $\Gamma_{l,A}$ of the left-dual basis
we found the expressions%
\begin{align}
\Gamma_{l,\text{e}}  &  =-\frac{1}{2}\lambda_{+}^{-1}\;\Gamma^{\text{e}}%
,\quad\Gamma_{l,1}=-\frac{1}{2}q^{-1}\;\Gamma^{4},\quad\Gamma_{l,2}=-\frac
{1}{2}\;\Gamma^{3},\nonumber\\
\Gamma_{l,3}  &  =-\frac{1}{2}\;\Gamma^{2},\quad\Gamma_{l,4}=-\frac{1}%
{2}q\;\Gamma^{1},\label{basisnew0}\\
& \nonumber\\
\Gamma_{l,12}  &  =-\;\Gamma^{34},\quad\Gamma_{l,13}=-\;\Gamma^{24}%
,\nonumber\\
\Gamma_{l,14}  &  =-q^{-2}(q^{4}+1)\lambda_{+}^{-1}\;\Gamma^{14}%
+\lambda\lambda_{+}^{-1}\;\Gamma^{23},\nonumber\\
\Gamma_{l,23}  &  =-2\lambda_{+}^{-1}\;\Gamma^{23}+\lambda\lambda_{+}%
^{-1}\;\Gamma^{14},\nonumber\\
\Gamma_{l,24}  &  =-q^{2}\;\Gamma^{13},\quad\Gamma_{34}=-q^{2}\Gamma^{12},\\
& \nonumber\\
\Gamma_{l,123}  &  =-2q\;\Gamma^{234},\quad\Gamma_{l,124}=-2q^{2}%
\;\Gamma^{134},\quad\Gamma_{l,134}=-2q^{2}\;\Gamma^{124},\nonumber\\
\Gamma_{l,234}  &  =-2q^{3}\;\Gamma^{123},\quad\Gamma_{l,1234}=-\frac{1}%
{8}q^{3}[[2]]_{q^{2}}^{-7}[[3]]_{q^{2}}^{-4}\;\Gamma^{1234}. \label{basisnew}%
\end{align}
Solving these equations for the $\Gamma^{B}$, $B\in\{$e, $1$, $2$, \ldots,
$1234\},$ shows that the $\Gamma_{l,A}$ indeed establish a basis of the
Clifford algebra to four-dimensional $q$-deformed Euclidean space.

Alternatively, we can ask for matrices $\Gamma_{r,A}$, $A\in\{$e, $1$, $2$,
\ldots, $1234\}$, subject to
\begin{equation}
\mbox{Tr}_{q}\,(\Gamma^{B}\,\Gamma_{r,A})=\delta^{B}{}_{A}.
\end{equation}
These matrices lead to a right-dual basis. We found that the $\Gamma_{r,A}$
are identical to the $\Gamma_{l,A}$ with the exception of the following
matrices:
\begin{gather}
\Gamma_{r,1}=q^{2}\;\Gamma_{l,1},\quad\Gamma_{r,4}=q^{-2}\;\Gamma
_{l,4},\nonumber\\[0.1in]
\Gamma_{r,12}=q^{2}\;\Gamma_{l,12},\quad\Gamma_{r,13}=q^{2}\;\Gamma
_{l,13},\Gamma_{r,24}=q^{-2}\;\Gamma_{l,24},\quad\Gamma_{r,34}=q^{-2}%
\;\Gamma_{l,34},\nonumber\\[0.1in]
\Gamma_{r,123}=q^{2}\;\Gamma_{l,123},\quad\Gamma_{r,234}=q^{-2}\;\Gamma
_{l,234}.
\end{gather}
The reader may wonder why the two dual bases are not the same. This has to do
with the fact that the quantum metric is not symmetrical.

As we know, the 16 matrices $\Gamma^{A}$, $A\in\{$e, $1$, $2$, \ldots,
$1234\}$, constitute a basis. Thus, we should be able to expand an arbitrary
$4\times4$ matrix $M$ in terms of these matrices, i.e.
\begin{equation}
M=\sum\nolimits_{A}c_{A}\Gamma^{A}.\label{ExpM}%
\end{equation}
In analogy to the undeformed case, each expansion coefficient is given by a
quantum trace over the product of $M$ with a dual matrix:
\begin{equation}
c_{B}=\mbox{Tr}_{q}(M\,\Gamma_{r,B})=\mbox{Tr}_{q}(\Gamma_{l,B}\,M).
\end{equation}
Plugging these expressions into the expansion of $M$ we can finally\ read off
as completeness relations
\begin{align}
\delta_{a}{}^{b}\delta_{c}{}^{d} &  =\sum_{B}(D_{R})_{cm}(D_{L}^{\text{T}%
})^{mk}(\Gamma_{l,B})_{k}{}^{b}(\Gamma^{B})_{a}{}^{d}\nonumber\\
&  =\sum_{B}(\Gamma^{B})_{a}{}^{d}(\Gamma_{r,B})_{c}{}^{k}(D_{R})_{km}%
(D_{L}^{\text{T}})^{mb}%
\end{align}
with the matrices $D_{R}$ and $D_{L}$ from (\ref{DMat}). To write this result
in a more convenient form we introduce
\begin{align}
(\Gamma_{l,B}^{\text{c}})_{a}{}^{b} &  \equiv(D_{R})_{am}(D_{L}^{\text{T}%
})^{mk}(\Gamma_{l,B})_{k}{}^{b},\nonumber\\
(\Gamma_{l,B}^{\text{c}})_{a}{}^{b} &  \equiv(\Gamma_{r,B})_{a}{}^{k}%
(D_{R})_{km}(D_{L}^{\text{T}})^{mb}.
\end{align}
With these new matrices at hand we finally arrive at
\begin{equation}
\sum_{B}(\Gamma_{l,B}^{\text{c}})_{c}{}^{b}\,(\Gamma^{B})_{a}{}^{d}=\delta
_{a}{}^{b}\delta_{c}{}^{d},\qquad\sum_{B}(\Gamma^{B})_{a}{}^{d}\,(\Gamma
_{r,B}^{\text{c}})_{c}{}^{b}=\delta_{a}{}^{b}\delta_{c}{}^{d}.\label{CompRel}%
\end{equation}
For the sake of completeness we would like to express the matrices
$\Gamma_{l,B}^{c}$ in terms of the $\Gamma^{B}$:%
\begin{align}
\Gamma_{\text{e}}^{\text{c}}= &  \;\frac{1}{2}(q^{4}+1)q^{-2}\lambda_{+}%
^{-2}\;\Gamma^{\text{e}}-\frac{1}{2}\lambda\lambda_{+}^{-1}\;\Gamma
^{14},\nonumber\\
\Gamma_{1}^{\text{c}}= &  \;\frac{1}{2}\;\Gamma^{4},\quad\Gamma_{2}^{\text{c}%
}=\frac{1}{4}\lambda_{+}\;\Gamma^{3}+\frac{q}{2}\lambda\;\Gamma^{134}%
,\nonumber\\
\Gamma_{3}^{\text{c}}= &  \;\frac{1}{4}\lambda_{+}\;\Gamma^{2}+\frac{q}%
{2}\lambda\;\Gamma^{124},\quad\Gamma_{4}^{\text{c}}=\frac{1}{2}\;\Gamma
^{1},\label{GaC1}\\
& \nonumber\\
\Gamma_{12}^{\text{c}}= &  \;q\;\Gamma^{34},\quad\Gamma_{13}^{\text{c}%
}=q\;\Gamma^{24},\nonumber\\
\Gamma_{14}^{\text{c}}= &  \;2q^{-2}(q^{4}+1)\lambda_{+}^{-2}\;\Gamma
^{14}-2\lambda\lambda_{+}^{-2}\;\Gamma^{23}-2\lambda\lambda_{+}^{-1}%
\;\Gamma^{e},\nonumber\\
&  +\frac{1}{4}q^{3}\lambda^{2}[[2]]_{q^{2}}^{-5}[[2]]_{q^{2}}^{-2}%
\;\Gamma^{1234},\nonumber\\
\Gamma_{23}^{\text{c}}= &  \;4\lambda_{+}^{-2}\;\Gamma^{23}-2\lambda
\lambda_{+}^{-2}\;\Gamma^{14}-\frac{1}{2}q^{3}\lambda\lbrack\lbrack2]]_{q^{2}%
}^{-5}[[3]]_{q^{2}}^{-2}\;\Gamma^{1234},\nonumber\\
\Gamma_{24}^{\text{c}}= &  \;q\;\Gamma^{13},\quad\Gamma_{34}^{\text{c}%
}=q\;\Gamma^{12},\label{GaC2}\\
& \nonumber\\
\Gamma_{123}^{\text{c}}= &  \;2q^{2}\;\Gamma^{234},\quad\Gamma_{124}%
^{\text{c}}=q^{2}\lambda_{+}\;\Gamma^{134}+\frac{1}{2}q\lambda\;\Gamma
^{3},\nonumber\\
\Gamma_{134}^{\text{c}}= &  \;q^{2}\lambda_{+}\;\Gamma^{124}+\frac{1}%
{2}q\lambda\;\Gamma^{2},\quad\Gamma_{234}^{\text{c}}=2q^{2}\;\Gamma^{123},\\
& \nonumber\\
\Gamma_{1234}^{\text{c}}= &  \;\frac{1}{8}q^{2}(q^{4}+1)[[2]]_{q^{2}}%
^{-8}[[3]]_{q^{2}}^{-4}\;\Gamma^{1234}\nonumber\\
&  +\frac{1}{2}q^{3}\lambda\lbrack\lbrack2]]_{q^{2}}^{-5}[[3]]_{q^{2}}%
^{-2}(\frac{1}{2}\lambda\;\Gamma^{14}-\Gamma^{23}).
\end{align}
Similar formulae hold for the matrices $\Gamma_{r,B}^{c}$

Now, we turn to a second basis of the Clifford algebra under consideration. It
consists of the elements
\begin{gather}
\mbox{1 \kern-.59em {\rm l}},\;\gamma^{1},\;\gamma^{2},\;\gamma^{3}%
,\;\gamma^{4},\;\gamma^{5},\gamma^{5}\gamma^{1},\;\gamma^{5}\gamma
^{2},\;\gamma^{5}\gamma^{3},\;\gamma^{5}\gamma^{4},\nonumber\\
\gamma^{\lbrack1}\gamma^{2]_{q}},\;\gamma^{\lbrack1}\gamma^{3]_{q}}%
,\;\gamma^{\lbrack1}\gamma^{4]_{q}},\;\gamma^{\lbrack2}\gamma^{3]_{q}%
},\;\gamma^{\lbrack2}\gamma^{4]_{q}},\gamma^{\lbrack3}\gamma^{4]_{q}}.
\label{gammabasis2eu4}%
\end{gather}
Instead of antisymmetrized products of three or four $\gamma$-matrices the new
basis contains the matrices $\Gamma^{5\mu}\equiv\gamma^{5}\gamma^{\mu}$,
$\mu=1,\ldots,4$, and $\Gamma^{5}\equiv\gamma^{5}$. That (\ref{gammabasis2eu4}%
) is indeed equivalent to our previous basis should become obvious from\ the
relations
\begin{align}
\gamma^{\lbrack\mu}\gamma^{\nu}\gamma^{\rho]_{q}}  &  =\frac{1}{2}%
q^{-1}[[2]]_{q^{2}}^{-1}[[3]]_{q^{2}}^{-1}\,\varepsilon^{\mu\nu\rho\sigma
}g_{\sigma\lambda}\,\gamma^{\lambda}\gamma^{5},\nonumber\\
\gamma^{5}\gamma^{\rho}  &  =\lambda_{+}\,\varepsilon^{\rho\lambda\nu\mu
}\,g_{\mu\mu^{\prime}}g_{\nu\nu^{\prime}}g_{\lambda\lambda^{\prime}}%
\,\gamma^{\lbrack\mu^{\prime}}\gamma^{\nu^{\prime}}\gamma^{\lambda^{\prime
}]_{q}}\nonumber\\
&  =-[[3]]_{q^{2}}^{-1}\,\varepsilon_{\sigma\lambda\rho\nu}\,g^{\sigma\mu
}\,\gamma^{\nu}\gamma^{\rho}\gamma^{\lambda}.
\end{align}

The above reasonings about left- and right-duals carry over to the new basis.
There are only some minor changes concerning the matrices $\Gamma^{5}$ and
$\Gamma^{5\mu}$. In this respect, we have to notice that their left-duals read
as%
\begin{align}
\Gamma_{l,5}  &  =-\frac{1}{2}\lambda_{+}^{-1}\Gamma^{5}=-\frac{1}{2}%
\lambda_{+}^{-1}\gamma^{5},\nonumber\\
\Gamma_{l,51}  &  =\frac{1}{2}q^{-1}\;\Gamma^{54},\quad\Gamma_{l,52}=\frac
{1}{2}\;\Gamma^{52},\nonumber\\
\Gamma_{l,53}  &  =\frac{1}{2}\;\Gamma^{52}\quad\Gamma_{l,54}=\frac{1}%
{2}q\;\Gamma^{51},
\end{align}
and the corresponding right-duals are of the same form apart from
\begin{equation}
\Gamma_{r,51}=q^{2}\Gamma_{51},\quad\Gamma_{r,54}=q^{-2}\Gamma_{54}.
\end{equation}

It should be emphasized that the formulae (\ref{ExpM})-(\ref{CompRel}) also
apply to the basis in (\ref{gammabasis2eu4}) if we take into account that
\begin{align}
\Gamma_{2}^{\text{c}}  &  =\frac{1}{4}\lambda_{+}\;\Gamma^{3}+\frac{1}%
{4}\lambda\;\Gamma^{53},\quad\Gamma_{3}^{\text{c}}=\frac{1}{4}\lambda
_{+}-\frac{1}{4}\lambda\;\Gamma^{52},\\[0.12in]
\Gamma_{14}^{\text{c}}  &  =2q^{-2}(q^{4}+1)\lambda_{+}^{-2}\;\Gamma
^{14}-2\lambda\lambda_{+}^{-2}\;\Gamma^{23}+\frac{1}{2}\lambda^{2}\lambda
_{+}^{-2}\;\Gamma^{5}-\frac{1}{2}\lambda\lambda_{+}^{-1}\;\Gamma^{\text{e}%
},\nonumber\\
\Gamma_{23}^{\text{c}}  &  =4\lambda_{+}^{-2}\;\Gamma^{23}-2\lambda\lambda
_{+}^{-1}\;\Gamma^{14}-\lambda^{2}\lambda_{+}^{-2}\;\Gamma^{5},\\[0.12in]
\Gamma_{51}^{\text{c}}  &  =-\frac{1}{2}\;\Gamma^{54},\quad\Gamma
_{52}^{\text{c}}=-\frac{1}{4}\lambda_{+}\;\Gamma^{53}-\frac{1}{4}%
\lambda\;\Gamma^{3},\nonumber\\
\Gamma_{53}^{\text{c}}  &  =-\frac{1}{4}\lambda_{+}\;\Gamma^{52}-\frac{1}%
{4}\lambda\;\Gamma^{2},\quad\Gamma_{54}^{\text{c}}=-\frac{1}{2}\;\Gamma
^{51},\\[0.12in]
\Gamma_{5}^{\text{c}}  &  =\frac{1}{2}q^{-2}(q^{4}+1)\;\Gamma^{5}+\lambda
^{2}\lambda_{+}^{-2}(\frac{1}{2}\;\Gamma^{14}-\Gamma^{23}),
\end{align}
whereas the expressions for the remaining matrices $\Gamma_{A}^{\text{c}}$
again follow from (\ref{GaC1}) and (\ref{GaC2}).

\subsubsection{Relations concerning $q$-deformed Dirac\ matrices
\label{RelConEuc}}

Now, we would like to present $q$-analogs of well-known relations including
$\gamma$-matrices. Let us first note that in the remainder of this section
multiplication of $\gamma$-matrices is always understood as
\begin{equation}
\gamma^{\mu}\gamma^{\nu}=(\gamma^{\mu}\gamma^{\nu})_{a}{}^{b}=(\gamma^{\mu
})_{a}{}^{a^{\prime}}(\gamma^{\nu})_{a^{\prime}}{}^{b}.
\end{equation}

Computations in quantum field theory often require to evaluate traces over
products of $\gamma$-matrices. We found as $q$-analogs of well-known trace
relations:%
\begin{align}
\mbox{Tr}_{q}\,\gamma^{\mu}  &  =0,\quad\mbox{Tr}_{q}\,\gamma^{\mu}\gamma
^{\nu}\gamma^{\rho}=0,\quad\mbox{Tr}_{q}\,\gamma^{\mu}\gamma^{\nu}=2\eta
^{\mu\nu},\nonumber\\
\mbox{Tr}_{q}\,\gamma^{\mu}\gamma^{\nu}\gamma^{\rho}\gamma^{\sigma}  &
=-q\,g^{\mu\nu}g^{\rho\sigma}+\;g^{\mu\kappa}\hat{R}^{\nu\rho}{}%
_{\kappa\lambda}\,g^{\lambda\sigma}-q^{-1}\,g^{\mu\sigma}g^{\nu\rho
}\nonumber\\
&  =-q^{-1}\,g^{\mu\nu}g^{\rho\sigma}+g^{\mu\kappa}(\hat{R}^{-1})^{\nu\rho}%
{}_{\kappa\lambda}\,g^{\lambda\sigma}-q\,g^{\mu\sigma}g^{\nu\rho},
\end{align}
for all $\mu$, $\nu$, $\rho$, $\sigma\in{\{}1,2,3,4{\}}$. Furthermore, we
have
\begin{gather}
\mbox{Tr}_{q}\,\gamma^{5}=\mbox{Tr}_{q}\,\gamma^{\mu}\gamma^{5}=\mbox{Tr}_{q}%
\,\gamma^{\mu}\gamma^{\nu}\gamma^{5}=\mbox{Tr}_{q}\,\gamma^{\mu}\gamma^{\nu
}\gamma^{\rho}\gamma^{5}=0,\nonumber\\
\mbox{Tr}_{q}\,\gamma^{\mu}\gamma^{\nu}\gamma^{\rho}\gamma^{\sigma}\gamma
^{5}=-q^{-2}\;\varepsilon^{\mu\nu\rho\sigma},
\end{gather}
and
\begin{align}
&  \mbox{Tr}_{q}\;\gamma^{5}\gamma^{\mu_{1}}\gamma^{\mu_{2}}\gamma^{\mu_{3}%
}\gamma^{\mu_{4}}\gamma^{\mu_{5}}\gamma^{\mu_{6}}=\nonumber\\
&  \qquad\qquad=\,-2q^{-1}\big[k_{1}\,\eta^{\mu_{1}\mu_{2}}\,\varepsilon
^{\mu_{3}\mu_{4}\mu_{5}\mu_{6}}+k_{2}\,\eta^{\mu_{1}\mu_{3}^{\prime}}\hat
{R}^{\mu_{2}\mu_{3}}{}_{\mu_{3}^{\prime}\mu_{2}^{\prime}}\,\varepsilon
^{\mu_{2}^{\prime}\mu_{4}\mu_{5}\mu_{6}}\nonumber\\
&  \qquad\qquad\hspace{0.19in}\,+k_{3}\,\eta^{\mu_{2}\mu_{3}}\,\varepsilon
^{\mu_{1}\mu_{4}\mu_{5}\mu_{6}}+k_{4}\,\eta^{\mu_{4}\mu_{5}}\,\varepsilon
^{\mu_{1}\mu_{2}\mu_{3}\mu_{6}}\nonumber\\
&  \qquad\qquad\hspace{0.19in}\,+k_{5}\,\eta^{\mu_{4}\mu_{6}^{\prime}}%
\,\hat{R}^{\mu_{5}\mu_{6}}{}_{\mu_{6}^{\prime}\mu_{5}^{\prime}}\,\varepsilon
^{\mu_{1}\mu_{2}\mu_{3}\mu_{5}^{\prime}}+k_{6}\,\eta^{\mu_{5}\mu_{6}%
}\,\varepsilon^{\mu_{1}\mu_{2}\mu_{3}\mu_{4}}\big], \label{splangeu4}%
\end{align}
where
\begin{equation}
k_{1}=1,\;k_{2}=-q^{-1},\;k_{3}=q^{-2},\;k_{4}=1,\;k_{5}=-q^{-1}%
,\;k_{6}=q^{-2}.
\end{equation}
If we replace $\hat{R}$ with $\hat{R}^{-1}$ the coefficients in
(\ref{splangeu4}) instead take on the values
\begin{equation}
k_{1}=q^{-2},\;k_{2}=-q^{-1},\;k_{3}=1,\;k_{4}=q^{-2},\;k_{5}=-q^{-1}%
,\;k_{6}=1.
\end{equation}

It is also worth recording here that quantum traces over products of $\gamma
$-matrices obey a kind of cyclicity:
\begin{align}
\mbox{Tr}_{q}\;\gamma^{\mu}\gamma^{\nu}  &  =q^{3}\hat{R}^{\mu\nu}{}%
_{\nu^{\prime}\mu^{\prime}}\,\mbox{Tr}_{q}\;\gamma^{\nu^{\prime}}\gamma
^{\mu^{\prime}},\nonumber\\
\mbox{Tr}_{q}\;\gamma^{\mu}\gamma^{\nu}\gamma^{\rho}\gamma^{\sigma}  &
=q^{3}\hat{R}^{\mu\nu}{}_{\nu^{\prime}\mu^{\prime}}\hat{R}^{\mu^{\prime}\rho
}{}_{\rho^{\prime}\mu^{\prime\prime}}\hat{R}^{\mu^{\prime\prime}\sigma}%
{}_{\sigma^{\prime}\mu^{\prime\prime\prime}}\,\mbox{Tr}_{q}\;\gamma
^{\nu^{\prime}}\gamma^{\rho^{\prime}}\gamma^{\sigma^{\prime}}\gamma
^{\mu^{\prime\prime\prime}}. \label{cycspeu4}%
\end{align}
Applying the substitutions $\hat{R}\rightarrow\hat{R}^{-1},$ $q\rightarrow
q^{-1}$ to the expressions in (\ref{cycspeu4}) leads us to alternative formulations.

Next, we turn to formulae with $\gamma$-matrices being contracted by the
quantum metric:
\begin{align}
g_{\mu\nu}\,\gamma^{\mu}\gamma^{\nu}  &  =\lambda_{+}%
\mbox{1 \kern-.59em {\rm l}},\nonumber\\
g_{\rho\mu}\,\hat{R}^{\mu\nu}{}_{\nu^{\prime}\mu^{\prime}}\,\gamma^{\rho
}\gamma^{\nu^{\prime}}\gamma^{\mu^{\prime}}  &  =-\gamma^{\nu},\nonumber\\
g_{\sigma\mu}\,\hat{R}^{\mu\nu}{}_{\nu^{\prime}\mu^{\prime}}\hat{R}%
^{\mu^{\prime}\rho}{}_{\rho^{\prime}\mu^{\prime\prime}}\,\gamma^{\sigma}%
\gamma^{\nu^{\prime}}\gamma^{\rho^{\prime}}\gamma^{\mu^{\prime\prime}}  &
=g^{\nu\rho}\mbox{1 \kern-.59em {\rm l}}. \label{gammarel1}%
\end{align}
These equations remain valid if we perform the substitution $\hat
{R}\rightarrow\hat{R}^{-1}$. Additionally, we have the little bit lengthy
relation
\begin{align}
&  g_{\delta\mu}\,\hat{R}^{\mu\nu}{}_{\nu^{\prime}\mu^{\prime}}\hat{R}%
^{\mu^{\prime}\rho}{}_{\rho^{\prime}\mu^{\prime\prime}}\hat{R}^{\mu
^{\prime\prime}\sigma}{}_{\sigma^{\prime}\mu^{\prime\prime\prime}}%
\,\gamma^{\delta}\gamma^{\nu^{\prime}}\gamma^{\rho^{\prime}}\gamma
^{\sigma^{\prime}}\gamma^{\mu^{\prime\prime\prime}}\nonumber\\
&  =-q^{3}\;\hat{R}^{\rho^{\prime\prime}\sigma^{\prime\prime}}{}%
_{\sigma^{\prime}\rho^{\prime}}\hat{R}^{\nu^{\prime\prime}\sigma}{}%
_{\sigma^{\prime\prime}\nu^{\prime}}\hat{R}^{\nu\rho}{}_{\rho^{\prime\prime
}\nu^{\prime\prime}}\gamma^{\sigma^{\prime}}\gamma^{\rho^{\prime}}\gamma
^{\nu^{\prime}}, \label{gammarel2N}%
\end{align}
which can be recognized as $q$-analog of $\gamma^{\mu}\gamma^{\nu}\gamma
^{\rho}\gamma^{\sigma}\gamma_{\mu}=-2\gamma^{\sigma}\gamma^{\rho}\gamma^{\nu
}.$ Once again, the replacements $\hat{R}\rightarrow\hat{R}^{-1}%
,q^{-3}\rightarrow-q^{3}$ lead us to an equally good relation.

We close with a decomposition formula for products of three $\gamma$-matrices.
Concretely, we have
\begin{align}
\gamma^{\mu}\gamma^{\nu}\gamma^{\rho}  &  =\gamma^{\lbrack\mu}\gamma^{\nu
}\gamma^{\rho]_{q}}+\frac{q}{2}\eta^{\mu\nu}\gamma^{\rho}\nonumber\\
&  -\;\frac{1}{2}\eta^{\mu\lambda^{\prime}}\hat{R}^{\nu\rho}{}_{\lambda
^{\prime}\nu^{\prime}}\,\gamma^{\nu^{\prime}}+\frac{1}{2q}\gamma^{\mu}%
\eta^{\nu\rho}.
\end{align}
There is also a version of this formula with $\hat{R}$ and $q$ being replaced
by $\hat{R}^{-1}$ and $q^{-1},$ respectively. The above result tells us that
products of three $\gamma$-matrices split into a totally antisymmetric part
and a linear combination of $\gamma$-matrices.

\subsubsection{The $q$-deformed spin\ matrices and some more relations
\label{SpinMatKapEuc}}

It is now our aim to discuss $q$-analogs of the four-dimensional $\Sigma
$-matrices. In analogy to the undeformed case they are defined by%
\begin{equation}
(\Sigma^{\mu\nu})_{a}{}^{b}=(P_{A})^{\mu\nu}{}_{\mu^{\prime}\nu^{\prime}%
}\,(\gamma^{\mu})_{a}{}^{a^{\prime}}(\gamma^{\nu})_{a^{\prime}}{}^{b}.
\label{defsigmaeu4}%
\end{equation}
The four-dimensional $\Sigma$-matrices are made up of their two-dimensional
counterparts $\sigma^{\mu\nu}$ and $\tilde{\sigma}^{\mu\nu}$ according to
\begin{equation}
(\Sigma^{\mu\nu})_{a}{}^{b}=\left(
\begin{array}
[c]{cc}%
(\sigma^{\mu\nu})_{\alpha}{}^{\beta} & 0\\
0 & (\tilde{\sigma}^{\mu\nu})_{\dot{\alpha}}{}^{{\dot{\beta}}}%
\end{array}
\right)  . \label{defgammaeu4}%
\end{equation}

Now, we come to useful relations involving $\gamma$- and $\Sigma$-matrices.
Making suitable ansaetze we could prove the following identities:%
\begin{align}
\gamma^{\mu}\gamma^{\nu}=  &  \;\Sigma^{\mu\nu}+\lambda_{+}^{-1}g^{\mu\nu
},\nonumber\\
\gamma^{5}\gamma^{\mu}\gamma^{\nu}=  &  \;\lambda_{+}^{-1}g^{\mu\nu}\gamma
^{5}+q^{-1}\lambda_{+}^{-1}\varepsilon^{\mu\nu\rho\sigma}g_{\rho\rho^{\prime}%
}g_{\sigma\sigma^{\prime}}\Sigma^{\sigma^{\prime}\rho^{\prime}},\\[0.16in]
\gamma^{5}\Sigma^{\mu\nu}=  &  \;q^{-1}\lambda_{+}^{-1}\varepsilon^{\mu\nu
\rho\sigma}g_{\rho\rho^{\prime}}g_{\sigma\sigma^{\prime}}\Sigma^{\sigma
^{\prime}\rho^{\prime}},\nonumber\\
\gamma^{\rho}\Sigma^{\mu\nu}=  &  \;\frac{1}{2}q^{-2}\varepsilon^{\rho\mu
\nu\lambda}g_{\lambda\delta}\,\gamma^{5}\gamma^{\delta}+\frac{1}{2}\lambda
_{+}(P_{A})^{\mu\nu}{}_{\mu^{\prime}\nu^{\prime}}\,g^{\rho\mu^{\prime}}%
\gamma^{\nu^{\prime}},\nonumber\\
\Sigma^{\mu\nu}\gamma^{\rho}=  &  \;\frac{1}{2}q^{-2}\varepsilon^{\mu\nu
\rho\lambda}\,g_{\lambda\delta}\,\gamma^{5}\gamma^{\delta}+\frac{1}{2}%
\lambda_{+}(P_{A})^{\mu\nu}{}_{\mu^{\prime}\nu^{\prime}}\,g^{\nu^{\prime}\rho
}\gamma^{\mu^{\prime}},\\[0.16in]
\gamma^{5}\gamma^{\rho}\Sigma^{\mu\nu}=  &  \;\frac{1}{2}q^{-2}\varepsilon
^{\rho\mu\nu\lambda}g_{\lambda\delta}\,\gamma^{\delta}+\frac{1}{2}\lambda
_{+}(P_{A})^{\mu\nu}{}_{\mu^{\prime}\nu^{\prime}}\,g^{\rho\mu^{\prime}}%
\gamma^{5}\gamma^{\nu^{\prime}},\nonumber\\
\Sigma^{\mu\nu}\gamma^{5}\gamma^{\rho}=  &  \;\frac{1}{2}q^{-2}\varepsilon
^{\mu\nu\rho\lambda}g_{\lambda\delta}\,\gamma^{\delta}+\frac{1}{2}\lambda
_{+}(P_{A})^{\mu\nu}{}_{\mu^{\prime}\nu^{\prime}}\,g^{\nu^{\prime}\rho}%
\gamma^{5}\gamma^{\mu^{\prime}},\\[0.16in]
\Sigma^{\mu\nu}\Sigma^{\kappa\lambda}=  &  \;\frac{1}{2}(P_{A})^{\mu\nu}%
{}_{\mu^{\prime}\nu^{\prime}}(P_{A})^{\kappa\lambda}{}_{\kappa^{\prime}%
\lambda^{\prime}}\,g^{\mu^{\prime}\lambda^{\prime}}g^{\nu^{\prime}%
\kappa^{\prime}}+\frac{1}{2}q^{-2}\lambda_{+}^{-1}\varepsilon^{\mu\nu
\kappa\lambda}\gamma^{5}\nonumber\\
&  +\lambda_{+}(P_{A})^{\mu\nu}{}_{\mu^{\prime}\nu^{\prime}}(P_{A}%
)^{\kappa\lambda}{}_{\kappa^{\prime}\lambda^{\prime}}\,g^{\nu^{\prime}%
\kappa^{\prime}}\Sigma^{\mu^{\prime}\lambda^{\prime}}.
\end{align}

For quantum traces over $\Sigma$-matrices we have%
\begin{align}
\mbox{Tr}_{q}\;\Sigma^{\mu\nu}  &  =0,\nonumber\\
\mbox{Tr}_{q}\;\Sigma^{\rho\sigma}\Sigma^{\mu\nu}  &  =q^{-1}\lambda_{+}%
(P_{A})^{\mu\nu}{}_{\mu^{\prime}\nu^{\prime}}\,\hat{R}^{\sigma\mu^{\prime}}%
{}_{\mu^{\prime\prime}\sigma^{\prime}}\,g^{\rho\mu^{\prime\prime}}%
g^{\sigma^{\prime}\nu^{\prime}},
\end{align}
where we are again allowed to apply the replacements $\hat{R}\rightarrow
\hat{R}^{-1}$, $q\rightarrow q^{-1}$.

Next we would like to present $q$-deformed analoga of some relations listed in
Ref. \cite{Core} (see p.13):
\begin{equation}
g_{\mu\mu^{\prime}}\hat{R}^{\mu^{\prime}\nu}{}_{\nu^{\prime}\mu^{\prime\prime
}}\hat{R}^{\mu^{\prime\prime}\rho}{}_{\rho^{\prime}\mu^{\prime\prime\prime}%
}\gamma^{\mu}\Sigma^{\nu^{\prime}\rho^{\prime}}\gamma^{\mu^{\prime\prime
\prime}}=0,
\end{equation}%
\begin{align}
&  \gamma^{\mu}\,\Sigma^{\nu^{\prime}\rho^{\prime}}\gamma^{\sigma^{\prime}%
}\gamma^{\mu^{\prime\prime\prime\prime}}\hat{R}^{\mu^{\prime}\nu}{}%
_{\nu^{\prime}\mu^{\prime\prime}}\hat{R}^{\mu^{\prime\prime}\rho}{}%
_{\rho^{\prime}\mu^{\prime\prime\prime}}\hat{R}^{\mu^{\prime\prime\prime
}\sigma}{}_{\sigma^{\prime}\mu^{\prime\prime\prime\prime}}g_{\mu\mu^{\prime}%
}=\nonumber\\
&  \qquad\qquad=\,q^{2}\,\hat{R}^{\nu\sigma^{\prime}}{}_{\sigma^{\prime\prime
}\nu^{\prime}}\hat{R}^{\rho\sigma}{}_{\sigma^{\prime}\rho^{\prime}}%
\gamma^{\sigma^{\prime\prime}}\Sigma^{\nu^{\prime}\rho^{\prime}}%
,\nonumber\\[0.06in]
&  \gamma^{\mu}\gamma^{\nu^{\prime}}\Sigma^{\rho^{\prime}\sigma^{\prime}%
}\gamma^{\mu^{\prime\prime\prime\prime\prime}}\hat{R}^{\mu^{\prime}\nu}{}%
_{\nu^{\prime}\mu^{\prime\prime}}\hat{R}^{\mu^{\prime\prime}\rho}{}%
_{\rho^{\prime}\mu^{\prime\prime\prime}}\hat{R}^{\mu^{\prime\prime\prime
}\sigma}{}_{\sigma^{\prime}\mu^{\prime\prime\prime\prime}}g_{\mu\mu^{\prime}%
}=\nonumber\\
&  \qquad\qquad=\,q^{2}\,\hat{R}^{\nu^{\prime}\sigma}{}_{\sigma^{\prime}%
\nu^{\prime\prime}}\hat{R}^{\nu\rho}{}_{\rho^{\prime}\nu^{\prime}}\Sigma
^{\rho^{\prime}\sigma^{\prime}}\gamma^{\nu^{\prime\prime}},
\end{align}%
\begin{align}
&  \gamma^{\mu}\gamma^{\rho_{1}^{\prime}}\gamma^{\rho_{2}^{\prime}}%
\gamma^{\rho_{3}^{\prime}}\gamma^{\rho_{4}^{\prime}}\gamma^{\nu}\,\hat{R}%
^{\mu^{\prime}\rho_{1}}{}_{\rho_{1}^{\prime}\mu^{\prime\prime}}\hat{R}%
^{\mu^{\prime\prime}\rho_{2}}{}_{\rho_{2}^{\prime}\mu^{\prime\prime\prime}%
}\hat{R}^{\mu^{\prime\prime\prime}\rho_{3}}{}_{\rho_{3}^{\prime}\mu
^{\prime\prime\prime\prime}}\hat{R}^{\mu^{\prime\prime\prime\prime}\rho_{4}}%
{}_{\rho_{4}^{\prime}\nu}\,g_{\mu\mu^{\prime}}=\nonumber\\
&  \qquad\qquad=\,q^{2}\,\gamma^{\rho_{4}^{\prime\prime\prime}}\gamma
^{\rho_{1}^{\prime}}\gamma^{\rho_{2}^{\prime}}\gamma^{\rho_{3}^{\prime}}%
\,\hat{R}^{\rho_{1}\rho_{4}^{\prime\prime}}{}_{\rho_{4}^{\prime\prime\prime
}\rho_{1}^{\prime}}\hat{R}^{\rho_{2}\rho_{4}^{\prime}}{}_{\rho_{4}%
^{\prime\prime}\rho_{2}^{\prime}}\hat{R}^{\rho_{3}\rho_{4}}{}_{\rho
_{4}^{\prime}\rho_{3}^{\prime}}\nonumber\\
&  \qquad\qquad\hspace{0.19in}+\,q^{4}\,\gamma^{\rho_{2}^{\prime}}\gamma
^{\rho_{3}^{\prime}}\gamma^{\rho_{4}^{\prime}}\gamma^{\rho_{1}^{\prime
\prime\prime}}\hat{R}^{\rho_{1}\rho_{2}}{}_{\rho_{2}^{\prime}\rho_{1}^{\prime
}}\hat{R}^{\rho_{1}^{\prime}\rho_{3}}{}_{\rho_{3}^{\prime}\rho_{1}%
^{\prime\prime}}\hat{R}^{\rho_{1}^{\prime\prime}\rho_{4}}{}_{\rho_{4}^{\prime
}\rho_{1}^{\prime\prime\prime}}\nonumber\\
&  \qquad\qquad=\,q^{2}\,\gamma^{\rho_{3}^{\prime\prime}}\gamma^{\rho
_{2}^{\prime\prime}}\gamma^{\rho_{1}^{\prime\prime}}\gamma^{\rho_{4}}\,\hat
{R}^{\rho_{2}^{\prime}\rho_{3}^{\prime}}{}_{\rho_{3}^{\prime\prime}\rho
_{2}^{\prime\prime}}\hat{R}^{\rho_{1}\rho_{2}}{}_{\rho_{2}^{\prime}\rho
_{1}^{\prime}}\hat{R}^{\rho_{1}^{\prime}\rho_{3}}{}_{\rho_{3}^{\prime}\rho
_{1}^{\prime\prime}}\nonumber\\
&  \qquad\qquad\hspace{0.19in}+\,q^{4}\,\gamma^{\rho_{1}}\gamma^{\rho
_{4}^{\prime\prime}}\gamma^{\rho_{3}^{\prime\prime}}\gamma^{\rho_{2}%
^{\prime\prime}}\,\hat{R}^{\rho_{2}\rho_{4}^{\prime}}{}_{\rho_{4}%
^{\prime\prime}\rho_{2}^{\prime}}\hat{R}^{\rho_{2}^{\prime}\rho_{3}^{\prime}%
}{}_{\rho_{3}^{\prime\prime}\rho_{2}^{\prime\prime}}\hat{R}^{\rho_{3}\rho_{4}%
}{}_{\rho_{4}^{\prime}\rho_{3}^{\prime}},
\end{align}
where the substitutions $\hat{R}\rightarrow\hat{R}^{-1},$ $q\rightarrow
q^{-1}$ lead to further relations.

It should also be mentioned that a product of a $\Sigma$-matrix with
$\gamma^{5}$ can be written as linear combination of $\Sigma$-matrices. For
the independent $\Sigma$-matrices we have
\begin{align}
\Sigma^{12}\gamma^{5}  &  =\Sigma^{12},\quad\Sigma^{13}\gamma^{5}=-\Sigma
^{13},\nonumber\\
\Sigma^{14}\gamma^{5}  &  =2\lambda_{+}^{-1}\Sigma^{23}-\lambda\lambda
_{+}^{-1}\Sigma^{14},\nonumber\\
\Sigma^{23}\gamma^{5}  &  =2\lambda_{+}^{-1}\Sigma^{14}+\lambda\lambda
_{+}^{-1}\Sigma^{23},\nonumber\\
\Sigma^{24}\gamma^{5}  &  =-\Sigma^{24},\quad\Sigma^{34}\gamma^{5}=\Sigma
^{34}.
\end{align}

As in the undeformed case the scalar $1$, the vector $\gamma^{\mu}$, the
pseudoscalar $\gamma^{5}$, the 'axial' vector $\gamma^{5}\gamma^{\mu}$, and
the antisymmetric tensor $\Sigma^{\mu\nu}$ can be taken as 16 independent
$4\times4$ matrices. In this manner, we have the following expansion for an
arbitrary $4\times4$ matrix $A$:%
\begin{equation}
A=a_{0}\mbox{1 \kern-.59em {\rm l}}+a_{5}\gamma^{5}+v_{\mu}\gamma^{\mu}%
+a_{\mu}\gamma^{5}\gamma^{\mu}+T_{\mu\nu}\Sigma^{\mu\nu},
\end{equation}
where%
\begin{align}
a_{0}  &  =\frac{1}{2}\lambda_{+}^{-1}\,\mbox{Tr}_{q}(A),\nonumber\\
a_{5}  &  =\frac{1}{2}\lambda_{+}^{-1}\,\mbox{Tr}_{q}(\gamma^{5}A),\nonumber\\
v_{\mu}  &  =\frac{1}{2}g_{\mu\nu}\,\mbox{Tr}_{q}(\gamma^{\nu}A),\nonumber\\
a_{\mu}  &  =-\frac{1}{2}g_{\mu\nu}\,\mbox{Tr}_{q}(\gamma^{5}\gamma^{\nu
}A),\nonumber\\
T_{\mu\nu}  &  =\lambda_{+}^{-1}g_{\mu\mu^{\prime}}g_{\nu\nu^{\prime}%
}\,\mbox{Tr}_{q}(\Sigma^{\nu^{\prime}\mu^{\prime}}A).
\end{align}
If we use instead the matrices in (\ref{gammabasis1eu4}) as basis the
expansion reads
\begin{equation}
A=a_{0}\mbox{1 \kern-.59em {\rm l}}+a_{16}\Gamma^{1234}+v_{\mu}\gamma^{\mu
}+a_{K}\Gamma^{K}+T_{\mu\nu}\Sigma^{\mu\nu},
\end{equation}
where%
\begin{align}
a_{0}  &  =\frac{1}{2}\lambda_{+}^{-1}\,\mbox{Tr}_{q}(A),\nonumber\\
a_{16}  &  =\frac{1}{8}q^{3}[[2]]_{q^{2}}^{-7}[[3]]_{q^{2}}^{-4}%
\,\mbox{Tr}_{q}(\Gamma^{1234}A),\nonumber\\
v_{\mu}  &  =\frac{1}{2}g_{\mu\nu}\,\mbox{Tr}_{q}(\gamma^{\nu}A),\nonumber\\
a_{K}  &  =-\mbox{Tr}_{q}(\Gamma_{K}A),\nonumber\\
T_{\mu\nu}  &  =\lambda_{+}^{-1}g_{\mu\mu^{\prime}}g_{\nu\nu^{\prime}%
}\,\mbox{Tr}_{q}(\Sigma^{\nu^{\prime}\mu^{\prime}}A),
\end{align}
and $K$ is an index running over $K\in{\{}123,$ $124,$ $134,$ $234{\}}$.

\subsection{$q$-Deformed Min\-kow\-ski space\label{gammaminkkap}}

In this section we deal with $\gamma$- and $\Sigma$-matrices for $q$-deformed
Min\-kow\-ski space. Our reasonings are rather similar to those for the
Euclidean case. However, there is one remarkable difference compared to the
presentation in the previous section. It stems from the fact that we now have
to distinguish between $\gamma$-matrices and their so-called 'inverse'
variants, which should not be confused with inverse matrices in the sense of
matrix multiplication.

\subsubsection{The $q$-deformed Dirac matrices\label{KapGammaDef}}

\noindent\textbf{Definition of }$\gamma$\textbf{-matrices:\ }The $\gamma
$-matrices are defined as
\begin{equation}
(\gamma^{\mu})_{a}{}^{b}\equiv\left(
\begin{array}
[c]{cc}%
0 & (\sigma^{\mu})_{\alpha}{}^{\dot{\beta}}\\
(\bar{\sigma}^{\mu})_{\dot{\alpha}}{}^{{\beta}} & 0
\end{array}
\right)  , \label{defgamma}%
\end{equation}
where $\sigma^{\mu}$ and $\bar{\sigma}^{\mu}$ denote the Pauli matrices of
$q$-deformed Min\-kow\-ski space, as they were introduced in part I of the
article. The $\gamma$-matrices with two upper indices and those with two lower
indices are given by
\begin{equation}
(\gamma^{\mu})_{ab}\equiv\left(
\begin{array}
[c]{cc}%
0 & (\sigma^{\mu})_{\alpha\dot{\beta}}\\
(\bar{\sigma}^{\mu})_{\dot{\alpha}\beta} & 0
\end{array}
\right)  ,\quad(\gamma^{\mu})^{ab}\equiv\left(
\begin{array}
[c]{cc}%
0 & (\sigma^{\mu})^{\alpha\dot{\beta}}\\
(\bar{\sigma}^{\mu})^{\dot{\alpha}\beta} & 0
\end{array}
\right)  . \label{defgamma2}%
\end{equation}
Again, we can find matrices for raising bispinor indices of $\gamma$-matrices.
In view of the properties of Pauli matrices for $q$-deformed Min\-kow\-ski
space we should have%
\begin{equation}
(\gamma^{\mu})^{ab}=(D_{L}^{\text{T}})^{aa^{\prime}}(\gamma^{\mu})_{a^{\prime
}}{}^{b},\quad(\gamma^{\mu})_{ab}=(\gamma^{\mu})_{a}{}^{b^{\prime}%
}(D_{R})_{b^{\prime}b},
\end{equation}
with%
\begin{equation}
(D_{L})^{aa^{\prime}}=\left(
\begin{array}
[c]{cc}%
\varepsilon^{\alpha\alpha^{\prime}} & 0\\
0 & -\varepsilon^{\dot{\alpha}\dot{\alpha}^{\prime}}%
\end{array}
\right)  ,\quad(D_{R})_{bb^{\prime}}=\left(
\begin{array}
[c]{cc}%
\varepsilon_{\beta\beta^{\prime}} & 0\\
0 & -\varepsilon_{\dot{\beta}\dot{\beta}^{\prime}}%
\end{array}
\right)  .
\end{equation}

Choosing Pauli matrices for symmetrized spinors [cf. part I] the $\gamma
$-matrices in (\ref{defgamma}) explicitly read as%
\begin{gather}
(\gamma^{+})_{\alpha}{}^{\beta}=\left(
\begin{array}
[c]{cccc}%
0 & 0 & 0 & 0\\
0 & 0 & -q^{3/2} & 0\\
0 & 0 & 0 & 0\\
q^{-1/2} & 0 & 0 & 0
\end{array}
\right)  ,(\gamma^{-})_{\alpha}{}^{\beta}=\left(
\begin{array}
[c]{cccc}%
0 & 0 & 0 & q^{1/2}\\
0 & 0 & 0 & 0\\
0 & -q^{-3/2} & 0 & 0\\
0 & 0 & 0 & 0
\end{array}
\right)  ,\nonumber\\[0.04in]
(\gamma^{3})_{\alpha}{}^{\beta}=\lambda_{+}^{-1/2}\left(
\begin{array}
[c]{cccc}%
0 & 0 & -q^{2} & 0\\
0 & 0 & 0 & 1\\
1 & 0 & 0 & 0\\
0 & -q^{-2} & 0 & 0
\end{array}
\right)  ,(\gamma^{0})_{\alpha}{}^{\beta}=\lambda_{+}^{-1/2}\left(
\begin{array}
[c]{cccc}%
0 & 0 & 1 & 0\\
0 & 0 & 0 & 1\\
1 & 0 & 0 & 0\\
0 & 1 & 0 & 0
\end{array}
\right)  . \label{gammaminkN}%
\end{gather}
These $\gamma$-matrices fulfill the Clifford algebra relations in
(\ref{Cliffmink123}) as can be checked by inserting.

The matrices in (\ref{gammaminkN}) enable us to calculate a $q$-analog of the
matrix $\gamma^{5}$. It is defined as the totally antisymmetric product of
four $\gamma$-matrices. In this manner, we get
\begin{equation}
\gamma^{5}\equiv\frac{q^{5}}{[[2]]_{q^{2}}[[3]]_{q^{2}}}\,\varepsilon_{\mu
\nu\rho\sigma}\,\gamma^{\sigma}\gamma^{\rho}\gamma^{\nu}\gamma^{\mu}=\left(
\begin{array}
[c]{cccc}%
-1 & 0 & 0 & 0\\
0 & -1 & 0 & 0\\
0 & 0 & 1 & 0\\
0 & 0 & 0 & 1
\end{array}
\right)  , \label{gamma5mink}%
\end{equation}
where a normalization factor has been introduced to give the above expression
for $\gamma^{5}$ its simple form. In analogy to the undeformed case
$\gamma^{5}$ anticommutes with all $\gamma$-matrices:%
\begin{equation}
{\{}\gamma^{5},\gamma^{\mu}{\}}=0,\quad\mu\in\{+,3,0,-\}. \label{AntGam5}%
\end{equation}

\noindent\textbf{Definition of 'inverse' }$\gamma$\textbf{-matrices:\ } In
part I it was stressed that in the case of $q$-deformed Min\-kow\-ski space we
have to distinguish between Pauli matrices and their so-called 'inverse'
counterparts. This observation can be used to introduce '\textit{inverse}'
$\gamma$-matrices:%
\begin{equation}
(\gamma_{\mu}^{-1})_{a}{}^{b}\equiv\left(
\begin{array}
[c]{cc}%
0 & (\sigma_{\mu}^{-1})_{\alpha}{}^{\dot{\beta}}\\
(\bar{\sigma}_{\mu}^{-1})_{\dot{\alpha}}{}^{{\beta}} & 0
\end{array}
\right)  . \label{gammainv}%
\end{equation}
For a detailed discussion of the 'inverse' Pauli matrices $\sigma_{\mu}^{-1}$
and $\bar{\sigma}_{\mu}^{-1}$ we refer the reader to part I. Here, let us
recall that the 'inverse' Pauli matrices $\sigma_{\mu}^{-1}$ and $\bar{\sigma
}_{\mu}^{-1}$ are not inverse in the sense of matrix multiplication and the
same holds for 'inverse' $\gamma$-matrices. Finally, it should be mentioned
that bispinor indices\ of 'inverse' $\gamma$-matrices are raised according to
the rules
\begin{equation}
(\gamma_{\mu}^{-1})^{ab}=(D_{L})^{aa^{\prime}}(\gamma_{\mu
}^{-1})_{a^{\prime}}{}^{b},\quad(\gamma_{\mu}^{-1})_{ab
}=(\gamma_{\mu}^{-1})_{a}{}^{b^{\prime}}(D_{R}^{\text{T}%
})_{b^{\prime}b}.
\end{equation}
These conventions follow directly from the rules for raising and lowering
spinor indices of 'inverse' Pauli matrices.

To underpin our statement that 'inverse' $\gamma$-matrices are not inverse in
the usual sense we give the expressions:
\begin{gather}
(\gamma_{+}^{-1})_{a}{}^{b}=\left(
\begin{array}
[c]{cccc}%
0 & 0 & 0 & q^{-3/2}\\
0 & 0 & 0 & 0\\
0 & -q^{1/2} & 0 & 0\\
0 & 0 & 0 & 0
\end{array}
\right)  ,(\gamma_{-}^{-1})_{a}{}^{b}=\left(
\begin{array}
[c]{cccc}%
0 & 0 & 0 & 0\\
0 & 0 & -q^{-1/2} & 0\\
0 & 0 & 0 & 0\\
q^{3/2} & 0 & 0 & 0
\end{array}
\right)  ,\nonumber\\[0.04in]
(\gamma_{3}^{-1})_{a}{}^{b}=\lambda_{+}^{-1/2}\left(
\begin{array}
[c]{cccc}%
0 & 0 & q^{-2} & 0\\
0 & 0 & 0 & -1\\
-1 & 0 & 0 & 0\\
0 & q^{2} & 0 & 0
\end{array}
\right)  ,(\gamma_{0}^{-1})_{a}{}^{b}=\lambda_{+}^{-1/2}\left(
\begin{array}
[c]{cccc}%
0 & 0 & 1 & 0\\
0 & 0 & 0 & 1\\
1 & 0 & 0 & 0\\
0 & 1 & 0 & 0
\end{array}
\right)  .
\end{gather}
Nothing prevents us from introducing a matrix $\gamma_{5}^{-1}$. In analogy to
(\ref{gamma5mink}) we have
\begin{equation}
\gamma_{5}^{-1}\equiv-\frac{q^{5}}{[[2]]_{q^{2}}[[3]]_{q^{2}}}\,\varepsilon
^{\mu\nu\rho\sigma}\,\gamma_{\mu}^{-1}\gamma_{\nu}^{-1}\gamma_{\rho}%
^{-1}\gamma_{\sigma}^{-1}=\gamma^{5}=\left(
\begin{array}
[c]{cccc}%
-1 & 0 & 0 & 0\\
0 & -1 & 0 & 0\\
0 & 0 & 1 & 0\\
0 & 0 & 0 & 1
\end{array}
\right)  .
\end{equation}

\noindent\textbf{Left- and right-dual bases:\ }Next, we are seeking a basis
for the Clifford algebra of $q$-deformed Min\-kow\-ski space. As in the
undeformed case, such a basis can be build up from $\gamma$-matrices. One
possibility to achieve this is given by the following set of matrices:
\begin{align}
&  1,\;\gamma^{+},\;\gamma^{3},\;\gamma^{0},\;\gamma^{-},\;\gamma^{\lbrack
+}\gamma^{3]_{q}},\;\gamma^{\lbrack+}\gamma^{0]_{q}},\;\gamma^{\lbrack+}%
\gamma^{-]_{q}},\;\gamma^{\lbrack3}\gamma^{0]_{q}},\;\gamma^{\lbrack3}%
\gamma^{-]_{q}},\nonumber\\
&  \gamma^{\lbrack0}\gamma^{-]_{q}},\;\gamma^{\lbrack+}\gamma^{3}%
\gamma^{0]_{q}},\;\gamma^{\lbrack+}\gamma^{3}\gamma^{-]_{q}},\;\gamma
^{\lbrack+}\gamma^{0}\gamma^{-]_{q}},\;\gamma^{\lbrack3}\gamma^{0}%
\gamma^{-]_{q}},\;\gamma^{\lbrack+}\gamma^{3}\gamma^{0}\gamma^{-]_{q}}.
\label{gammabasis}%
\end{align}
In the order of their appearance the matrices in (\ref{gammabasis}) are
denoted by $\Gamma^{\text{e}}$, $\Gamma^{+}$,$...$, $\Gamma^{+30-}$. The
antisymmetrized products of two, three, or four $\gamma$-matrices are
calculated from the expressions:
\begin{align}
\gamma^{\lbrack\mu}\gamma^{\nu]_{q}}  &  \equiv(P_{A})^{\mu\nu}{}_{\mu
^{\prime}\nu^{\prime}}\gamma^{\mu^{\prime}}\gamma^{\nu^{\prime}},\nonumber\\
\gamma^{\lbrack\mu}\gamma^{\nu}\gamma^{\rho]_{q}}  &  \equiv\frac{q^{6}%
}{2[[3]]_{q^{2}}}\,\varepsilon^{\mu\nu\rho\sigma}\,\varepsilon_{\mu^{\prime
}\nu^{\prime}\rho^{\prime}\sigma}\,\gamma^{\rho^{\prime}}\gamma^{\nu^{\prime}%
}\gamma^{\mu^{\prime}}\nonumber\\
\gamma^{\lbrack\mu}\gamma^{\nu}\gamma^{\rho}\gamma^{\sigma]_{q}}  &
\equiv-2q^{-8}[[2]]_{q^{2}}^{2}[[3]]_{q^{2}}\,\varepsilon^{+30-}%
\,\varepsilon_{\mu\nu\rho\sigma}\,\gamma^{\sigma}\gamma^{\rho}\gamma^{\nu
}\gamma^{\mu}\nonumber\\
&  =-2q^{-13}[[2]]_{q^{2}}^{3}[[3]]_{q^{2}}^{2}\;\gamma^{5}. \label{prgamma}%
\end{align}
Clearly, the antisymmetrizer $P_{A}$ and the $q$-deformed epsilon tensor now
refer to $q$-deformed Min\-kow\-ski space.

The quantum trace for Pauli matrices (for its definition see part I) can be
generalized to $\gamma$-matrices in the following manner:%
\begin{equation}
\mbox{Tr}_{q}\;\gamma^{\mu}\gamma^{\nu}\equiv\mbox{Tr}\;D_{L}^{\text{T}%
}\,\gamma^{\mu}\gamma^{\nu}D_{R}. \label{trgamma}%
\end{equation}
This quantum trace enables us to introduce a left-dual basis of
(\ref{gammabasis}). Its elements $\Gamma_{l,A}$, $A\in\{$e, $+$,\ldots,
$+30-\}$, are determined by
\begin{equation}
\mbox{Tr}_{q}\;(\Gamma_{l,A}\,\Gamma^{B})=\delta^{B}{}_{A}. \label{SkalGamLef}%
\end{equation}
Exploiting these conditions we found%
\begin{align}
\Gamma_{l,\text{e}}  &  =-\frac{1}{2}\lambda_{+}^{-1}\;\Gamma^{\text{e}}%
,\quad\Gamma_{l,+}=-\frac{q}{2}\;\Gamma^{-},\quad\Gamma_{l,3}=\frac{1}%
{2}\;\Gamma^{3},\nonumber\\
\Gamma_{l,0}  &  =-\frac{1}{2}\;\Gamma^{0},\quad\Gamma_{l,-}=-\frac{1}%
{2q}\;\Gamma^{+},\label{basisnew1230}\\[0.16in]
\Gamma_{l,+3}  &  =2q^{-1}\lambda_{+}^{-1}\;\Gamma^{3-}+\lambda\lambda
_{+}^{-1}\;\Gamma^{0-},\nonumber\\
\Gamma_{l,+0}  &  =-2q\lambda_{+}^{-1}\;\Gamma^{0-}+\lambda\lambda_{+}%
^{-1}\;\Gamma^{3-},\nonumber\\
\Gamma_{l,+-}  &  =-(q^{2}+q^{-2})\lambda_{+}^{-1}\;\Gamma^{+-}+\lambda
\lambda_{+}^{-1}\;\Gamma^{30},\nonumber\\
\Gamma_{l,30}  &  =-2\lambda_{+}^{-1}\;\Gamma^{30}+\lambda\lambda_{+}%
^{-1}\;\Gamma^{+-},\nonumber\\
\Gamma_{l,3-}  &  =2q^{-3}\lambda_{+}^{-1}\;\Gamma^{+3}+q^{-2}\lambda
\lambda_{+}^{-1}\;\Gamma^{+0},\nonumber\\
\Gamma_{l,0-}  &  =-2q^{-1}\lambda_{+}^{-1}\;\Gamma^{+0}+q^{-2}\lambda
\lambda_{+}^{-1}\;\Gamma^{+3},\\[0.16in]
\Gamma_{l,+30}  &  =-2q^{-1}\;\Gamma^{30-},\quad\Gamma_{l,+3-}=2q^{-4}%
\;\Gamma^{+3-},\quad\Gamma_{l,+0-}=-\frac{1}{2}\;\Gamma^{+0-},\nonumber\\
\Gamma_{l,30-}  &  =-2q^{-3}\;\Gamma^{+30},\quad\Gamma_{l,+30-}=-\frac{1}%
{8}q^{27}[[2]]_{q^{2}}^{-7}[[3]]_{q^{2}}^{-4}\;\Gamma^{+30-}.
\label{basisnew123}%
\end{align}

There is also the notion of a right-dual basis of (\ref{gammabasis}). Its
elements $\Gamma_{r,A}$, $A\in\{$e, $+$,\ldots, $+30-\}$ are subject to
\begin{equation}
\mbox{Tr}_{q}\;(\Gamma^{A}\,\Gamma_{r,B})=\delta^{A}{}_{B}.
\label{SkalGamRight}%
\end{equation}
The elements of the right-dual basis are obtained from those of the left-dual
one. Concretely, we have the correspondences%
\begin{gather}
\Gamma_{r,+}=q^{-2}\,\Gamma_{l,+},\quad\Gamma_{r,-}=q^{2}\,\Gamma
_{l,-},\nonumber\\
\Gamma_{r,+3}=q^{-2}\,\Gamma_{l,+3},\quad\Gamma_{r,+0}=q^{-2}\,\Gamma
_{l,+0},\quad\Gamma_{r,3-}=q^{2}\,\Gamma_{l,3-},\quad\Gamma_{r,0-}%
=q^{2}\,\Gamma_{l,0-},\nonumber\\
\Gamma_{r,+30}=q^{-2}\,\Gamma_{l,+30},\quad\Gamma_{r,30-}=q^{2}\,\Gamma
_{l,30-}.
\end{gather}
For the remaining matrices it holds $\Gamma_{r,A}=\Gamma_{l,A}.$

Now, we can repeat the same reasonings as above for 'inverse' matrices. Let us
recall that the 'inverse' $\gamma$-matrices are representations of the
elements $\tilde{\xi}_{\mu},$ $\mu\in\{+,0,3,-\}$. In this sense, they fulfill
the relations in (\ref{CliffCov}). The 16 matrices
\begin{align}
&  1,\;\gamma_{\mu}^{-1},\;\mu\in\{+,3,0,-\},\nonumber\\
&  \gamma_{\lbrack\mu}^{-1}\gamma_{\nu]_{q}}^{-1},\;(\mu,\nu)\in
\{(+,3),(+,0),(+,-),(3,-),(3,0),(0,-)\},\nonumber\\
&  \gamma_{\lbrack\mu}^{-1}\gamma_{\nu}^{-1}\gamma_{\rho]_{q}}^{-1},\;(\mu
,\nu,\rho)\in\{(+,3,-),(+,0,-),(3,0,-)\},\nonumber\\
&  \gamma_{\lbrack+}^{-1}\gamma_{3}^{-1}\gamma_{0}^{-1}\gamma_{-]_{q}}^{-1},
\label{BasGammInvMin}%
\end{align}
where%
\begin{align}
\gamma_{\lbrack\mu}^{-1}\gamma_{\nu]_{q}}^{-1}  &  \equiv\left(  P_{A}\right)
^{\mu^{\prime}\nu^{\prime}}{}_{\mu\nu}\gamma_{\mu^{\prime}}^{-1}\gamma
_{\nu^{\prime}}^{-1},\nonumber\\
\gamma_{\lbrack\mu}^{-1}\gamma_{\nu}^{-1}\gamma_{\rho]_{q}}^{-1}  &
\equiv\frac{q^{6}}{2[[3]]_{q^{2}}}\,\varepsilon_{\sigma\rho\nu\mu
}\,\varepsilon^{\sigma\rho^{\prime}\nu^{\prime}\mu^{\prime}}\gamma
_{\rho^{\prime}}^{-1}\gamma_{\nu^{\prime}}^{-1}\gamma_{\mu^{\prime}}%
^{-1},\nonumber\\
\gamma_{\lbrack\mu}^{-1}\gamma_{\nu}^{-1}\gamma_{\rho}^{-1}\gamma_{\sigma
]_{q}}^{-1}  &  \equiv-2q^{-8}[[2]]_{q^{2}}^{2}[[3]]_{q^{2}}\,\varepsilon
_{-03+}\,\varepsilon^{\sigma\rho\nu\mu}\gamma_{\sigma}^{-1}\gamma_{\rho}%
^{-1}\gamma_{\nu}^{-1}\gamma_{\mu}^{-1}\nonumber\\
&  =2q^{-13}[[2]]_{q^{2}}^{3}[[3]]_{q^{2}}^{2}\;\gamma_{5}^{-1},
\end{align}
establish a basis of the Clifford algebra subject to the relations in
(\ref{CliffCov}). We refer to the matrices in (\ref{BasGammInvMin}) as
$\Gamma_{A}^{-1}$, $A$ $\in{\{}1,+,3,0,...,+30-{\}}$.

To determine a dual basis of (\ref{BasGammInvMin}) we first have to adjust the
definition of the quantum trace to 'inverse' $\gamma$-matrices. As a
consequence of the quantum trace for 'inverse' Pauli matrices [cf. part I]
this task can be achieved by%
\begin{equation}
\mbox{Tr}_{q}\,\gamma_{\mu}^{-1}\gamma_{\nu}^{-1}\equiv\mbox{Tr}\;D_{L}%
\,\gamma_{\mu}^{-1}\gamma_{\nu}^{-1}D_{R}^{\text{T}}. \label{trgammainv}%
\end{equation}

We require that\ the matrices $(\Gamma^{-1})^{l,A}$ of a left-dual basis
satisfy
\begin{equation}
\mbox{Tr}_{q}\,(\Gamma^{-1})^{A}\,\Gamma_{B}^{-1}=\delta^{A}{}_{B}.
\end{equation}
We simply get the matrices $(\Gamma^{-1})^{l,A}$, $A$ $\in{\{}%
1,+,3,0,...,+30-{\}}$, by applying the replacements
\begin{equation}
\Gamma^{A}\rightarrow\Gamma_{A}^{-1},\quad\Gamma_{l,A}\rightarrow(\Gamma
^{-1})^{l,A},%
\end{equation}
together with
\begin{equation}
\lambda\rightarrow-\lambda,
\end{equation}
to the formulae in (\ref{basisnew1230})-(\ref{basisnew123}). This procedure,
for example, yields%
\begin{equation}
(\Gamma^{-1})^{l,+-}=-q^{-2}(q^{4}+1)\lambda_{+}^{-1}\Gamma_{+-}^{-1}%
-\lambda\lambda_{+}^{-1}\Gamma_{30}^{-1}.
\end{equation}

\noindent\textbf{Completeness relations:\ }The $\Gamma^{A}$ constitute a basis
in the space of $4\times4$ matrices, so we are able to expand an arbitrary
matrix $M$ as
\begin{equation}
M=\sum\nolimits_{A}c_{A}\Gamma^{A}.
\end{equation}
The coefficients are obtained from $M$ by the trace formula
\begin{equation}
c_{B}=\mbox{Tr}_{q}(M\,\Gamma_{r,B})=\mbox{Tr}_{q}(\Gamma_{l,B}\,M),
\end{equation}
which is a direct consequence of (\ref{SkalGamLef}) and (\ref{SkalGamRight}).

We can use this trace expansion to deduce the completeness relation
\begin{align}
\delta_{a}{}^{b}\delta_{c}{}^{d}  &  =\sum_{B}(D_{R})_{cm}(D_{L}^{\text{T}%
})^{mk}(\Gamma_{l,B})_{k}{}^{b}(\Gamma^{B})_{a}{}^{d}\nonumber\\
&  =\sum_{B}(\Gamma^{B})_{a}{}^{d}(\Gamma_{r,B})_{c}{}^{k}(D_{R})_{km}%
(D_{L}^{\text{T}})^{mb},
\end{align}
Introducing the new matrices
\begin{align}
(\Gamma_{l,B}^{\text{c}})_{a}{}^{b}  &  \equiv(D_{R})_{am}(D_{L}^{\text{T}%
})^{mk}(\Gamma_{l,B})_{k}{}^{b},\nonumber\\
(\Gamma_{l,B}^{\text{c}})_{a}{}^{b}  &  \equiv(\Gamma_{r,B})_{a}{}^{k}%
(D_{R})_{km}(D_{L}^{\text{T}})^{mb}. \label{GammaC}%
\end{align}
we finally arrive at the relations
\begin{equation}
\sum_{B}(\Gamma_{l,B}^{\text{c}})_{c}{}^{b}(\Gamma^{B})_{a}{}^{d}=\delta_{a}%
{}^{b}\delta_{c}{}^{d},\qquad\sum_{B}(\Gamma^{B})_{a}{}^{d}(\Gamma
_{r,B}^{\text{c}})_{c}{}^{b}=\delta_{a}{}^{b}\delta_{c}{}^{d}.
\label{ComRelMin}%
\end{equation}
Incidentally, the matrices $\Gamma_{B}^{\text{c}}$ are related to the
$\Gamma^{B},$ $B$ $\in{\{}1,$ $+,$ $3,$ $0,...,$ $+30-{\}}$, via the
identities
\begin{align}
\Gamma_{1}^{\text{c}}=  &  \;\frac{1}{2}q^{-2}(q^{4}+1)\lambda_{+}^{-2}%
\Gamma^{\text{e}}+\frac{1}{2}\lambda\lambda_{+}^{-1}\Gamma^{+-},\quad
\Gamma_{+}^{\text{c}}=\frac{1}{2}\Gamma^{-},\nonumber\\
\Gamma_{3}^{\text{c}}=  &  \;-\lambda_{+}^{-1}\Gamma^{3}-\frac{1}{4}%
\lambda^{2}\lambda_{+}^{-1}\Gamma^{0}+\frac{1}{2}q^{-2}\lambda\Gamma
^{+30},\nonumber\\
\Gamma_{0}^{\text{c}}=  &  \;\frac{1}{2}q^{-2}(q^{4}+1)\lambda_{+}^{-1}%
\Gamma^{0}-\frac{1}{4}\lambda^{2}\lambda_{+}^{-1}\Gamma^{3}+\frac{1}{2}%
q^{-1}\lambda^{2}\Gamma^{+30}-\frac{1}{2}\lambda\Gamma^{+0-},\nonumber\\
\Gamma_{-}^{\text{c}}=  &  \;\frac{1}{2}\Gamma^{+},\quad\Gamma_{+3}^{\text{c}%
}=-2q^{-2}\lambda_{+}^{-1}\Gamma^{3-}-q^{-1}\lambda\lambda_{+}^{-1}\Gamma
^{0-},\\[0.16in]
\Gamma_{+0}^{\text{c}}=  &  \;2\lambda_{+}^{-1}\Gamma^{0-}-q^{-1}%
\lambda\lambda_{+}^{-1}\Gamma^{3-},\nonumber\\
\Gamma_{+-}^{\text{c}}=  &  \;2q^{-2}(q^{4}+1)\lambda_{+}^{-2}\Gamma
^{+-}-2\lambda\lambda_{+}^{-2}\Gamma^{+30-}\nonumber\\
&  -\frac{1}{4}q^{10}\lambda^{2}\lambda_{+}^{-5}[[3]]_{q^{2}}^{-2}%
\Gamma^{+30-}+\frac{1}{2}\lambda\lambda_{+}^{-1}\Gamma^{\text{e}},\nonumber\\
\Gamma_{30}^{\text{c}}=  &  \;4\lambda_{+}^{-2}\Gamma^{30}-2\lambda\lambda
_{+}^{-2}\Gamma^{+-}+\frac{1}{2}q^{11}\lambda^{2}\lambda_{+}^{-5}[[3]]_{q^{2}%
}^{-2}\Gamma^{+30-},\nonumber\\
\Gamma_{3-}^{\text{c}}=  &  \;-2q^{-2}\lambda_{+}^{-1}\Gamma^{+3}%
-q^{-1}\lambda\lambda_{+}^{-1}\Gamma^{0-},\nonumber\\
\Gamma_{0-}^{\text{c}}=  &  \;2\lambda_{+}^{-1}\Gamma^{+0}-q^{-1}%
\lambda\lambda_{+}^{-1}\Gamma^{+3},\quad\Gamma_{+30}^{\text{c}}=2q^{-2}%
\Gamma^{30-},\\[0.16in]
\Gamma_{+3-}^{\text{c}}=  &  \;2q^{-2}\lambda_{+}^{-1}(\lambda\lambda
_{+}-2)\Gamma^{+3-}\frac{1}{2}q^{-2}\lambda\Gamma^{3}+\frac{1}{2}q^{-1}%
\lambda^{2}\Gamma^{0}\nonumber\\
&  -q^{-3}(3q^{2}+1)\lambda\lambda_{+}^{-1}\Gamma^{+0-},\nonumber\\
\Gamma_{+0-}^{\text{c}}=  &  \;4\lambda_{+}^{-1}\Gamma^{+0-}-q^{-3}%
(3q^{2}+1)\lambda\lambda_{+}^{-1}\Gamma^{+3-}-\frac{1}{2}\Gamma^{0}%
,\nonumber\\
\Gamma_{30-}^{\text{c}}=  &  \;2q^{-2}\Gamma^{+30},\\[0.16in]
\Gamma_{+30-}^{\text{c}}=  &  \;\frac{1}{8}q^{18}(q^{4}+1)\lambda_{+}%
^{-8}[[3]]_{q^{2}}^{-4}\Gamma^{+30-}-\frac{1}{4}q^{10}\lambda^{2}\lambda
_{+}^{-5}[[3]]_{q^{2}}^{-2}\Gamma^{+-}\nonumber\\
&  +\frac{1}{2}q^{11}\lambda^{2}\lambda_{+}^{-5}[[3]]_{q^{2}}^{-2}\Gamma^{30}.
\label{gammacompmink123}%
\end{align}

It is not very difficult to adjust the above reasonings to 'inverse' $\gamma
$-matrices. The set of $\Gamma_{A}^{-1},$ $A$ $\in{\{}1,+,3,0,...,+30-{\}}$,
again give us a basis of $4\times4$ matrices:%
\begin{equation}
M=\sum\nolimits_{A}c^{A}\Gamma_{A}^{-1},
\end{equation}
where
\begin{equation}
c^{B}=\mbox{Tr}_{q}(M\,(\Gamma^{-1})^{r,B})=\mbox{Tr}_{q}((\Gamma^{-1}%
)^{l,B}\,M).
\end{equation}
These identities imply
\begin{align}
\delta_{a}{}^{b}\delta_{c}{}^{d}  &  =\sum_{B}(D_{R}^{\text{T}})_{cm}%
(D_{L})^{mk}((\Gamma^{-1})^{l,B})_{k}{}^{b}((\Gamma_{B}^{-1})_{a}{}%
^{d}\nonumber\\
&  =\sum_{B}(\Gamma_{B}^{-1})_{a}{}^{d}((\Gamma^{-1})^{r,B})_{c}{}^{k}%
(D_{R})_{km}(D_{L}^{\text{T}})^{mb},
\end{align}
or, more compactly,%
\begin{align}
\sum_{B}((\Gamma_{\text{c}}^{-1})^{l,B})_{c}{}^{b}\,(\Gamma_{B}^{-1})_{a}%
{}^{d}  &  =\delta_{a}{}^{b}\delta_{c}{}^{d},\nonumber\\
\sum_{B}(\Gamma_{B}^{-1})_{a}{}^{d}\,((\Gamma_{c}^{-1})^{l,B})_{c}{}^{b}  &
=\delta_{a}{}^{b}\delta_{c}{}^{d},
\end{align}
with
\begin{align}
((\Gamma_{\text{c}}^{-1})^{l,B})_{a}{}^{b}  &  \equiv(D_{R}^{\text{T}}%
)_{am}(D_{L})^{mk}((\Gamma^{-1})^{l,B})_{k}{}^{b},\nonumber\\
((\Gamma_{c}^{-1})^{l,B})_{a}{}^{b}  &  \equiv((\Gamma^{-1})^{r,B})_{a}{}%
^{k}(D_{R}^{\text{T}})_{km}(D_{L})^{mb}.
\end{align}
Once again, we can express the matrices $(\Gamma_{\text{c}}^{-1})^{A}$ by the
$\Gamma_{A}^{-1}$, $A$ $\in{\{}1$, $+$, $\ 3$, $0$,$...$, $+30-{\}}$:%
\begin{align}
(\Gamma_{\text{c}}^{-1})^{\text{e}}=  &  \;\frac{1}{2}q^{-2}(q^{4}%
+1)\lambda_{+}^{-2}\Gamma_{\text{e}}^{-1}+\frac{1}{2}\lambda\lambda_{+}%
^{-1}\Gamma_{+-}^{-1},\quad(\Gamma_{\text{c}}^{-1})^{+}=\frac{1}{2}\Gamma
_{-}^{-1},\nonumber\\
(\Gamma_{\text{c}}^{-1})^{3}=  &  -\lambda_{+}^{-1}\Gamma_{3}^{-1}-\frac{1}%
{4}\lambda^{2}\lambda_{+}^{-1}\Gamma_{0}^{-1}+\frac{1}{2}q^{-2}\lambda
\Gamma_{+30}^{-1},\nonumber\\
(\Gamma_{\text{c}}^{-1})^{0}=  &  \;\frac{1}{2}q^{-2}(q^{4}+1)\lambda_{+}%
^{-1}\Gamma_{0}^{-1}+\frac{1}{4}\lambda^{2}\lambda_{+}^{-1}\Gamma_{3}%
^{-1}-\frac{1}{2}q^{-1}\lambda^{2}\Gamma_{+3-}^{-1}-\frac{1}{2}\lambda
\Gamma_{+0-}^{-1},\nonumber\\
(\Gamma_{\text{c}}^{-1})^{-}=  &  \;\frac{1}{2}\Gamma_{+}^{-1},\quad
(\Gamma_{\text{c}}^{-1})^{+3}=-2q^{-2}\lambda_{+}^{-1}\Gamma_{3-}^{-1}%
+q^{-1}\lambda\lambda_{+}^{-1}\Gamma_{0-}^{-1},\\[0.16in]
(\Gamma_{\text{c}}^{-1})^{+0}=  &  \;2\lambda_{+}^{-1}\Gamma_{0-}^{-1}%
+q^{-1}\lambda\lambda_{+}^{-1}\Gamma_{3-}^{-1},\nonumber\\
(\Gamma_{\text{c}}^{-1})^{+-}=  &  \;2q^{-2}(q^{4}+1)\lambda_{+}^{-2}%
\Gamma_{+-}^{-1}+2\lambda\lambda_{+}^{-2}\Gamma_{30}^{-1}\nonumber\\
&  +\frac{1}{4}q^{10}\lambda^{2}\lambda_{+}^{-5}[[3]]_{q^{2}}^{-2}%
\Gamma_{+30-}^{-1}+\frac{1}{2}\lambda\lambda_{+}^{-1}\Gamma_{\text{e}}%
^{-1},\nonumber\\
(\Gamma_{\text{c}}^{-1})^{30}=  &  \;4\lambda_{+}^{-2}\Gamma_{30}%
^{-1}+2\lambda\lambda_{+}^{-2}\Gamma_{+-}^{-1}+\frac{1}{2}q^{11}\lambda
^{2}\lambda_{+}^{-5}[[3]]_{q^{2}}^{-2}\Gamma_{+30-}^{-1},\nonumber\\
(\Gamma_{\text{c}}^{-1})^{3-}=  &  -2q^{-2}\lambda_{+}^{-1}\Gamma_{+3}%
^{-1}+q^{-1}\lambda\lambda_{+}^{-1}\Gamma_{+0}^{-1},\nonumber\\
(\Gamma_{\text{c}}^{-1})^{0-}=  &  \;2\lambda_{+}^{-1}\Gamma_{+0}^{-1}%
+q^{-1}\lambda\lambda_{+}^{-1}\Gamma_{+3}^{-1},\quad(\Gamma_{\text{c}}%
^{-1})^{+30}=2q^{-2}\Gamma_{30-}^{-1},\\[0.16in]
(\Gamma_{\text{c}}^{-1})^{+3-}=  &  \;2q^{-2}\lambda_{+}^{-1}(\lambda
\lambda_{+}-2)\Gamma_{+3-}^{-1}\frac{1}{2}q^{-2}\lambda\Gamma_{3}^{-1}%
-\frac{1}{2}q^{-1}\lambda^{2}\Gamma_{0}^{-1}\nonumber\\
&  +q^{-3}(3q^{2}+1)\lambda\lambda_{+}^{-1}\Gamma_{+0-}^{-1},\nonumber\\
(\Gamma_{\text{c}}^{-1})^{+0-}=  &  \;4\lambda_{+}^{-1}\Gamma_{+0-}%
^{-1}-q^{-3}(3q^{2}+1)\lambda\lambda_{+}^{-1}\Gamma_{+3-}^{-1}-\frac{1}%
{2}\Gamma_{0}^{-1},\nonumber\\
(\Gamma_{\text{c}}^{-1})^{30-}=  &  \;2q^{-2}\Gamma_{+30}^{-1},\nonumber\\
(\Gamma_{\text{c}}^{-1})^{+30-}=  &  \;\frac{1}{8}q^{18}(q^{4}+1)\lambda
_{+}^{-8}[[3]]_{q^{2}}^{-4}\Gamma_{+30-}^{-1}+\frac{1}{4}q^{10}\lambda
^{2}\lambda_{+}^{-5}[[3]]_{q^{2}}^{-2}\Gamma_{+-}^{-1}\nonumber\\
&  +\frac{1}{2}q^{11}\lambda^{2}\lambda_{+}^{-5}[[3]]_{q^{2}}^{-2}\Gamma
_{30}^{-1}.
\end{align}

\noindent\textbf{A more physical oriented basis:\ }In physical applications it
is more convenient to work with a basis that uses the matrices $\Gamma
^{5}\equiv\gamma^{5}$ and $\Gamma^{5\mu}\equiv\gamma^{5}\gamma^{\mu}$, $\mu
\in\{+$, $3$, $0$, $-\}$, instead of antisymmetrized products of three and
four $\gamma$-matrices:%
\begin{align}
&  1,\;\gamma^{+},\;\gamma^{3},\;\gamma^{0},\;\gamma^{-},\;\gamma^{5}%
,\gamma^{5}\gamma^{+},\;\gamma^{5}\gamma^{3},\;\gamma^{5}\gamma^{0}%
,\;\gamma^{5}\gamma^{-},\nonumber\\
&  \gamma^{\lbrack+}\gamma^{3]_{q}},\;\gamma^{\lbrack+}\gamma^{0]_{q}%
},\;\gamma^{\lbrack+}\gamma^{-]_{q}},\;\gamma^{\lbrack3}\gamma^{0]_{q}%
},\;\gamma^{\lbrack3}\gamma^{-]_{q}},\gamma^{\lbrack0}\gamma^{-]_{q}}.
\label{basisgamma5mink}%
\end{align}
From (\ref{SkalGamLef}) we find that the left-duals of $\Gamma^{5}$ and
$\Gamma^{5\mu}$ become
\begin{gather}
\Gamma_{l,5+}=\frac{q}{2}\Gamma^{5-},\quad\Gamma_{l,53}=-\frac{1}{2}%
\Gamma^{53},\nonumber\\
\Gamma_{l,50}=\frac{1}{2}\Gamma^{50},\quad\Gamma_{l,5-}=\frac{1}{2}%
q^{-1}\Gamma^{5+},\quad\Gamma_{l,5}=\frac{1}{2}\lambda_{+}^{-1}\Gamma^{5}.
\label{LefDualGa5}%
\end{gather}
Furthermore, if we work with the basis of (\ref{basisgamma5mink}) the
completeness relations in (\ref{ComRelMin}) require to deal with the matrices%
\begin{align}
\Gamma_{3}^{\text{c}}  &  =-\lambda_{+}^{-1}\Gamma^{3}-\frac{1}{4}\lambda
^{2}\lambda_{+}^{-1}\Gamma^{0}+\frac{1}{4}\lambda\Gamma^{50},\nonumber\\
\Gamma_{0}^{\text{c}}  &  =\frac{1}{2}q^{-2}(q^{4}+1)\lambda_{+}^{-1}%
\Gamma^{0}-\frac{1}{4}\lambda^{2}\lambda_{+}^{-1}\Gamma^{3}-\frac{1}{4}%
\lambda\Gamma^{53},\nonumber\\
\Gamma_{5}^{\text{c}}  &  =\frac{1}{2}q^{-2}(q^{4}+1)\lambda_{+}^{-2}%
\Gamma^{5}+\frac{1}{2}\lambda^{2}\lambda_{+}^{2}\Gamma^{+-}-\lambda\lambda
_{+}^{-1}\Gamma^{30},\label{GamC2Bas}\\[0.16in]
\Gamma_{5+}^{\text{c}}  &  =-\frac{1}{2}\Gamma^{5-},\nonumber\\
\Gamma_{53}^{\text{c}}  &  =\lambda_{+}^{-1}\Gamma^{53}-\frac{1}{4}%
\lambda\Gamma^{0}+\frac{1}{4}\lambda^{2}\lambda_{+}^{-1}\Gamma^{50}%
,\nonumber\\
\Gamma_{50}^{\text{c}}  &  =-\frac{1}{2}q^{-2}(q^{4}+1)\lambda_{+}^{-1}%
\Gamma^{50}+\frac{1}{4}\lambda\Gamma^{3}+\frac{1}{4}\lambda^{2}\lambda
_{+}^{-1}\Gamma^{53},\nonumber\\
\Gamma_{5-}^{\text{c}}  &  =-\frac{1}{2}\Gamma^{5+}.
\end{align}

Replacing in (\ref{basisgamma5mink}) all $\gamma$-matrices by their 'inverse'
counterparts we are led to a basis of the Clifford algebra subject to the
relations in (\ref{CliffCov}):%
\begin{align}
&  1,\;\gamma^{5},\;\gamma_{\mu}^{-1},\;\gamma_{5}^{-1}\gamma_{\mu}^{-1}%
,\;\mu\in\{+,3,0,-\},\nonumber\\
&  \gamma_{\lbrack\mu}^{-1}\gamma_{\nu]_{q}}^{-1},\;(\mu,\nu)\in
\{(+,3),(+,0),(+,-),(3,-),(3,0),(0,-)\}. \label{InvBasGam5}%
\end{align}
Applying the substitutions $\Gamma^{A}\rightarrow\Gamma_{A}^{-1}$,
$\Gamma_{l,A}\rightarrow(\Gamma^{-1})^{l,A}$ to (\ref{LefDualGa5}) gives us
the expressions for the left-duals $\Gamma_{l,5}^{-1}$ and $\Gamma_{l,5\mu
}^{-1}$, $\mu\in\{+,3,0,-\}$. If we want to work with the matrices
$(\Gamma_{\text{c}}^{-1})^{A}$ corresponding to the basis (\ref{InvBasGam5})
we also need%
\begin{align}
(\Gamma_{\text{c}}^{-1})^{3}  &  =-\lambda_{+}^{-1}\Gamma_{3}^{-1}+\frac{1}%
{4}\lambda^{2}\lambda_{+}^{-1}\Gamma_{0}^{-1}-\frac{1}{4}\lambda\Gamma
_{50}^{-1},\nonumber\\
(\Gamma_{\text{c}}^{-1})^{0}  &  =\,\frac{1}{2}q^{-2}(q^{4}+1)\lambda_{+}%
^{-1}\Gamma_{0}^{-1}+\frac{1}{4}\lambda^{2}\lambda_{+}^{-1}\Gamma_{3}%
^{-1}+\frac{1}{4}\lambda\Gamma_{53}^{-1},\nonumber\\
(\Gamma_{\text{c}}^{-1})^{5}  &  =\,\frac{1}{2}q^{-2}(q^{4}+1)\lambda_{+}%
^{-2}\Gamma_{5}^{-1}+\frac{1}{2}\lambda^{2}\lambda_{+}^{2}\Gamma_{+-}%
^{-1}+\lambda\lambda_{+}^{-1}\Gamma_{30}^{-1},\\[0.16in]
(\Gamma_{\text{c}}^{-1})^{5+}  &  =-\frac{1}{2}\Gamma_{5-}^{-1},\nonumber\\
(\Gamma_{\text{c}}^{-1})^{53}  &  =\,\lambda_{+}^{-1}\Gamma_{53}^{-1}+\frac
{1}{4}\lambda\Gamma_{0}^{-1}-\frac{1}{4}\lambda^{2}\lambda_{+}^{-1}\Gamma
_{50}^{-1},\nonumber\\
(\Gamma_{\text{c}}^{-1})^{50}  &  =-\frac{1}{2}q^{-2}(q^{4}+1)\lambda_{+}%
^{-1}\Gamma_{50}^{-1}-\frac{1}{4}\lambda\Gamma_{3}^{-1}-\frac{1}{4}\lambda
^{2}\lambda_{+}^{-1}\Gamma_{53}^{-1},\nonumber\\
(\Gamma_{\text{c}}^{-1})^{5-}  &  =-\frac{1}{2}\Gamma_{5+}^{-1}.
\end{align}

\subsubsection{Relations concerning $q$-deformed Dirac
matrices\label{RelConMin}}

In the literature one can find numerous identities for $\gamma$-matrices (see,
for example, Refs. \cite{bailin, Core, Peskin, Wessbagger, Landau, Grimm}).
Many of them play an important role in quantum field theory. This subsection
is devoted to $q$-analogs of such relations. It should be emphasized that the
objects ($\hat{R}$-matrix, projectors, quantum metric, epsilon tensor,
$\gamma$-matrices etc.) appearing in the following formulae all refer to
$q$-deformed Min\-kow\-ski space and the matrix product of $\gamma$-matrices
is understood as $\gamma^{\mu}\gamma^{\nu}=(\gamma^{\mu})_{a}{}^{a^{\prime}%
}(\gamma^{\nu})_{a^{\prime}}{}^{b}$.

We begin with trace relations for $\gamma$-matrices:
\begin{align}
\mbox{Tr}_{q}\,\gamma^{\mu}  &  =0,\qquad\mbox{Tr}_{q}\,\gamma^{\mu}%
\gamma^{\nu}=2\eta^{\mu\nu},\qquad\mbox{Tr}_{q}\,\gamma^{\mu}\gamma^{\nu
}\gamma^{\rho}=0,\nonumber\\
\mbox{Tr}_{q}\,\gamma^{\mu}\gamma^{\nu}\gamma^{\rho}\gamma^{\sigma}  &
=-q^{-1}\eta^{\mu\nu}\eta^{\rho\sigma}+q\,\eta^{\mu\kappa}\hat{R}^{\nu\rho}%
{}_{\kappa\lambda}\eta^{\lambda\sigma}-q\,\eta^{\mu\sigma}\eta^{\nu\rho
}\nonumber\\
&  =-q\,\eta^{\mu\nu}\eta^{\rho\sigma}+q^{-1}\eta^{\mu\kappa}(\hat{R}%
^{-1})^{\nu\rho}{}_{\kappa\lambda}\eta^{\lambda\sigma}-q^{-1}\eta^{\mu\sigma
}\eta^{\nu\rho}.
\end{align}
In addition to this, we could also check the validity of the following trace
formula:%
\begin{align}
&  \mbox{Tr}_{q}\;\gamma^{5}\gamma^{\mu_{1}}\gamma^{\mu_{2}}\gamma^{\mu_{3}%
}\gamma^{\mu_{4}}\gamma^{\mu_{5}}\gamma^{\mu_{6}}=\nonumber\\
&  \qquad\qquad=\,-2q^{-1}\big [\,k_{1}\,\eta^{\mu_{1}\mu_{2}}\,\varepsilon
^{\mu_{3}\mu_{4}\mu_{5}\mu_{6}}+k_{2}\,\eta^{\mu_{1}\mu_{3}^{\prime}}\,\hat
{R}^{\mu_{2}\mu_{3}}{}_{\mu_{3}^{\prime}\mu_{2}^{\prime}}\,\varepsilon
^{\mu_{2}^{\prime}\mu_{4}\mu_{5}\mu_{6}}\nonumber\\
&  \qquad\qquad\hspace{0.19in}\,+k_{3}\,\eta^{\mu_{2}\mu_{3}}\,\varepsilon
^{\mu_{1}\mu_{4}\mu_{5}\mu_{6}}+k_{4}\,\eta^{\mu_{4}\mu_{5}}\,\varepsilon
^{\mu_{1}\mu_{2}\mu_{3}\mu_{6}}\nonumber\\
&  \qquad\qquad\hspace{0.19in}\,+k_{5}\,\eta^{\mu_{4}\mu_{6}^{\prime}}%
\,\hat{R}^{\mu_{5}\mu_{6}}{}_{\mu_{6}^{\prime}\mu_{5}^{\prime}}\,\varepsilon
^{\mu_{1}\mu_{2}\mu_{3}\mu_{5}^{\prime}}+k_{6}\,\eta^{\mu_{5}\mu_{6}%
}\,\varepsilon^{\mu_{1}\mu_{2}\mu_{3}\mu_{4}}\big], \label{splangMin}%
\end{align}
where the values of the coefficients are
\begin{equation}
k_{1}=1,\;k_{2}=-q^{-1},\;k_{3}=q^{-2},\;k_{4}=1,\;k_{5}=-q^{-1}%
,\;k_{6}=q^{-2}.
\end{equation}
Notice that in (\ref{splangMin}) we are allowed to replace $\hat{R}$ with
$\hat{R}^{-1}$, if we instead set
\begin{equation}
k_{1}=q^{-2},\;k_{2}=-q^{-1},\;k_{3}=1,\;k_{4}=q^{-2},\;k_{5}=-q^{-1}%
,\;k_{6}=1.
\end{equation}
Finally, we could also calculate trace formulae with $\gamma^{5}$:%
\begin{equation}
\mbox{Tr}_{q}\,\gamma^{5}=\mbox{Tr}_{q}\,\gamma^{\mu}\gamma^{5}=\mbox{Tr}_{q}%
\,\gamma^{\mu}\gamma^{\nu}\gamma^{5}=\mbox{Tr}_{q}\,\gamma^{\mu}\gamma^{\nu
}\gamma^{\rho}\gamma^{5}=0,
\end{equation}
and%
\begin{equation}
\mbox{Tr}_{q}\,\gamma^{\mu}\gamma^{\nu}\gamma^{\rho}\gamma^{\sigma}\gamma
^{5}=q^{2}\;\varepsilon^{\mu\nu\rho\sigma}.
\end{equation}

Next, we turn our attention to relations in which two $\gamma$-matrices are
contracted by the quantum metric (possibly via $\hat{R}$-matrices). We found
\begin{align}
\eta_{\mu\nu}\,\gamma^{\mu}\gamma^{\nu}  &  =-\lambda_{+}\nonumber\\
q^{4}[[3]]_{q^{2}}^{-1}\,\varepsilon_{\lambda\rho\nu\kappa}\,\eta^{\kappa\mu
}\,\gamma^{\nu}\gamma^{\rho}\gamma^{\lambda}  &  =\gamma^{5}\gamma^{\nu
},\nonumber\\
\hat{R}^{\mu\nu}{}_{\nu^{\prime}\mu^{\prime}}\,\eta_{\rho\mu}\,\gamma^{\rho
}\gamma^{\nu^{\prime}}\gamma^{\mu^{\prime}}  &  =q^{-1}\gamma^{\nu
},\nonumber\\
\hat{R}^{\mu\nu}{}_{\nu^{\prime}\mu^{\prime}}\hat{R}^{\mu^{\prime}\rho}%
{}_{\rho^{\prime}\mu^{\prime\prime}}\,\eta_{\lambda\mu}\,\gamma^{\lambda
}\gamma^{\nu^{\prime}}\gamma^{\rho^{\prime}}\gamma^{\mu^{\prime\prime}}  &
=q^{-2}\eta^{\nu\rho}, \label{gammarel1N}%
\end{align}
and the more lengthy relation
\begin{align}
&  \hat{R}^{\mu\nu}{}_{\nu^{\prime}\mu^{\prime}}\hat{R}^{\mu^{\prime}\rho}%
{}_{\rho^{\prime}\mu^{\prime\prime}}\hat{R}^{\mu^{\prime\prime}\sigma}%
{}_{\sigma^{\prime}\mu^{\prime\prime\prime}}\,\eta_{\lambda\mu}\,\gamma
^{\lambda}\gamma^{\nu^{\prime}}\gamma^{\rho^{\prime}}\gamma^{\sigma^{\prime}%
}\gamma^{\mu^{\prime\prime\prime}}\nonumber\\
&  =q^{-3}\,\hat{R}^{\rho^{\prime\prime}\sigma^{\prime\prime}}{}%
_{\sigma^{\prime}\rho^{\prime}}\hat{R}^{\nu^{\prime\prime}\sigma}{}%
_{\sigma^{\prime\prime}\nu^{\prime}}\hat{R}^{\nu\rho}{}_{\rho^{\prime\prime
}\nu^{\prime\prime}}\,\gamma^{\sigma^{\prime}}\gamma^{\rho^{\prime}}%
\gamma^{\nu^{\prime}}, \label{gammarel2}%
\end{align}
which can be viewed as q-analog of the undeformed relation $\gamma^{\mu}%
\gamma^{\nu}\gamma^{\rho}\gamma^{\sigma}\gamma_{\mu}=-2\gamma^{\sigma}%
\gamma^{\rho}\gamma^{\nu}$ \cite{Peskin}. Let us note that we get variants of
the relations in (\ref{gammarel1N}) and (\ref{gammarel2}) through the
replacements $\hat{R}\rightarrow\hat{R}^{-1}$, $q\rightarrow q^{-1}$.

A product of three $\gamma$-matrices can be decomposed as follows:
\begin{align}
\gamma^{\mu}\gamma^{\nu}\gamma^{\rho}=  &  \;\gamma^{\lbrack\mu}\gamma^{\nu
}\gamma^{\rho]_{q}}-\frac{1}{2}q^{-1}\,\eta^{\mu\nu}\gamma^{\rho}\nonumber\\
&  +\frac{q}{2}\,\eta^{\mu\lambda^{\prime}}\hat{R}^{\nu\lambda}{}%
_{\lambda^{\prime}\nu^{\prime}}\,\gamma^{\nu^{\prime}}-\frac{q}{2}%
\,\gamma^{\mu}\eta^{\nu\lambda}\nonumber\\
=  &  \;\gamma^{\lbrack\mu}\gamma^{\nu}\gamma^{\rho]_{q}}-\frac{1}{2}%
q\,\eta^{\mu\nu}\gamma^{\rho}\nonumber\\
&  +\frac{1}{2}q^{-1}\,\eta^{\mu\lambda^{\prime}}(\hat{R}^{-1})^{\nu\lambda}%
{}_{\lambda^{\prime}\nu^{\prime}}\,\gamma^{\nu^{\prime}}-\frac{1}{2}%
q^{-1}\,\gamma^{\mu}\,\eta^{\nu\lambda}. \label{gammarel4}%
\end{align}

All relations in this subsection can alternatively be formulated in terms of
'inverse' $\gamma$-matrices. We simply have to make the replacements
\begin{gather}
\gamma^{\mu}\rightarrow\gamma_{\mu}^{-1},\qquad\gamma^{5}\rightarrow\gamma
_{5}^{-1},\nonumber\\
\eta^{\mu\nu}\rightarrow\eta_{\mu\nu},\qquad\varepsilon^{\mu\nu\rho\sigma
}\rightarrow\varepsilon_{\sigma\rho\nu\mu},\qquad\varepsilon_{\mu\nu\rho
\sigma}\rightarrow\varepsilon^{\sigma\rho\nu\mu},\nonumber\\
\hat{R}^{\mu\nu}{}_{\rho\sigma}\rightarrow\hat{R}^{\rho\sigma}{}_{\mu\nu
},\qquad(\hat{R}^{-1})^{\mu\nu}{}_{\rho\sigma}\rightarrow(\hat{R}^{-1}%
)^{\rho\sigma}{}_{\mu\nu}. \label{replinv}%
\end{gather}
In doing so, we have to be aware of the fact that the quantum trace for
'inverse' matrices is given by (\ref{trgammainv}).

\subsubsection{The $q$-deformed spin matrices and some more
relations\label{SpinMatKapMin}}

In analogy to the definition of two-dimensional spin matrices [cf. part I] we
define the four-dimensional spin matrices as antisymmetrized product of two
$\gamma$-matrices:
\begin{align}
(\Sigma^{\mu\nu})_{a}{}^{b}  &  \equiv(P_{A})^{\mu\nu}{}_{\mu^{\prime}%
\nu^{\prime}}\,(\gamma^{\mu^{\prime}})_{a}{}^{a^{\prime}}(\gamma^{\nu^{\prime
}})_{a^{\prime}}{}^{b},\nonumber\\
(\Sigma_{\mu\nu}^{-1})_{a}{}^{b}  &  \equiv(P_{A})^{\mu^{\prime}\nu^{\prime}%
}{}_{\mu\nu}\,(\gamma_{\mu^{\prime}}^{-1})_{a}{}^{a^{\prime}}(\gamma
_{\nu^{\prime}}^{-1})_{a^{\prime}}{}^{b}.
\end{align}
Raising of indices works in the same manner as for the $\gamma$-matrices. The
$\Sigma$-matrices can also be written in the form%
\begin{align}
(\Sigma^{\mu\nu})_{a}{}^{b}  &  =\left(
\begin{array}
[c]{cc}%
(\sigma^{\mu\nu})_{\alpha}{}^{\beta} & 0\\
0 & (\bar{\sigma}^{\mu\nu})_{\dot{\alpha}}{}^{{\dot{\beta}}}%
\end{array}
\right)  ,\nonumber\\
(\Sigma_{\mu\nu}^{-1})_{a}{}^{b}  &  =\left(
\begin{array}
[c]{cc}%
(\sigma_{\mu\nu}^{-1})_{\alpha}{}^{\beta} & 0\\
0 & (\bar{\sigma}_{\mu\nu}^{-1})_{\dot{\alpha}}{}^{{\dot{\beta}}}%
\end{array}
\right)  , \label{MatFormSpin}%
\end{align}
with $\sigma^{\mu\nu}$, $\bar{\sigma}^{\mu\nu}$, $\sigma_{\mu\nu}^{-1}$, and
$\bar{\sigma}_{\mu\nu}^{-1}$ standing for the two-dimensional spin matrices of
$q$-de\-formed Minkowski space. Exploiting the properties of two-dimensional
spin matrices as they were discussed in part I, one can find from
(\ref{MatFormSpin}) that
\begin{equation}
(\Sigma^{\mu\nu})_{ab}(\Sigma_{\rho\sigma}^{-1})^{ab}=(\Sigma_{\rho\sigma
}^{-1})^{ab}(\Sigma^{\mu\nu})_{ab}=\lambda_{+}(P_{A})^{\mu\nu}{}_{\rho\sigma}.
\end{equation}

Next, we come to useful decomposition formulae involving products of $\gamma$-
and $\Sigma$-matrices. For their undeformed counterparts we refer to Ref.
\cite{Core}, for example. Inserting the expressions for $\gamma$- and $\Sigma
$-matrices into suitable ansaetze, we could derive the identities%
\begin{align}
\gamma^{\mu}\gamma^{\nu}=  &  \;\Sigma^{\mu\nu}-\lambda_{+}^{-1}\eta^{\mu\nu
},\nonumber\\
\gamma^{5}\gamma^{\mu}\gamma^{\nu}=  &  -\lambda_{+}^{-1}\eta^{\mu\nu}%
\gamma^{5}-q^{2}\lambda_{+}^{-1}\,\varepsilon^{\mu\nu\rho\sigma}\eta_{\rho
\rho^{\prime}}\eta_{\sigma\sigma^{\prime}}\Sigma^{\rho^{\prime}\sigma^{\prime
}},\label{sigmarelminkgr0}\\[0.16in]
\gamma^{5}\Sigma^{\mu\nu}=  &  -q^{-2}\lambda_{+}^{-1}\,\varepsilon^{\mu
\nu\rho\sigma}\,\eta_{\rho\rho^{\prime}}\eta_{\sigma\sigma^{\prime}}%
\Sigma^{\rho^{\prime}\sigma^{\prime}},\nonumber\\
\gamma^{\rho}\Sigma^{\mu\nu}=  &  \;\frac{q^{2}}{2}\,\varepsilon^{\rho\mu
\nu\lambda}\,\eta_{\lambda\delta}\,\gamma^{5}\gamma^{\delta}-\frac{1}%
{2}\lambda_{+}^{-1}(P_{A})^{\mu\nu}{}_{\mu^{\prime}\nu^{\prime}}\,\eta
^{\rho\mu^{\prime}}\gamma^{nu^{\prime}},\nonumber\\
\Sigma^{\mu\nu}\gamma^{\rho}=  &  \;\frac{q^{2}}{2}\,\varepsilon^{\mu\nu
\rho\lambda}\,\eta_{\lambda\delta}\,\gamma^{5}\gamma^{\delta}-\frac{1}%
{2}\lambda_{+}^{-1}(P_{A})^{\mu\nu}{}_{\mu^{\prime}\nu^{\prime}}\,\eta
^{\nu^{\prime}\rho}\gamma^{\mu^{\prime}},\\[0.16in]
\gamma^{5}\gamma^{\rho}\Sigma^{\mu\nu}=  &  \;\frac{q^{2}}{2}\,\varepsilon
^{\rho\mu\nu\lambda}\,\eta_{\lambda\delta}\,\gamma^{\delta}-\frac{1}{2}%
\lambda_{+}^{-1}(P_{A})^{\mu\nu}{}_{\mu^{\prime}\nu^{\prime}}\,\eta^{\rho
\mu^{\prime}}\gamma^{5}\gamma^{\nu^{\prime}},\nonumber\\
\Sigma^{\mu\nu}\gamma^{5}\gamma^{\rho}=  &  \;\frac{q^{2}}{2}\,\varepsilon
^{\mu\nu\rho\lambda}\,\eta_{\lambda\delta}\,\gamma^{\delta}-\frac{1}{2}%
\lambda_{+}^{-1}(P_{A})^{\mu\nu}{}_{\mu^{\prime}\nu^{\prime}}\,\eta
^{\nu^{\prime}\rho}\gamma^{5}\gamma^{\mu^{\prime}},\\[0.16in]
\Sigma^{\mu\nu}\Sigma^{\kappa\lambda}=  &  \;\frac{1}{2}(P_{A})^{\mu\nu}%
{}_{\mu^{\prime}\nu^{\prime}}(P_{A})^{\kappa\lambda}{}_{\kappa^{\prime}%
\lambda^{\prime}}\,\eta^{\mu^{\prime}\lambda^{\prime}}\eta^{\nu^{\prime}%
\kappa^{\prime}}-\frac{q^{2}}{2}\lambda_{+}^{-1}\,\varepsilon^{\mu\nu
\kappa\lambda}\gamma^{5},\nonumber\\
&  -\lambda_{+}(P_{A})^{\mu\nu}{}_{\mu^{\prime}\nu^{\prime}}(P_{A}%
)^{\kappa\lambda}{}_{\kappa^{\prime}\lambda^{\prime}}\,\eta^{\nu^{\prime
}\kappa^{\prime}}\Sigma^{\mu^{\prime}\lambda^{\prime}}.
\label{sigmarelminkgr1}%
\end{align}

Taking the quantum trace over\ $\Sigma$-matrices and their products yields
\begin{align}
\mbox{Tr}_{q}\;\Sigma^{\mu\nu}  &  =0,\nonumber\\
\mbox{Tr}_{q}\;\Sigma^{\rho\sigma}\Sigma^{\mu\nu}  &  =q^{-1}\lambda_{+}%
(P_{A})^{\mu\nu}{}_{\mu^{\prime}\nu^{\prime}}\hat{R}^{\sigma\mu^{\prime}}%
{}_{\mu^{\prime\prime}\sigma^{\prime}}\,\eta^{\rho\mu^{\prime\prime}}%
\eta^{\sigma^{\prime}\nu^{\prime}}, \label{TracSpin}%
\end{align}
where in the last identity we are allowed to apply the substitutions $\hat
{R}\rightarrow\hat{R}^{-1},$ $q\rightarrow q^{-1}$. Notice that the first
relation in (\ref{TracSpin}) conforms with the observation that the
two-dimensional spin matrices $\sigma^{\mu\nu}$ and $\bar{\sigma}^{\mu\nu}$
are traceless (in a $q$-deformed sense).

Now, we write down $q$-deformed analoga of some relations given on page 13
of\ Ref. \cite{Core}, for example:
\begin{equation}
\eta_{\mu\mu^{\prime}}\hat{R}^{\mu^{\prime}\nu}{}_{\nu^{\prime}\mu
^{\prime\prime}}\hat{R}^{\mu^{\prime\prime}\rho}{}_{\rho^{\prime}\mu
^{\prime\prime\prime}}\,\gamma^{\mu}\,\Sigma^{\nu^{\prime}\rho^{\prime}}%
\gamma^{\mu^{\prime\prime\prime}}=0, \label{CoreP130}%
\end{equation}%
\begin{align}
&  \gamma^{\mu}\Sigma^{\nu^{\prime}\rho^{\prime}}\gamma^{\sigma^{\prime}%
}\gamma^{\mu^{\prime\prime\prime\prime}}\hat{R}^{\mu^{\prime}\nu}{}%
_{\nu^{\prime}\mu^{\prime\prime}}\hat{R}^{\mu^{\prime\prime}\rho}{}%
_{\rho^{\prime}\mu^{\prime\prime\prime}}\hat{R}^{\mu^{\prime\prime\prime
}\sigma}{}_{\sigma^{\prime}\mu^{\prime\prime\prime\prime}}\eta_{\mu\mu
^{\prime}}=\nonumber\\
&  \qquad\qquad=\,q^{2}\,\hat{R}^{\nu\sigma^{\prime}}{}_{\sigma^{\prime\prime
}\nu^{\prime}}\hat{R}^{\rho\sigma}{}_{\sigma^{\prime}\rho^{\prime}}%
\,\gamma^{\sigma^{\prime\prime}}\Sigma^{\nu^{\prime}\rho^{\prime}},\\[0.06in]
&  \gamma^{\mu}\gamma^{\nu^{\prime}}\Sigma^{\rho^{\prime}\sigma^{\prime}%
}\gamma^{\mu^{\prime\prime\prime\prime\prime}}\hat{R}^{\mu^{\prime}\nu}{}%
_{\nu^{\prime}\mu^{\prime\prime}}\hat{R}^{\mu^{\prime\prime}\rho}{}%
_{\rho^{\prime}\mu^{\prime\prime\prime}}\hat{R}^{\mu^{\prime\prime\prime
}\sigma}{}_{\sigma^{\prime}\mu^{\prime\prime\prime\prime}}\eta_{\mu\mu
^{\prime}}=\nonumber\\
&  \qquad\qquad=\,q^{2}\,\hat{R}^{\nu^{\prime}\sigma}{}_{\sigma^{\prime}%
\nu^{\prime\prime}}\hat{R}^{\nu\rho}{}_{\rho^{\prime}\nu^{\prime}}\Sigma
^{\rho^{\prime}\sigma^{\prime}}\gamma^{\nu^{\prime\prime}}.
\end{align}%
\begin{align}
&  \gamma^{\mu}\gamma^{\rho_{1}^{\prime}}\gamma^{\rho_{2}^{\prime}}%
\gamma^{\rho_{3}^{\prime}}\gamma^{\rho_{4}^{\prime}}\gamma^{\nu}\hat{R}%
^{\mu^{\prime}\rho_{1}}{}_{\rho_{1}^{\prime}\mu^{\prime\prime}}\hat{R}%
^{\mu^{\prime\prime}\rho_{2}}{}_{\rho_{2}^{\prime}\mu^{\prime\prime\prime}%
}\hat{R}^{\mu^{\prime\prime\prime}\rho_{3}}{}_{\rho_{3}^{\prime}\mu
^{\prime\prime\prime\prime}}\hat{R}^{\mu^{\prime\prime\prime\prime}\rho_{4}}%
{}_{\rho_{4}^{\prime}\nu}\eta_{\mu\mu^{\prime}}=\nonumber\\
&  \qquad\qquad=\,q^{2}\,\gamma^{\rho_{4}^{\prime\prime\prime}}\gamma
^{\rho_{1}^{\prime}}\gamma^{\rho_{2}^{\prime}}\gamma^{\rho_{3}^{\prime}}%
\hat{R}^{\rho_{1}\rho_{4}^{\prime\prime}}{}_{\rho_{4}^{\prime\prime\prime}%
\rho_{1}^{\prime}}\hat{R}^{\rho_{2}\rho_{4}^{\prime}}{}_{\rho_{4}%
^{\prime\prime}\rho_{2}^{\prime}}\hat{R}^{\rho_{3}\rho_{4}}{}_{\rho
_{4}^{\prime}\rho_{3}^{\prime}}\nonumber\\
&  \qquad\qquad\hspace{0.19in}+\,q^{4}\,\gamma^{\rho_{2}^{\prime}}\gamma
^{\rho_{3}^{\prime}}\gamma^{\rho_{4}^{\prime}}\gamma^{\rho_{1}^{\prime
\prime\prime}}\hat{R}^{\rho_{1}\rho_{2}}{}_{\rho_{2}^{\prime}\rho_{1}^{\prime
}}\hat{R}^{\rho_{1}^{\prime}\rho_{3}}{}_{\rho_{3}^{\prime}\rho_{1}%
^{\prime\prime}}\hat{R}^{\rho_{1}^{\prime\prime}\rho_{4}}{}_{\rho_{4}^{\prime
}\rho_{1}^{\prime\prime\prime}}=\nonumber\\
&  \qquad\qquad=\,q^{2}\,\gamma^{\rho_{3}^{\prime\prime}}\gamma^{\rho
_{2}^{\prime\prime}}\gamma^{\rho_{1}^{\prime\prime}}\gamma^{\rho_{4}}\hat
{R}^{\rho_{2}^{\prime}\rho_{3}^{\prime}}{}_{\rho_{3}^{\prime\prime}\rho
_{2}^{\prime\prime}}\hat{R}^{\rho_{1}\rho_{2}}{}_{\rho_{2}^{\prime}\rho
_{1}^{\prime}}\hat{R}^{\rho_{1}^{\prime}\rho_{3}}{}_{\rho_{3}^{\prime}\rho
_{1}^{\prime\prime}}\nonumber\\
&  \qquad\qquad\hspace{0.19in}+\,q^{4}\,\gamma^{\rho_{1}}\gamma^{\rho
_{4}^{\prime\prime}}\gamma^{\rho_{3}^{\prime\prime}}\gamma^{\rho_{2}%
^{\prime\prime}}\hat{R}^{\rho_{2}\rho_{4}^{\prime}}{}_{\rho_{4}^{\prime\prime
}\rho_{2}^{\prime}}\hat{R}^{\rho_{2}^{\prime}\rho_{3}^{\prime}}{}_{\rho
_{3}^{\prime\prime}\rho_{2}^{\prime\prime}}\hat{R}^{\rho_{3}\rho_{4}}{}%
_{\rho_{4}^{\prime}\rho_{3}^{\prime}}. \label{CoreP13}%
\end{align}
Substituting $\hat{R}$ and $q$ by $\hat{R}^{-1}$ and $q^{-1}$, respectively,
leads to further relations. To get the undeformed counterparts of
(\ref{CoreP130})-(\ref{CoreP13}) one simple has to replace the $\hat{R}%
$-matrices by a normal twist, i.e.
\begin{equation}
\hat{R}^{\mu\nu}{}_{\rho\sigma}=\delta_{\sigma}^{\mu}\,\delta_{\rho}^{\nu}.
\end{equation}

The corresponding relations for 'inverse' $\gamma$- and $\Sigma$-matrices
follow from the formulae in (\ref{CoreP130})-(\ref{CoreP13}) via the
replacements (\ref{replinv}) together with%
\begin{equation}
\Sigma^{\mu\nu}\rightarrow\Sigma_{\mu\nu}^{-1},\qquad(P_{A})^{\mu\nu}{}%
_{\rho\sigma}\rightarrow(P_{A})^{\rho\sigma}{}_{\mu\nu}.
\end{equation}
By the way this correspondence also applies to the formulae in
(\ref{sigmarelminkgr0})-(\ref{TracSpin}).

Remarkably, products of $\Sigma$-matrices with $\gamma^{5}$ reduce to linear
combinations of $\Sigma$-matrices:
\begin{align}
\Sigma^{+3}\gamma^{5}  &  =-2q\lambda_{+}^{-1}\Sigma^{+0}+\lambda\lambda
_{+}^{-1}\Sigma^{+3},\nonumber\\
\Sigma^{+0}\gamma^{5}  &  =-2q^{-1}\lambda_{+}^{-1}\Sigma^{+3}-\lambda
\lambda_{+}^{-1}\Sigma^{+0},\nonumber\\
\Sigma^{+-}\gamma^{5}  &  =-2\lambda_{+}^{-1}\Sigma^{30}+\lambda\lambda
_{+}^{-1}\Sigma^{+-},\nonumber\\
\Sigma^{30}\gamma^{5}  &  =-2\lambda_{+}^{-1}\Sigma^{+-}-\lambda\lambda
_{+}^{-1}\Sigma^{30},\nonumber\\
\Sigma^{3-}\gamma^{5}  &  =2q\lambda_{+}^{-1}\Sigma^{0-}-\lambda\lambda
_{+}^{-1}\Sigma^{3-},\nonumber\\
\Sigma^{0-}\gamma^{5}  &  =2q^{-1}\lambda_{+}^{-1}\Sigma^{3-}+\lambda
\lambda_{+}^{-1}\Sigma^{0-}. \label{sgamma5minkex}%
\end{align}
The corresponding formulae for 'inverse' $\gamma$- and $\Sigma$-matrices we
get by the substitutions $\Sigma^{\mu\nu}\rightarrow\Sigma_{\mu\nu}^{-1}$
together with $\lambda\rightarrow-\lambda$.

The matrices $1$, $\gamma^{5}$, $\gamma^{\mu}$, $\gamma^{5}\gamma^{\mu}$, and
$\Sigma^{\mu\nu}$ play an important role in physics, since they lead to
bilinear forms with definite transformation properties (see also the
discussion in the succeeding section). On these grounds, the following
expansion of an arbitrary matrix could prove useful:
\begin{align}
A  &  =a_{0}\,\mbox{1 \kern-.59em {\rm l}}+a_{5}\gamma^{5}+v_{\mu}\gamma^{\mu
}+a_{\mu}\gamma^{5}\gamma^{\mu}+T_{\mu\nu}\Sigma^{\mu\nu},\nonumber\\
a_{0}  &  =-\frac{1}{2}\lambda_{+}^{-1}\,\mbox{Tr}_{q}(A),\nonumber\\
a_{5}  &  =-\frac{1}{2}\lambda_{+}^{-1}\,\mbox{Tr}_{q}(\gamma^{5}%
A),\nonumber\\
v_{\mu}  &  =\frac{1}{2}\eta_{\mu\nu}\,\mbox{Tr}_{q}(\gamma^{\nu
}A),\nonumber\\
a_{\mu}  &  =-\frac{1}{2}\eta_{\mu\nu}\,\mbox{Tr}_{q}(\gamma^{5}\gamma^{\nu
}A),\nonumber\\
T_{\mu\nu}  &  =-\lambda_{+}^{-1}\eta_{\mu\mu^{\prime}}\eta_{\nu\nu^{\prime}%
}\,\mbox{Tr}_{q}(\Sigma^{\nu^{\prime}\mu^{\prime}}A). \label{expandmink1}%
\end{align}
There is also a version with 'inverse' matrices:
\begin{align}
A  &  =a^{0}\mbox{1 \kern-.59em {\rm l}}+a^{5}\gamma_{5}^{-1}+v^{\mu}%
\gamma_{\mu}^{-1}+a^{\mu}\gamma_{5}^{-1}\gamma_{\mu}^{-1}+T^{\mu\nu}%
\Sigma_{\mu\nu}^{-1},\nonumber\\
a^{0}  &  =-\frac{1}{2}\lambda_{+}^{-1}\,\mbox{Tr}_{q}(A),\nonumber\\
a^{5}  &  =-\frac{1}{2}\lambda_{+}^{-1}\,\mbox{Tr}_{q}(\gamma_{5}%
^{-1}A),\nonumber\\
v^{\mu}  &  =\frac{1}{2}\eta^{\mu\nu}\,\mbox{Tr}_{q}(\gamma_{\nu}%
^{-1}A),\nonumber\\
a^{\mu}  &  =-\frac{1}{2}\eta^{\mu\nu}\,\mbox{Tr}_{q}(\gamma_{5}^{-1}%
\gamma_{\nu}^{-1}A),\nonumber\\
T^{\mu\nu}  &  =-\lambda_{+}^{-1}\eta^{\mu\mu^{\prime}}\eta^{\nu\nu^{\prime}%
}\,\mbox{Tr}_{q}(\Sigma_{\nu^{\prime}\mu^{\prime}}^{-1}A), \label{expandmink2}%
\end{align}
but now with the quantum trace from Eq.\thinspace(\ref{trgammainv}).

The expansion in (\ref{expandmink1}) refers to the basis in
(\ref{basisgamma5mink}). Additionally, we could also find an expansion for the
basis\ in (\ref{gammabasis}):
\begin{align}
A  &  =a_{0}\mbox{1 \kern-.59em {\rm l}}+a_{16}\Gamma^{+30-}+v_{\mu}%
\gamma^{\mu}+a_{K}\Gamma^{K}+T_{\mu\nu}\Sigma^{\mu\nu},\nonumber\\
a_{0}  &  =-\frac{1}{2}\lambda_{+}^{-1}\,\mbox{Tr}_{q}(A),\nonumber\\
a_{16}  &  =-\frac{1}{8}q^{27}[[2]]_{q^{2}}^{-7}[[3]]_{q^{2}}^{-4}%
\,\mbox{Tr}_{q}(\Gamma^{1234}A),\nonumber\\
v_{\mu}  &  =\frac{1}{2}\eta_{\mu\nu}\mbox{Tr}_{q}(\gamma^{\nu}A),\nonumber\\
a_{K}  &  =\mbox{Tr}_{q}(\Gamma_{K}A),\nonumber\\
T_{\mu\nu}  &  =-\lambda_{+}^{-1}\eta_{\mu\mu^{\prime}}\eta_{\nu\nu^{\prime}%
}\mbox{Tr}_{q}(\Sigma^{\nu^{\prime}\mu^{\prime}}A), \label{expandmink3}%
\end{align}
where $K$ stands for an index running over $K\in{\{}+30,$ $+3-,$ $+0-,$
$30-{\}}$. The expansion with 'inverse' matrices now reads
\begin{align}
A  &  =a^{0}\mbox{1 \kern-.59em {\rm l}}+a^{16}\Gamma_{+30-}^{-1}+v^{\mu
}\gamma_{\mu}^{-1}+a^{K}\Gamma_{K}^{-1}+T^{\mu\nu}\Sigma_{\mu\nu}%
^{-1},\nonumber\\
a^{0}  &  =-\frac{1}{2}\lambda_{+}^{-1}\,\mbox{Tr}_{q}(A),\nonumber\\
a^{16}  &  =-\frac{1}{8}q^{27}[[2]]_{q^{2}}^{-7}[[3]]_{q^{2}}^{-4}%
\,\mbox{Tr}_{q}(\Gamma_{+30-}^{-1}A),\nonumber\\
v^{\mu}  &  =\frac{1}{2}\eta^{\mu\nu}\,\mbox{Tr}_{q}(\gamma_{\nu}%
^{-1}A),\nonumber\\
a^{K}  &  =\mbox{Tr}_{q}(\Gamma_{K}^{-1}A),\nonumber\\
T^{\mu\nu}  &  =-\lambda_{+}^{-1}\eta^{\mu\mu^{\prime}}\eta^{\nu\nu^{\prime}%
}\,\mbox{Tr}_{q}(\Sigma_{\nu^{\prime}\mu^{\prime}}^{-1}A), \label{expandmink4}%
\end{align}
where we again have to take the quantum trace from Eq.\thinspace
(\ref{trgammainv}).

\section{$q$-Deformed Dirac spinors and bilinear covariants\label{CovKap}}

Now, we should have everything together to introduce $q$-deformed Dirac
spinors together with their bilinear covariants. It is also our aim to discuss
the behavior of these objects under $q$-deformed Lorentz transformations
(notice that the reasonings in this section refer to $q$-deformed
Min\-kow\-ski space). For this reason, we start our reasonings with a short
review of the $q$-deformed Lorentz group \cite{CSSW90, PW90}. Finally, we
should remark that in this section we work with Pauli matrices that are
multiplied by an overall factor $\lambda_{+}^{1/2}$ compared to the
expressions given in part I. In doing so, we avoid additional factors in our
formulae and the matrix $\gamma^{0}$ takes on the very simple form
\begin{equation}
\gamma^{0}=\left(
\begin{array}
[c]{cccc}%
0 & 0 & 1 & 0\\
0 & 0 & 0 & 1\\
1 & 0 & 0 & 0\\
0 & 1 & 0 & 0
\end{array}
\right)  .
\end{equation}

\subsection{The quantum group $SL_{q}(2)$ and its coactions on the quantum
plane}

In view of their importance for what follows, we repeat some facts given
already in part I. First, the algebra of $SL_{q}(2)$ is generated by four
non-commuting generators $a$, $b$, $c$, and $d$, subject to the relations
\begin{gather}
ab=qba,\quad ac=qca,\quad bc=cb,\quad bd=qdb,\nonumber\\
cd=qdc,\quad ad-da=(q-q^{-1})bc. \label{Mrel}%
\end{gather}
These generators can be recognized as entries of a matrix
\begin{equation}
M^{i}{}_{j}=\left(
\begin{array}
[c]{cc}%
a & b\\
c & d
\end{array}
\right)  .
\end{equation}
This notation enables us to write the relations in (\ref{Mrel}) with the
$\hat{R}$-matrix of $U_{q}(su(2))$ \cite{Schmidke, Wess00}:
\begin{equation}
\hat{R}^{ij}{}_{kl}\;M^{k}{}_{r}M^{l}{}_{s}=M^{i}{}_{k}M^{j}{}_{l}\hat{R}%
^{kl}{}_{rs}, \label{MrelR}%
\end{equation}
and the Hopf structure of $SL_{q}(2)$ takes on the form%
\begin{align}
\Delta(M^{i}{}_{j})  &  =M^{i}{}_{k}\otimes M^{k}{}_{j},\nonumber\\
S(M^{i}{}_{j})  &  =\varepsilon_{lj}M^{l}{}_{k}\varepsilon^{ik},\nonumber\\
S^{-1}(M^{i}{}_{j})  &  =\varepsilon_{jl}M^{l}{}_{k}\varepsilon^{ki}%
,\nonumber\\
\epsilon(M^{i}{}_{j})  &  =\delta^{i}{}_{j}, \label{hopfsl2}%
\end{align}
where the non-zero entries of the $q$-deformed spinor metric are given by%
\begin{equation}
\varepsilon_{12}=-q^{-1/2},\qquad\varepsilon_{21}=q^{1/2}.
\end{equation}

The quantum group $SL_{q}(2)$ gives rise to coactions on the so-called quantum
plane. (If the reader is not familiar with the notion of a coaction we
recommend to consult Refs. \cite{Maj95, ChDe96, Klimyk}). Remember that the
coordinates on the quantum plane are viewed as $q$-analogs of spinors. The
commutation relations for symmetrized spinors read as
\begin{equation}
x^{1}x^{2}=qx^{2}x^{1}\qquad\text{and}\qquad x_{1}x_{2}=q^{-1}x_{2}x_{1}.
\end{equation}
(There is also an antisymmetrized version of the quantum plane \cite{MSW04,
SW04}. However, our reasonings in this section do not depend on whether we
work with symmetrized or antisymmetrized spinors.) Notice that indices of
spinors are raised and lowered by the spinor metric, i.e., for example,
$x_{i}=\varepsilon_{ij}x^{j}$. Finally, let us recall the left- and
right-coactions on covariant as well as contravariant\ spinor coordinates [see
also part I]:%
\begin{align}
\beta_{L}(x^{i}) &  =M^{i}{}_{j}\otimes x^{j},\nonumber\\
\beta_{L}(x_{i}) &  =S^{-1}(M^{j}{}_{i})\otimes x_{j},\\[0.08in]
\beta_{R}(x^{i}) &  =x^{j}\otimes(M^{T})^{i}{}_{j},\nonumber\\
\beta_{R}(x_{i}) &  =x_{j}\otimes S((M^{T})^{j}{}_{i}).\label{coact3}%
\end{align}

\subsection{The quantum group $SL_{q}(2,\mathbb{C})$}

Next, we would like to describe the quantum group $SL_{q}(2,\mathbb{C})$, as
it is necessary for the discussion of the transformation properties of Dirac
spinors and their bilinear covariants. In part I we clarified the structure of
the quantum groups $SL_{q}(2)$ and $SU_{q}(2)$. The quantum groups $SU_{q}(2)$
and $SL_{q}(2,\mathbb{C})$ obey the same algebra relations and Hopf structure
as $SL_{q}(2)$, but differ in their $\ast$-structure. In the case of
$SU_{q}(2)$, conjugation maps the quantum group $SL_{q}(2)$ onto itself. In
the case of $SL_{q}(2,\mathbb{C})$, however, the conjugated generators of
$SL_{q}(2)$ live in a second copy of $SL_{q}(2\dot{)}$.

For this to become more clear, we denote the generators of the first copy of
$SL_{q}(2)$ by $M^{i}{}_{j}$ and those of the second copy by $\tilde{M}^{i}%
{}_{j}$. The two copies of $SL_{q}(2)$ can be combined to form one algebra if
we assume for the two sets of generators that%
\begin{equation}
\hat{R}^{ij}{}_{kl}\,M^{k}{}_{s}\tilde{M}^{l}{}_{t}=\tilde{M}^{i}{}_{m}M^{j}%
{}_{n}\,\hat{R}^{mn}{}_{st},\label{MMtilderel}%
\end{equation}
where $\hat{R}^{ij}{}_{kl}$ again stands for the $\hat{R}$-matrix of
$U_{q}(su(2))$.

If we work with left-coactions of the two copies of $SL_{q}(2)$, their
generators $M^{i}{}_{j}$ and $\tilde{M}^{i}{}_{j}$ are related to each other
by
\begin{equation}
\overline{M^{i}{}_{j}}\equiv\bar{M}^{i}{}_{j}=S^{-1}(\tilde{M}^{j}{}_{i}%
)\quad\Rightarrow\quad(M^{\dagger})^{i}{}_{j}=S^{-1}(\tilde{M}^{i}{}%
_{j}),\label{slq2ceins}%
\end{equation}
where $(M^{\text{T}})^{\alpha}{}_{\beta}=M^{\beta}{}_{\alpha}$ and
$M^{\dagger}=\overline{M^{\text{T}}}=\bar{M}^{\text{T}}$. However, this $\ast
$-structure is not consistent with right-coactions of the two copies of
$SL_{q}(2)$, since it would violate the $\ast$-coalgebra axiom%
\begin{equation}
\beta_{R}(\overline{x^{i}})=\overline{\beta_{R}(x^{i})}.
\end{equation}
If we deal with right-coactions, we instead have to chose the $\ast$-structure%
\begin{equation}
\bar{M}^{i}{}_{j}=S(\tilde{M}^{j}{}_{i})\quad\Rightarrow\quad(M^{\dagger}%
)^{i}{}_{j}=S(\tilde{M}^{i}{}_{j}).\label{slq2czwei}%
\end{equation}

Finally, it should be mentioned that the two copies of $SL_{q}(2)$ play
symmetrical roles. With the help of the $\ast$-Hopf algebra axiom $\ast\circ
S^{-1}=S\circ\ast$ we can invert the identities in (\ref{slq2ceins}) and
(\ref{slq2czwei}). This way, we respectively arrive at
\begin{equation}
\bar{\tilde{M}}^{i}{}_{j}=S^{-1}(M^{j}{}_{i})\quad\text{and}\quad\bar
{\tilde{M}}^{i}{}_{j}=S(M^{j}{}_{i}). \label{slq2cdrei}%
\end{equation}

\subsection{Relationship between $q$-Lorentz group and $SL_{q}(2,\mathbb{C})$}

In part I we introduced the generators of the quantum Lorentz group in spinor
basis, i.e.%
\begin{equation}
\Lambda^{\alpha\dot{\beta}}{}_{\gamma\dot{\delta}}=M^{\alpha}{}_{\gamma}%
{}\otimes\tilde{M}^{\dot{\beta}}{}_{\dot{\delta}}\in SL_{q}(2)\otimes
SL_{q}(2),\quad\alpha,\dot{\beta},\gamma,\dot{\delta}\in\{1,2\}.
\end{equation}
It arises the question how to define the generators of quantum Lorentz group
in a vector basis, i.e. $\Lambda^{\mu}{}_{\nu}$, $\mu$, $\nu\in\{+$, $3$, $0$,
$-\}$? To answer this question we consider the coaction [cf. part I]%
\begin{align}
\beta_{L}(X^{\mu})=  &  \,(\sigma^{\mu})_{\alpha\dot{\beta}}\,\beta
_{L}(X^{\alpha\dot{\beta}})\nonumber\\
=  &  \,(\sigma^{\mu})_{\alpha\dot{\beta}}\,M^{\alpha}{}_{\gamma}\tilde
{M}^{\dot{\beta}}{}_{\dot{\delta}}\otimes X^{\gamma\dot{\delta}}\nonumber\\
=  &  \,(\sigma^{\mu})_{\alpha\dot{\beta}}\,M^{\alpha}{}_{\gamma}\tilde
{M}^{\dot{\beta}}{}_{\dot{\delta}}\,(\sigma_{\nu}^{-1})^{\gamma\dot{\delta}%
}\otimes X^{\nu}\nonumber\\
\overset{!}{=}  &  \,\Lambda^{\mu}{}_{\nu}\otimes X^{\nu}.
\end{align}
From the last equality of the above calculation we read off that
\begin{equation}
\Lambda^{\mu}{}_{\nu}=(\sigma^{\mu})_{\alpha\dot{\beta}}\,M^{\alpha}{}%
_{\gamma}\tilde{M}^{\dot{\beta}}{}_{\dot{\delta}}\,(\sigma_{\nu}^{-1}%
)^{\gamma\dot{\delta}}. \label{gammatransmink1}%
\end{equation}
Similar arguments lead us to
\begin{equation}
\Lambda^{\mu}{}_{\nu}=(\bar{\sigma}^{\mu})_{\dot{\alpha}\beta}\,(\tilde
{M})^{\dot{\alpha}}{}_{\dot{\gamma}}M^{\beta}{}_{\delta}\,(\bar{\sigma}_{\nu
}^{-1})^{\dot{\gamma}\delta}. \label{gammatransmink2}%
\end{equation}

The generators in (\ref{gammatransmink1}) as well as (\ref{gammatransmink2})
indeed fulfill the crucial equation
\begin{equation}
\eta_{\mu\nu}\,\Lambda^{\mu}{}_{\mu^{\prime}}\Lambda^{\nu}{}_{\nu^{\prime}%
}=\eta_{\mu^{\prime}\nu^{\prime}}.
\end{equation}
To prove this relation we first transform the quantum metric and the group
generators into a spinor basis. Exploiting the properties of the Pauli
matrices together with those of the generators\ $M^{\alpha}{}_{\gamma}$ and
$\tilde{M}^{\dot{\beta}}{}_{\dot{\delta}}$ then should show us the validity of
the above identity.

The relations in (\ref{gammatransmink1}) and (\ref{gammatransmink2}) can also
be written in a slightly different form. To this end, let us return to the
coaction $\beta_{L}(X^{\mu})$:
\begin{align}
\beta_{L}(X^{\mu})=  &  \,\beta_{L}(X^{\alpha}{}_{\dot{\alpha}}(\sigma^{\mu
})_{\alpha}{}^{\dot{\alpha}})\nonumber\\
=  &  \,(\sigma^{\mu})_{\alpha}{}^{\dot{\alpha}}M^{\alpha}{}_{\beta}%
\,S^{-1}(\tilde{M}^{\dot{\beta}}{}_{\dot{\alpha}})\otimes X^{\beta}{}%
_{\dot{\beta}}\nonumber\\
=  &  \,(\sigma^{\mu})_{\alpha}{}^{\dot{\alpha}}M^{\alpha}{}_{\beta
}\,(M^{\dagger})^{\dot{\beta}}{}_{\dot{\alpha}}\otimes X^{\beta}{}_{\dot
{\beta}}\nonumber\\
\overset{!}{=}  &  \,\Lambda^{\mu}{}_{\nu}\otimes X^{\nu}=\Lambda^{\mu}{}%
_{\nu}\,(\sigma^{\nu})_{\beta}{}^{\dot{\beta}}\otimes X^{\beta}{}_{\dot{\beta
}}.
\end{align}
The last calculation should tell us that
\begin{equation}
(\Lambda_{L})^{\mu}{}_{\nu}\,(\sigma^{\nu})_{\beta}{}^{\dot{\beta}}=M^{\alpha
}{}_{\beta}\,(\sigma^{\mu})_{\alpha}{}^{\dot{\alpha}}\bar{M}^{\dot{\alpha}}%
{}_{\dot{\beta}},
\end{equation}
or, for short,%
\begin{equation}
(\Lambda_{L})^{\mu}{}_{\nu}\,\sigma^{\nu}=M^{\text{T}}\sigma^{\mu}\bar{M}.
\end{equation}
Similar arguments yield the following list of matrix identities:%
\begin{align}
(\Lambda_{L})^{\mu}{}_{\nu}\,\sigma^{\nu}  &  =M_{L}^{\text{T}}\sigma^{\mu
}\bar{M}_{L}, & (\Lambda_{L})^{\mu}{}_{\nu}\,\bar{\sigma}^{\nu}  &  =\tilde
{M}_{L}^{\text{T}}\bar{\sigma}^{\mu}\bar{\tilde{M}}_{L},\nonumber\\
(\Lambda_{R}^{\text{T}})^{\mu}{}_{\nu}\,\sigma^{\nu}  &  =M_{R}\,\sigma^{\mu
}M_{R}^{\dagger}, & (\Lambda_{R}^{\text{T}})^{\mu}{}_{\nu}\,\bar{\sigma}%
^{\nu}  &  =\tilde{M}_{R}\,\bar{\sigma}^{\mu}\tilde{M}_{R}^{\dagger
},\label{TransPaul1}\\[0.1in]
(\Lambda_{L})^{\mu}{}_{\nu}\,\sigma_{\mu}^{-1}  &  =\bar{\tilde{M}}%
_{L}\,\sigma_{\nu}^{-1}\tilde{M}_{L}^{\text{T}}, & (\Lambda_{L})^{\mu}{}_{\nu
}\,\bar{\sigma}_{\mu}^{-1}  &  =\bar{M}_{L}\,\bar{\sigma}_{\nu}^{-1}%
M_{L}^{\text{T}},\nonumber\\
(\Lambda_{R}^{\text{T}})^{\mu}{}_{\nu}\,\sigma_{\mu}^{-1}  &  =\tilde{M}%
_{R}^{\dagger}\,\sigma_{\nu}^{-1}\tilde{M}_{R}, & (\Lambda_{R}^{\text{T}%
})^{\mu}{}_{\nu}\,\bar{\sigma}_{\mu}^{-1}  &  =M_{R}^{\dagger}\,\bar{\sigma
}_{\nu}^{-1}M_{R}. \label{TransPaul2}%
\end{align}
Notice that the two labels $L$ and $R$ were introduced to indicate the type of
coaction (left- or right-coaction) to which each of the above relations
corresponds to. Furthermore, one should realize that the derivation of the
identities in (\ref{TransPaul2}) starts from applying coactions to the
defining relations of 'inverse' Pauli matrices, i.e.%
\begin{equation}
X^{\mu}(\sigma_{\mu}^{-1})_{\alpha}{}^{\dot{\beta}}=X_{\alpha}{}^{\dot{\beta}%
},\qquad X^{\mu}(\bar{\sigma}_{\mu}^{-1})_{\dot{\alpha}}{}^{\beta}%
=X_{\dot{\alpha}}{}^{\beta}.
\end{equation}

\subsection{Definition of Dirac spinors and bilinear covariants}

We introduce the \textit{Dirac spinor} as
\begin{equation}
\Psi_{a}\equiv\left(
\begin{array}
[c]{c}%
\psi_{\alpha}\\
\bar{\lambda}_{\dot{\alpha}}%
\end{array}
\right)  ,
\end{equation}
i.e. it is a $4$-spinor written with the two $2$-spinors
\begin{equation}
\psi_{\alpha}=\left(
\begin{array}
[c]{c}%
\psi_{1}\\
\psi_{2}%
\end{array}
\right)  ,\qquad\bar{\lambda}_{\dot{\alpha}}=\left(
\begin{array}
[c]{c}%
\bar{\lambda}_{\dot{1}}\\
\bar{\lambda}_{\dot{2}}%
\end{array}
\right)  .
\end{equation}
Then the \textit{Hermitian conjugate} of a Dirac spinor should become
\begin{equation}
(\Psi^{\dagger})^{a}\equiv(\bar{\psi}^{\dot{\alpha}},\lambda^{\alpha}),
\label{HerConDir}%
\end{equation}
where
\begin{equation}
\bar{\psi}^{\dot{\alpha}}=(\bar{\psi}^{\dot{1}},\bar{\psi}^{\dot{2}%
})=-\varepsilon^{\dot{\alpha}\dot{\alpha}^{\prime}}\bar{\psi}_{\dot{\alpha
}^{\prime}},\qquad\lambda^{\alpha}=(\lambda^{1},\lambda^{2})=\varepsilon
^{\alpha\alpha^{\prime}}\lambda_{\alpha^{\prime}}.
\end{equation}
In physics, however, it is more convenient to work with the so-called
\textit{Dirac conjugated }$4$\textit{-spinor}:
\begin{equation}
\bar{\Psi}^{a}\equiv(\lambda^{\alpha},\bar{\psi}^{\dot{\alpha}})=(\Psi
^{\dagger})\gamma^{0}.
\end{equation}

Next, we would like to introduce covariant bilinear forms. In doing so, we
need a second Dirac spinor
\begin{equation}
\Phi_{a}\equiv\left(
\begin{array}
[c]{c}%
\chi_{\alpha}\\
\bar{\phi}_{\dot{\alpha}}%
\end{array}
\right)  .
\end{equation}
Its Hermitian and Dirac conjugate are defined in complete analogy to those for
the Dirac spinor $\Psi$. With the two Dirac spinors $\Psi$ and $\Phi$ at hand
we are now in a position to introduce the bilinear covariants
\begin{align}
\bar{\Phi}\Psi &  =\phi^{\beta}\psi_{\beta}+\bar{\chi}^{\dot{\beta}}%
\bar{\lambda}_{\dot{\beta}},\nonumber\\
\bar{\Phi}\,\gamma^{5}\Psi &  =-\phi^{\beta}\psi_{\beta}+\bar{\chi}%
^{\dot{\beta}}\bar{\lambda}_{\dot{\beta}},\nonumber\\
\bar{\Phi}\,\gamma^{\mu}\Psi &  =\phi^{\alpha}(\sigma^{\mu})_{\alpha}{}%
^{\dot{\beta}}\bar{\lambda}_{\dot{\beta}}+\bar{\chi}^{\dot{\alpha}}%
(\bar{\sigma}^{\mu})_{\dot{\alpha}}{}^{\beta}\psi_{\beta},\nonumber\\
\bar{\Phi}\,\gamma^{\mu}\gamma^{5}\Psi &  =\phi^{\alpha}(\sigma^{\mu}%
)_{\alpha}{}^{\dot{\beta}}\bar{\lambda}_{\dot{\beta}}-\bar{\chi}^{\dot{\alpha
}}(\bar{\sigma}^{\mu})_{\dot{\alpha}}{}^{\beta}\psi_{\beta},\nonumber\\
\bar{\Phi}\Sigma^{\mu\nu}\Psi &  =\phi^{\alpha}(\sigma^{\mu\nu})_{\alpha}%
{}^{\beta}\psi_{\beta}+\bar{\chi}^{\dot{\alpha}}(\bar{\sigma}^{\mu\nu}%
)_{\dot{\alpha}}{}^{\dot{\beta}}\bar{\lambda}_{\dot{\beta}},\nonumber\\
\bar{\Phi}\Sigma^{\mu\nu}\gamma^{5}\Psi &  =-\phi^{\alpha}(\sigma^{\mu\nu
})_{\alpha}{}^{\beta}\psi_{\beta}+\bar{\chi}^{\dot{\alpha}}(\bar{\sigma}%
^{\mu\nu})_{\dot{\alpha}}{}^{\dot{\beta}}\bar{\lambda}_{\dot{\beta}%
}.\label{BiCov}%
\end{align}
Notice, that the first five covariants are sufficient to constitute a basis,
since $\Sigma^{\mu\nu}\gamma^{5}$ can be expressed as a linear combination of
$\Sigma$-matrices [cf. the relations in (\ref{sgamma5minkex})]. The
expressions in terms of $2$-spinors are a direct consequence of the definition
of Dirac spinors and the explicit form of the matrices $\gamma^{5}$,
$\gamma^{\mu}$, and $\Sigma^{\mu\nu}$.

\subsection{Conjugation properties}

Before we discuss the transformation properties of the bilinear forms in
(\ref{BiCov}) we would like to say a few words about their behavior under
conjugation. To begin with, we list the conjugation properties of the
different types of $2$-spinors:%
\begin{equation}
\overline{\psi_{\beta}}=\bar{\psi}^{\dot{\beta}},\quad\overline{\bar{\psi
}^{\dot{\beta}}}=\psi_{\beta},\quad\overline{\lambda^{\beta}}=\bar{\lambda
}_{\dot{\beta}},\quad\overline{\bar{\lambda}_{\dot{\beta}}}=\lambda^{\beta}.
\label{Con2Spin}%
\end{equation}
It should be clear that this conjugation assignment is consistent with
(\ref{HerConDir}).

Next, we come to the conjugation properties of the Pauli matrices, which can
be derived most easily from\ their defining relations, i.e.%
\begin{equation}
X^{\mu}=(\sigma^{\mu})_{\alpha\dot{\beta}}\,X^{\alpha\dot{\beta}}=(\bar
{\sigma}^{\mu})_{\dot{\alpha}{\beta}}\,X^{\dot{\alpha}{\beta}}.
\label{anspeu31}%
\end{equation}
Conjugating the expressions in (\ref{anspeu31}) and exploiting the identities
\begin{equation}
\overline{X^{\mu}}=(-1)^{\delta_{0\mu}}X_{\mu},\text{\quad}\overline
{X^{\alpha\dot{\beta}}}=X_{\beta\dot{\alpha}},
\end{equation}
we finally get
\begin{align}
\overline{(\sigma^{\mu})_{\alpha\dot{\beta}}}  &  =(\sigma^{\mu})_{\dot{\beta
}\alpha}^{\dagger}=(-1)^{\delta_{0\mu}}(\sigma_{\mu})^{\beta\dot{\alpha}%
},\nonumber\\
\overline{(\bar{\sigma}^{\mu})_{\dot{\alpha}\beta}}  &  =(\bar{\sigma}^{\mu
})_{\dot{\beta}\alpha}^{\dagger}=(-1)^{\delta_{0\mu}}(\bar{\sigma}_{\mu
})^{\dot{\beta}\alpha}. \label{KonPaul}%
\end{align}

Starting from the defining relations for 'inverse' Pauli matrices and
repeating similar arguments as above one should be able to show that
\begin{align}
\overline{(\sigma_{\mu}^{-1})^{\alpha\dot{\beta}}}  &  =(\sigma_{\mu}%
^{-1})^{\dag\,\beta\dot{\alpha}}=(-1)^{\delta_{0\mu}}(\sigma^{-1\,\mu}%
)_{\beta\dot{\alpha}},\nonumber\\
\overline{(\bar{\sigma}_{\mu}^{-1})^{\dot{\alpha}\beta}}  &  =(\bar{\sigma
}_{\mu}^{-1})^{\dag\,\dot{\beta}\alpha}=(-1)^{\delta_{0\mu}}(\bar{\sigma
}^{-1\,\mu})_{\dot{\beta}\alpha}. \label{KonPaulInv}%
\end{align}

Next, we turn to the conjugation properties of two-dimensional spin matrices.
To this end, we consider their definitions\ [cf. part I]
\begin{align}
(\sigma^{\mu\nu})_{\alpha}{}^{\beta}  &  =(P_{A})^{\mu\nu}{}_{\kappa\lambda
}(\sigma^{\kappa}\bar{\sigma}^{\lambda})_{\alpha}{}^{\beta}.\nonumber\\
(\bar{\sigma}^{\mu\nu})_{\dot{\alpha}}{}^{\dot{\beta}}  &  =(P_{A})^{\mu\nu}%
{}_{\kappa\lambda}(\bar{\sigma}^{\kappa}{\sigma}^{\lambda})_{\dot{\alpha}}%
{}^{\dot{\beta}},\\[0.1in]
(\sigma_{\mu\nu}^{-1})_{\alpha}{}^{\beta}  &  =(P_{A})^{\kappa\lambda}{}%
_{\mu\nu}(\sigma_{\kappa}^{-1}\bar{\sigma}_{\lambda}^{-1})_{\alpha}{}^{\beta
},\nonumber\\
(\bar{\sigma}_{\mu\nu}^{-1})_{\dot{\alpha}}{}^{\dot{\beta}}  &  =(P_{A}%
)^{\kappa\lambda}{}_{\mu\nu}(\bar{\sigma}_{\kappa}^{-1}{\sigma}_{\lambda}%
^{-1})_{\dot{\alpha}}{}^{\dot{\beta}}. \label{sigmamnmink2}%
\end{align}
We conjugate both sides of each equality in (\ref{sigmamnmink2}). Using the
identities in (\ref{KonPaul}) and (\ref{KonPaulInv}) together with%
\begin{equation}
(P_{A})^{\mu\nu}{}_{\kappa\lambda}=(P_{A})^{\kappa\lambda}{}_{\mu\nu},
\end{equation}
we should finally arrive at
\begin{align}
\overline{(\sigma^{\mu\nu})_{\alpha\beta}}  &  =(\sigma^{\mu\nu})_{\dot{\beta
}\dot{\alpha}}^{\dag}=(-1)^{\delta_{0\mu}+\delta_{0\nu}}(\bar{\sigma}_{\nu\mu
})^{\dot{\beta}\dot{\alpha}},\nonumber\\
\overline{(\bar{\sigma}^{\mu\nu})_{\dot{\alpha}\dot{\beta}}}  &  =(\bar
{\sigma}^{\mu\nu})_{\beta\alpha}^{\dag}=(-1)^{\delta_{0\mu}+\delta_{0\nu}%
}(\sigma_{\nu\mu})^{\beta\alpha},\\[0.1in]
\overline{(\sigma_{\mu\nu}^{-1})^{\alpha\beta}}  &  =(\sigma_{\mu\nu}%
^{-1})_{\dot{\beta}\dot{\alpha}}^{\dag}=(-1)^{\delta_{0\mu}+\delta_{0\nu}%
}(\bar{\sigma}^{-1\,\nu\mu})_{\dot{\beta}\dot{\alpha}},\nonumber\\
\overline{(\bar{\sigma}_{\mu\nu}^{-1})^{\dot{\alpha}\dot{\beta}}}  &
=(\bar{\sigma}_{\mu\nu}^{-1})_{\beta\alpha}^{\dag}=(-1)^{\delta_{0\mu}%
+\delta_{0\nu}}(\sigma^{-1\,\nu\mu})_{\beta\alpha}. \label{KonSpinMat}%
\end{align}

Now, we are ready to give the conjugation properties of the bilinear
covariants:%
\begin{align}
\overline{\bar{\Phi}\Psi}  &  =\bar{\Psi}\Phi,\nonumber\\
\overline{\bar{\Phi}\,\gamma^{5}\Psi}  &  =-\bar{\Psi}\,\gamma^{5}%
\Phi,\nonumber\\
\overline{\bar{\Phi}\,\gamma^{\mu}\Psi}  &  =(-1)^{\delta_{0\mu}}\,\bar{\Psi
}\,\gamma_{\mu}\Phi,\nonumber\\
\overline{\bar{\Phi}\,\gamma^{\mu}\gamma^{5}\Psi}  &  =(-1)^{\delta_{0\mu}%
}\,\bar{\Psi}\,\gamma_{\mu}\gamma^{5}\Phi,\nonumber\\
\overline{\bar{\Phi}\Sigma^{\mu\nu}\Psi}  &  =(-1)^{\delta_{0\mu}+\delta
_{0\nu}}\,\bar{\Psi}\Sigma_{\nu\mu}\Phi,\nonumber\\
\overline{\bar{\Phi}\Sigma^{\mu\nu}\gamma^{5}\Psi}  &  =(-1)^{\delta_{0\mu
}+\delta_{0\nu}+1}\,\bar{\Psi}\Sigma_{\nu\mu}\gamma^{5}\Phi, \label{KonCov}%
\end{align}
where%
\begin{equation}
\gamma_{\mu}=\eta_{\mu\nu}\gamma^{\nu},\qquad\Sigma_{\mu\nu}=\eta_{\mu
\mu^{\prime}}\eta_{\nu\nu^{\prime}}\,\Sigma^{\mu^{\prime}\nu^{\prime}}.
\end{equation}
The identities in (\ref{KonCov}) are readily checked by means of the
definitions in (\ref{BiCov}) together with the results in (\ref{Con2Spin}),
(\ref{KonPaul}), and (\ref{KonSpinMat}). An example shall illustrate this:%
\begin{align}
\overline{\bar{\Phi}\,\gamma^{\mu}\Psi}  &  =\overline{\phi^{\alpha}%
(\sigma^{\mu})_{\alpha}{}^{\dot{\beta}}\bar{\lambda}_{\dot{\beta}}}%
+\overline{\bar{\chi}^{\dot{\alpha}}(\bar{\sigma}^{\mu})_{\dot{\alpha}}%
{}^{\beta}\,\psi_{\beta}}\nonumber\\
&  =\lambda^{\beta}\,\overline{(\sigma^{\mu})_{\alpha}{}^{\dot{\beta}}}%
\,\bar{\phi}_{\dot{\alpha}}+\bar{\psi}^{\dot{\beta}}\,\overline{(\bar{\sigma
}^{\mu})_{\dot{\alpha}}{}^{\beta}}\,\chi_{\dot{\alpha}}\nonumber\\
&  =(-1)^{\delta_{\mu0}}\,\lambda^{\beta}(\sigma_{\mu})_{\beta}{}^{\dot
{\alpha}}\,\bar{\phi}_{\dot{\alpha}}+(-1)^{\delta_{\mu0}}\,\bar{\psi}%
^{\dot{\beta}}(\bar{\sigma}_{\mu})_{\dot{\beta}}{}^{\dot{\alpha}}\chi
_{\dot{\alpha}}\nonumber\\
&  =(-1)^{\delta_{0\mu}}\,\bar{\Psi}\gamma_{\mu}\Phi.
\end{align}

In analogy to the undeformed case the second relation in (\ref{KonCov}) should
be consistent with the identities
\begin{equation}
(\gamma^{5})^{\dag}=\gamma^{5}\quad\text{and}\quad\gamma^{0}(\gamma^{5}%
)^{\dag}\gamma^{0}=-\gamma^{5},
\end{equation}
which are a direct consequence of (\ref{gamma5mink}) and (\ref{AntGam5}).
Indeed we have
\begin{align}
\overline{\bar{\Phi}\,\gamma^{5}\Psi}  &  =\overline{\Phi^{\dag}\gamma
^{0}\gamma^{5}\Psi}=\Psi^{\dag}(\gamma^{5})^{\dag}(\gamma^{0})^{\dag}%
\Phi\nonumber\\
&  =\Psi^{\dag}\gamma^{0}\gamma^{0}(\gamma^{5})^{\dag}\gamma^{0}\Phi
=-\Psi^{\dag}\gamma^{0}\gamma^{5}\Phi\nonumber\\
&  =-\bar{\Psi}\,\gamma^{5}\Phi.
\end{align}
Moreover, inspection of the explicit form of the $\gamma$-matrices shows that
\begin{equation}
\gamma^{0}(\gamma^{\mu})^{\dag}\gamma^{0}=(-1)^{\delta_{\mu0}}\gamma_{\mu},
\label{KonGam5}%
\end{equation}
which, in turn, implies%
\begin{align}
\gamma^{0}(\gamma^{\mu}\gamma^{5})^{\dag}\gamma^{0}  &  =\gamma^{0}\gamma
^{5}(\gamma^{\mu})^{\dag}\gamma^{0}\nonumber\\
&  =-\gamma^{5}\gamma^{0}(\gamma^{\mu})^{\dag}\gamma^{0}=(-1)^{\delta_{\mu
0}+1}\,\gamma^{5}\gamma_{\mu}\nonumber\\
&  =(-1)^{\delta_{\mu0}}\,\gamma_{\mu}\gamma^{5}. \label{KonPseuVec}%
\end{align}
Once again, (\ref{KonGam5}) and (\ref{KonPseuVec}) are in agreement with
(\ref{KonCov}). In a similar way the last two identities in (\ref{KonCov})
require for the $\Sigma$-matrices that%
\begin{align}
\gamma^{0}(\Sigma^{\mu\nu})^{\dag}\gamma^{0}  &  =(-1)^{\delta_{\mu0}%
+\delta_{\nu0}}\,\Sigma_{\nu\mu},\nonumber\\
\gamma^{0}(\Sigma^{\mu\nu}\gamma^{5})^{\dag}\gamma^{0}  &  =(-1)^{\delta
_{\mu0}+\delta_{\nu0}+1}\,\Sigma_{\nu\mu}\gamma^{5}.
\end{align}

\subsection{Transformation properties of Dirac spinors}

Next, we discuss the behavior of Dirac spinors under $q$-deformed Lorentz
transformations. Each Dirac spinor is made up of two 2-spinors, so its
transformation properties are completely determined by the coactions of
$q$-deformed Lorentz group on its 2-spinors:%
\begin{align}
\beta_{L}(\lambda^{\alpha})  &  =M^{\alpha}{}_{\beta}\otimes\lambda^{\beta
},\nonumber\\
\beta_{L}(\psi_{\alpha})  &  =S^{-1}((M^{\text{T}})^{\alpha}{}_{\beta}%
)\otimes\psi_{\beta},\label{LeftCoactSpin}\\[0.1in]
\beta_{R}(\lambda^{\alpha})  &  =\lambda^{\beta}\otimes M^{\beta}{}_{\alpha
},\nonumber\\
\beta_{R}(\psi_{\alpha})  &  =\psi_{\beta}\otimes S((M^{\text{T}})^{\beta}%
{}_{\alpha}), \label{RightCoactSpin}%
\end{align}
and
\begin{align}
\beta_{L}(\bar{\lambda}^{\dot{\alpha}})  &  =\tilde{M}^{\dot{\alpha}}{}%
_{\dot{\beta}}\otimes\bar{\lambda}^{\dot{\beta}},\nonumber\\
\beta_{L}(\bar{\lambda}_{\dot{\alpha}})  &  =S^{-1}((\tilde{M}^{\text{T}%
})^{\dot{\alpha}}{}_{\dot{\beta}})\otimes\bar{\lambda}_{\dot{\beta}},\\[0.1in]
\beta_{R}(\bar{\lambda}^{\dot{\alpha}})  &  =\bar{\lambda}^{\dot{\beta}%
}\otimes\tilde{M}^{\dot{\beta}}{}_{\dot{\alpha}},\nonumber\\
\beta_{R}(\bar{\lambda}_{\dot{\alpha}})  &  =\bar{\lambda}_{\dot{\beta}%
}\otimes S((\tilde{M}^{\text{T}})^{\dot{\beta}}{}_{\dot{\alpha}}).
\label{RightCoactSpinCon}%
\end{align}

With these formulae at hand the left-coaction of the $q$-deformed Lorentz
group on a Dirac spinor becomes%
\begin{equation}
\beta_{L}(\Psi)=\beta_{L}\left(
\begin{array}
[c]{c}%
\psi_{\alpha}\\
\bar{\lambda}_{\dot{\alpha}}%
\end{array}
\right)  =\left(
\begin{array}
[c]{c}%
S^{-1}(M^{\text{T}})^{\alpha}{}_{\beta}\otimes\psi_{\beta}\\
S^{-1}(\tilde{M}^{\text{T}})^{\dot{\alpha}}{}_{\dot{\beta}}\otimes\bar
{\lambda}_{\dot{\beta}}%
\end{array}
\right)  . \label{linkscoN}%
\end{equation}
Notice that $M$ and $\tilde{M}$ denote the matrices to the two copies of
$SL_{q}(2)$, i.e.%
\begin{equation}
M^{\alpha}{}_{\beta}\otimes\tilde{M}^{\dot{\alpha}}{}_{\dot{\beta}}\in
SL_{q}(2,\mathbb{C})\simeq SL_{q}(2)\otimes SL_{q}(2).
\end{equation}
Further, we work with the transposed matrices $M^{\text{T}}$ and $\tilde
{M}^{\text{T}}$. This enables us to think in terms of matrix multiplication,
as is indicated through the position of spinor indices (in our convention the
first index of a matrix labels its rows and the second one its columns).

Now, we want to get rid of the generators with a tilde. We achieve this by the
$\ast$-structures of $SL_{q}(2,\mathbb{C})$ that are given through the
identifications in (\ref{slq2ceins}) and (\ref{slq2czwei}). From the $\ast
$-structure being relevant for left-coactions [cf. Eq. (\ref{slq2ceins})] we
find%
\begin{equation}
S^{-1}(\tilde{M}^{\text{T}})^{\dot{\alpha}}{}_{\dot{\beta}}=\overline
{(M^{\text{T}})}^{\,\dot{\beta}}{}_{\dot{\alpha}}=(M^{\dagger})^{\dot{\alpha}%
}{}_{\dot{\beta}},
\end{equation}
or, for short,
\begin{equation}
S^{-1}(\tilde{M}^{\text{T}})=\bar{M}. \label{AntMtild}%
\end{equation}

We can also avoid the antipodes in (\ref{linkscoN}). From a comparison of
\begin{equation}
\varepsilon_{\alpha\beta}\,M^{\alpha}{}_{\alpha^{\prime}}M^{\beta}{}%
_{\beta^{\prime}}=\varepsilon_{\alpha^{\prime}\beta^{\prime}}%
\end{equation}
with the identities
\begin{equation}
S^{-1}(M^{\beta^{\prime}}{}_{\alpha})M^{\beta}{}_{\beta^{\prime}}=S(M^{\beta
}{}_{\beta^{\prime}})M^{\beta^{\prime}}{}_{\alpha}=\delta^{\beta}{}_{\alpha},
\end{equation}
we find
\begin{align}
S^{-1}(M^{\beta}{}_{\alpha})  &  =\varepsilon_{\alpha\alpha^{\prime}}%
M^{\alpha^{\prime}}{}_{\beta^{\prime}}\varepsilon^{\beta^{\prime}\beta
}=-(\varepsilon^{\text{T}})^{\beta\beta^{\prime}}(M^{\text{T}})^{\beta
^{\prime}}{}_{\alpha^{\prime}}(\varepsilon^{\text{T}})^{\alpha^{\prime}\alpha
},\nonumber\\
S(M^{\beta}{}_{\alpha})  &  =\varepsilon^{\beta\beta^{\prime}}M^{\alpha
^{\prime}}{}_{\beta^{\prime}}\varepsilon_{\alpha^{\prime}\alpha}%
=-\varepsilon^{\beta\beta^{\prime}}(M^{\text{T}})^{\beta^{\prime}}{}%
_{\alpha^{\prime}}\varepsilon^{\alpha^{\prime}\alpha}, \label{defmatmult345}%
\end{align}
where we have used $\varepsilon^{\alpha\beta}=-\varepsilon_{\alpha\beta}$.
From now on matrix multiplication between $\varepsilon$ and $M$ will always be
understood as defined in (\ref{defmatmult345}) with two adjacent indices. On
these grounds we write the results in (\ref{defmatmult345}) as%
\begin{equation}
S^{-1}(M^{\text{T}})=-\varepsilon M\varepsilon,\qquad S(M)=-\varepsilon
M^{\text{T}}\varepsilon. \label{AntM}%
\end{equation}

Before proceeding any further let us sum up that for left coactions the
relationship between the generators of the two tensor factors of
$SL_{q}(2,\mathbb{C})$ is given by
\begin{equation}
\tilde{M}=S(M^{\dag})=-\varepsilon^{\text{T}}\bar{M}\,\varepsilon^{\text{T}},
\end{equation}
whereas for right coactions we have%
\begin{equation}
\tilde{M}=S^{-1}(M^{\dag})=-\varepsilon\bar{M}\varepsilon.
\end{equation}

With (\ref{AntMtild}) and (\ref{AntM}) the left-coaction of a Dirac spinor
finally becomes%
\begin{align}
\beta_{L}(\Psi)  &  =\left(
\begin{array}
[c]{cc}%
-\varepsilon^{\alpha\alpha^{\prime\prime}}M^{\alpha^{\prime\prime}}{}%
_{\alpha^{\prime\prime\prime}}\varepsilon^{\alpha^{\prime\prime\prime}%
\alpha^{\prime}} & 0\\
0 & \bar{M}^{\dot{\alpha}}{}_{\dot{\alpha}^{\prime}}%
\end{array}
\right)  \otimes\left(
\begin{array}
[c]{c}%
\psi_{\alpha^{\prime}}\\
\bar{\lambda}_{\dot{\alpha}^{\prime}}%
\end{array}
\right)  ,\nonumber\\[0.04in]
&  =\left(
\begin{array}
[c]{cc}%
-\varepsilon M\varepsilon & 0\\
0 & \bar{M}%
\end{array}
\right)  \otimes\left(
\begin{array}
[c]{c}%
\psi\\
\bar{\lambda}%
\end{array}
\right) \nonumber\\[0.04in]
&  \equiv L_{S}(\Lambda)\otimes\Psi,
\end{align}
where we introduced the Lorentz matrix $L_{S}$ for notational convenience. Of
course, these ideas carry over to right coactions of Dirac spinors. With the
help of (\ref{RightCoactSpin}) and (\ref{RightCoactSpinCon}) we get, at once,
\begin{align}
\beta_{R}(\Psi)  &  =\beta_{R}\left(
\begin{array}
[c]{c}%
\psi_{\alpha}\\
\bar{\lambda}_{\dot{\alpha}}%
\end{array}
\right)  =\left(
\begin{array}
[c]{c}%
\psi_{\alpha^{\prime}}\otimes S(M^{\text{T}})^{\alpha^{\prime}}{}_{\alpha}\\
\bar{\lambda}_{\dot{\alpha}^{\prime}}\otimes S(\tilde{M}^{\text{T}}%
)^{\dot{\alpha}^{\prime}}{}_{\dot{\alpha}}%
\end{array}
\right) \nonumber\\
&  =\left(
\begin{array}
[c]{cc}%
1\otimes S(M)^{\alpha}{}_{\alpha^{\prime}} & 0\\
0 & 1\otimes S(\tilde{M})^{\dot{\alpha}}{}_{\dot{\alpha}^{\prime}}%
\end{array}
\right)  \left(
\begin{array}
[c]{c}%
\psi_{\alpha^{\prime}}\otimes1\\
\bar{\lambda}_{\dot{\alpha}^{\prime}}\otimes1
\end{array}
\right)  .
\end{align}
Making use of (\ref{slq2czwei}) and (\ref{AntM}) the last expression can be
rewritten as%
\begin{align}
\beta_{R}(\Psi)  &  =\left(
\begin{array}
[c]{cc}%
-1\otimes\varepsilon^{\alpha\alpha^{\prime\prime\prime}}(M^{\text{T}}%
)^{\alpha^{\prime\prime\prime}}{}_{\alpha^{\prime\prime}}\varepsilon
^{\alpha^{\prime\prime}\alpha^{\prime}} & 0\\
0 & 1\otimes(M^{\dagger})^{\dot{\alpha}}{}_{\dot{\alpha}^{\prime}}%
\end{array}
\right)  \left(
\begin{array}
[c]{c}%
\psi_{\alpha^{\prime}}\otimes1\\
\bar{\lambda}_{\dot{\alpha}^{\prime}}\otimes1
\end{array}
\right) \nonumber\\[0.04in]
&  =\left(
\begin{array}
[c]{cc}%
-1\otimes\varepsilon M^{\text{T}}\varepsilon & 0\\
0 & 1\otimes M^{\dagger}%
\end{array}
\right)  \left(
\begin{array}
[c]{c}%
\psi\otimes1\\
\bar{\lambda}\otimes1
\end{array}
\right) \nonumber\\[0.04in]
&  \equiv\,(1\otimes R_{S}(\Lambda))(\Psi\otimes1),
\end{align}
where the Lorentz matrix $R_{S}$ is defined in an obvious way.

It is not very difficult to repeat the above reasonings for the Hermitian
conjugate as well as the Dirac conjugate. Thus, we left the details to the
reader and restrict ourselves to stating the results, only. Instead, we give a
summary of all coactions on the different types of 4-spinors. We write our
results in an index-free notation.\textbf{ }From what we have done so far it
should be rather clear how to write them in a form with all indices being displayed.

We begin with left coactions on Dirac spinors. In doing so, we need the
matrices
\begin{equation}
L_{S}(\Lambda)=\left(
\begin{array}
[c]{cc}%
-\varepsilon M\varepsilon & 0\\
0 & \bar{M}%
\end{array}
\right)  ,\qquad L_{S}^{\dagger}(\Lambda)=\left(
\begin{array}
[c]{cc}%
-\varepsilon^{\text{T}}M^{\dagger}\varepsilon^{\text{T}} & 0\\
0 & M^{\text{T}}%
\end{array}
\right)  .
\end{equation}
With these matrices the left coactions of $q$-deformed Lorentz group on
4-spinors take on the rather intuitive form%
\begin{align}
\beta_{L}(\Psi^{a})  &  =L_{S}(\Lambda)^{a}{}_{b}\otimes\Psi^{b},\nonumber\\
\beta_{L}(\Psi_{a}^{\dagger})  &  =(1\otimes\Psi_{b}^{\dagger})(L_{S}%
^{\dagger}(\Lambda)^{b}{}_{a}\otimes1),\nonumber\\
\beta_{L}(\bar{\Psi}_{a})  &  =(1\otimes\bar{\Psi}_{b})((\gamma^{0}%
L_{S}^{\dagger}(\Lambda)\gamma^{0})^{b}{}_{a}\otimes1).
\label{spinortransleft}%
\end{align}

In dealing with right coactions we need the matrices
\begin{equation}
R_{S}(\Lambda)=\left(
\begin{array}
[c]{cc}%
-\varepsilon M^{\text{T}}\varepsilon & 0\\
0 & M^{\dagger}%
\end{array}
\right)  ,\qquad R_{S}^{\dagger}(\Lambda)=\left(
\begin{array}
[c]{cc}%
-\varepsilon^{\text{T}}\bar{M}\,\varepsilon^{\text{T}} & 0\\
0 & M
\end{array}
\right)  .
\end{equation}
In analogy to (\ref{spinortransleft}) we have
\begin{align}
\beta_{R}(\Psi^{a})  &  =(1\otimes R_{S}(\Lambda)^{a}{}_{b})(\Psi^{b}%
\otimes1),\nonumber\\
\beta_{R}(\Psi_{a}^{\dagger})  &  =\Psi_{b}^{\dagger}\otimes R_{S}^{\dagger
}(\Lambda)^{b}{}_{a},\nonumber\\
\beta_{R}(\bar{\Psi}_{a})  &  =\bar{\Psi}_{b}\otimes(\gamma^{0}R_{S}^{\dagger
}(\Lambda)\gamma^{0})^{b}{}_{a}. \label{spinortransright}%
\end{align}

The reader may have realized that we are working with a kind of Weyl
representation. A characteristic feature of the Weyl representation is that
4-spinors are written as two 2-spinors that transform independently from each
other under Lorentz transformations. A short look at (\ref{LeftCoactSpin}%
)-(\ref{RightCoactSpinCon}) tells us that the transformation properties of a
two-spinor are completely determined by the type of its index (dotted or
undotted, covariant or contravariant). On these grounds, it should be obvious
that the 4-spinor
\begin{equation}
\Psi^{c}\equiv\left(
\begin{array}
[c]{c}%
\lambda_{\alpha}\\
\bar{\psi}_{\dot{\alpha}}%
\end{array}
\right)  =%
\begin{pmatrix}
\varepsilon_{\alpha\beta} & 0\\
0 & -\varepsilon_{\dot{\alpha}\dot{\beta}}%
\end{pmatrix}
\left(
\begin{array}
[c]{c}%
\lambda^{\beta}\\
\bar{\psi}^{\dot{\beta}}%
\end{array}
\right)  =C\bar{\Psi}^{\text{T}}%
\end{equation}
behaves under $q$-deformed Lorentz transformations in very much the same way
as $\Psi$. It can be viewed as the 4-spinor one obtains by charge conjugation.
In analogy to the undeformed case the Majorana condition
\begin{equation}
\Psi^{c}=\Psi,
\end{equation}
gives $\lambda=\psi$, i.e.%
\begin{equation}
\Psi^{c}=\Psi=\left(
\begin{array}
[c]{c}%
\psi_{\alpha}\\
\bar{\psi}_{\dot{\alpha}}%
\end{array}
\right)  .
\end{equation}

\subsection{Transformation properties of bilinear covariants}

Now, we are well prepared to discuss the transformation properties of the
bilinear covariants. We can assume that in the expressions of (\ref{BiCov})
the transformation properties are assigned to the spinors, i.e. we simply have
to transform the 4-spinors according to (\ref{spinortransleft}) and
(\ref{spinortransright}), while the $\gamma$-matrices remain unchanged.

First of all, we consider the transformation of $\bar{\Phi}\Psi$:%
\begin{align}
\beta_{L}(\bar{\Phi}\Psi)  &  =\beta_{L}(\bar{\Phi})\,\beta_{L}(\Psi
)\nonumber\\
&  =(1\otimes\bar{\Phi})\,(\gamma^{0}L_{S}^{\dagger}(\Lambda)\,\gamma^{0}%
L_{S}(\Lambda)\otimes\Psi)\nonumber\\
&  =\mbox{1 \kern-.59em {\rm l}}\otimes\bar{\Phi}\Psi.
\end{align}
First, we apply the homomorphism property of the coaction. Then we insert the
expressions from (\ref{spinortransleft}). The last equality follows from the
calculation
\begin{align}
(\gamma^{0}L_{S}^{\dagger}(\Lambda)\,\gamma^{0})\,L_{S}(\Lambda)  &  =\left(
\begin{array}
[c]{cc}%
M^{T} & 0\\
0 & -\varepsilon^{T}M^{\dagger}\varepsilon^{T}%
\end{array}
\right)  \left(
\begin{array}
[c]{cc}%
-\varepsilon M\varepsilon & 0\\
0 & \bar{M}%
\end{array}
\right) \nonumber\\[0.03in]
&  =\left(
\begin{array}
[c]{cc}%
M^{T}S^{-1}(M^{T}) & 0\\
0 & S(\bar{M})\bar{M}%
\end{array}
\right)  =\mbox{1 \kern-.59em {\rm l}}, \label{QuasUniL}%
\end{align}
The second step uses (\ref{AntM}) and the third one is a direct consequence of
the Hopf algebra axioms%
\begin{equation}
M_{(2)}\,S^{-1}(M)_{(1)}=M_{(1)}\,S(M)_{(2)}=1.
\end{equation}
Likewise, we can show that%
\begin{equation}
\gamma^{0}R_{S}^{\dagger}(\Lambda)\,\gamma^{0}R_{S}(\Lambda
)=\mbox{1 \kern-.59em {\rm l}}, \label{QuasUniR}%
\end{equation}
from which we conclude
\begin{equation}
\beta_{R}(\bar{\Phi}\Psi)=\bar{\Phi}\Psi\otimes\mbox{1 \kern-.59em {\rm l}}.
\end{equation}
To sum up, $\bar{\Phi}\Psi$ transforms like a scalar under the coactions of
$q$-deformed Lorentz group. Finally, a short look at (\ref{QuasUniL}) and
(\ref{QuasUniR}) tells us that $L_{S}(\Lambda)$ and $R_{S}(\Lambda)$ are not
unitary matrices.

Next, let us examine how the bicovariants with $\gamma$-matrices behave under
Lorentz transformations:
\begin{align}
\beta_{L}(\bar{\Phi}\,\gamma^{\mu}\Psi)  &  =\beta_{L}(\bar{\Phi}%
)\,\gamma^{\mu}\beta_{L}(\Psi)\nonumber\\
&  =(\mbox{1 \kern-.59em {\rm l}}\otimes\bar{\Phi})(\gamma^{0}L_{S}^{\dagger
}(\Lambda)\,\gamma^{0}\gamma^{\mu}L_{S}(\Lambda)\otimes\Psi)\nonumber\\
&  =\Lambda^{\mu}{}_{\nu}\otimes\bar{\Phi}\,\gamma^{\nu}\Psi. \label{TrafoVec}%
\end{align}
The last equality results from the identity%
\begin{equation}
(\Lambda_{L})^{\mu}{}_{\nu}\,\gamma^{\nu}=(\gamma^{0}L_{S}^{\dagger}%
(\Lambda)\,\gamma^{0})\,\gamma^{\mu}L_{S}(\Lambda), \label{TrafoGammaL}%
\end{equation}
which can be recognized as a four-dimensional generalization of
(\ref{TransPaul1}). Its proof reads as follows:%
\begin{align}
(\gamma^{0}L_{S}^{\dagger}(\Lambda)\,\gamma^{0})\,\gamma^{\mu}L_{S}(\Lambda)
&  =%
\begin{pmatrix}
M^{T} & 0\\
0 & S(\bar{M})
\end{pmatrix}%
\begin{pmatrix}
0 & \sigma^{\mu}\\
\bar{\sigma}^{\mu} & 0
\end{pmatrix}%
\begin{pmatrix}
S^{-1}(M^{T}) & 0\\
0 & \bar{M}%
\end{pmatrix}
\nonumber\\[0.03in]
&  =%
\begin{pmatrix}
0 & M^{T}\sigma^{\mu}\bar{M}\\
S(\bar{M})\bar{\sigma}^{\mu}S^{-1}(M^{T}) & 0
\end{pmatrix}
\nonumber\\[0.03in]
&  =%
\begin{pmatrix}
0 & M^{T}\sigma^{\mu}\bar{M}\\
\tilde{M}^{T}\bar{\sigma}^{\mu}\bar{\tilde{M}} & 0
\end{pmatrix}
=%
\begin{pmatrix}
0 & (\Lambda_{L})^{\mu}{}_{\nu}\,\sigma^{\nu}\\
(\Lambda_{L})^{\mu}{}_{\nu}\,\bar{\sigma}^{\nu} & 0
\end{pmatrix}
\nonumber\\[0.03in]
&  =(\Lambda_{L})^{\mu}{}_{\nu}\,\gamma^{\nu}.
\end{align}
Similar arguments lead us to
\begin{equation}
(\Lambda_{R}^{T})^{\mu}{}_{\nu}\,\gamma^{\nu}=(\gamma^{0}R_{S}^{\dagger
}(\Lambda)\,\gamma^{0})\,\gamma^{\mu}R_{S}(\Lambda), \label{TrafoGammaR}%
\end{equation}
which, in turn, means%
\begin{equation}
\beta_{R}(\bar{\Phi}\,\gamma^{\mu}\Psi)=(\Lambda^{\text{T}})^{\mu}{}_{\nu
}\otimes\bar{\Phi}\,\gamma^{\nu}\Psi.
\end{equation}
In summary, $\bar{\Phi}\gamma^{\mu}\Psi$ transforms under $q$-deformed Lorentz
transformations as a 4-vector.

As already mentioned, (\ref{TrafoGammaL}) and (\ref{TrafoGammaR}) can be seen
as four-dimensional generalizations of (\ref{TransPaul1}). There are also
four-dimensional generalizations of (\ref{TransPaul2}):%
\begin{align}
(\Lambda_{L})^{\mu}{}_{\nu}\gamma_{\mu}^{-1}  &  =L_{S}(\Lambda)\,\gamma_{\nu
}^{-1}(\gamma_{0}^{-1}L_{S}^{\dagger}(\Lambda)\,\gamma_{0}^{-1}),\nonumber\\
(\Lambda_{R}^{\text{T}})^{\mu}{}_{\nu}\gamma_{\mu}^{-1}  &  =R_{S}%
(\Lambda)\,\gamma_{\nu}^{-1}(\gamma_{0}^{-1}R_{S}^{\dagger}(\Lambda
)\,\gamma_{0}^{-1}).
\end{align}
To prove these relations one can proceed in the following manner:%
\begin{align}
\beta_{L}(X^{\mu}\gamma_{\mu}^{-1})  &  =(\Lambda_{L})^{\mu}{}_{\nu}\otimes
X^{\nu}\gamma_{\mu}^{-1}=%
\begin{pmatrix}
0 & \beta_{L}(X^{\mu}\sigma_{\mu}^{-1})\\
\beta_{L}(X^{\mu}\bar{\sigma}_{\mu}^{-1}) & 0
\end{pmatrix}
\nonumber\\[0.03in]
&  =%
\begin{pmatrix}
0 & \bar{\tilde{M}}\sigma_{\nu}^{-1}\tilde{M}^{\text{T}}\\
\bar{M}\bar{\sigma}_{\nu}^{-1}M^{\text{T}} & 0
\end{pmatrix}
\nonumber\\[0.03in]
&  =%
\begin{pmatrix}
\bar{\tilde{M}} & 0\\
0 & \bar{M}%
\end{pmatrix}%
\begin{pmatrix}
0 & \sigma_{\nu}^{-1}\\
\bar{\sigma}_{\nu}^{-1} & 0
\end{pmatrix}%
\begin{pmatrix}
M^{\text{T}} & 0\\
0 & \tilde{M}^{\text{T}}%
\end{pmatrix}
\nonumber\\[0.03in]
&  =L_{S}(\Lambda)\,\gamma_{\nu}^{-1}(\gamma_{0}^{-1}L_{S}^{\dagger}%
(\Lambda)\,\gamma_{0}^{-1}).
\end{align}

The covariants with $\Sigma$-matrices transform like antisymmetric tensors
under $q$-deformed Lorentz transformations:%
\begin{align}
\beta_{L}(\bar{\Phi}\Sigma^{\mu\nu}\Psi)  &  =\Lambda^{\mu}{}_{\mu^{\prime}%
}\Lambda^{\nu}{}_{\nu^{\prime}}(\bar{\Phi}\Sigma^{\mu^{\prime}\nu^{\prime}%
}\Psi),\nonumber\\
\beta_{R}(\bar{\Phi}\Sigma^{\mu\nu}\Psi)  &  =(\Lambda^{T})^{\mu}{}%
_{\mu^{\prime}}(\Lambda^{T})^{\nu}{}_{\nu^{\prime}}(\bar{\Phi}\Sigma
^{\mu^{\prime}\nu^{\prime}}\Psi).
\end{align}
We illustrate the ideas leading to these relations by the following
calculation:%
\begin{align}
\beta_{L}(\bar{\Phi}\Sigma^{\mu\nu}\Psi)  &  =(P_{A})^{\mu\nu}{}_{\mu^{\prime
}\nu^{\prime}}\,\beta_{L}(\bar{\Phi}\,\gamma^{\mu}\gamma^{\nu}\Psi)\nonumber\\
&  =(P_{A})^{\mu\nu}{}_{\mu^{\prime}\nu^{\prime}}\Lambda^{\mu^{\prime}}{}%
_{\mu^{\prime\prime}}\Lambda^{\nu^{\prime}}{}_{\nu^{\prime\prime}}\otimes
(\bar{\Phi}\,\gamma^{\mu^{\prime\prime}}\gamma^{\nu^{\prime\prime}}%
\Psi)\nonumber\\
&  =\Lambda^{\mu}{}_{\mu^{\prime}}\Lambda^{\nu}{}_{\nu^{\prime}}(P_{A}%
)^{\mu^{\prime}\nu^{\prime}}{}_{\mu^{\prime\prime}\nu^{\prime\prime}}%
\otimes(\bar{\Phi}\,\gamma^{\mu^{\prime\prime}}\gamma^{\nu^{\prime\prime}}%
\Psi)\nonumber\\
&  =\Lambda^{\mu}{}_{\mu^{\prime}}\Lambda^{\nu}{}_{\nu^{\prime}}\otimes
(\bar{\Phi}\Sigma^{\mu^{\prime}\nu^{\prime}}\Psi).
\end{align}
The first equality is the definition of the $\Sigma$-matrices. The second
equality can be seen as an application of (\ref{TrafoVec}). To understand the
third identity one has to realize that the generators of $q$-deformed Lorentz
group obey the relations%
\begin{equation}
\hat{R}^{\mu\nu}{}_{\mu^{\prime}\nu^{\prime}}\Lambda^{\mu^{\prime}}{}%
_{\mu^{\prime\prime}}\Lambda^{\nu^{\prime}}{}_{\nu^{\prime\prime}}%
=\Lambda^{\mu}{}_{\mu^{\prime}}\Lambda^{\nu}{}_{\nu^{\prime}}\hat{R}%
^{\mu^{\prime}\nu^{\prime}}{}_{\mu^{\prime\prime}\nu^{\prime\prime}},
\end{equation}
where the $\hat{R}$-matrix refers to $q$-deformed Lorentz group. It should be
clear that these relations remain valid if the $\hat{R}$-matrix is substituted
by any of its projectors.

To describe the transformation properties of the covariants with $\gamma^{5}$
we need the definition of the $q$-determinant for the quantum Lorentz group
{\cite{Umeyer, FioDet}}. In our notation it is determined by
\begin{equation}
(\det\nolimits_{q}\Lambda)\,\varepsilon_{\mu^{\prime}\nu^{\prime}\rho^{\prime
}\sigma^{\prime}}=\varepsilon_{\mu\nu\rho\sigma}\,\Lambda^{\sigma}{}%
_{\sigma^{\prime}}\Lambda^{\rho}{}_{\rho^{\prime}}\Lambda^{\nu}{}_{\nu
^{\prime}}\Lambda^{\mu}{}_{\mu^{\prime}}. \label{DefqDet}%
\end{equation}
Contracting both sides of this relation with $\varepsilon^{\sigma^{\prime}%
\rho^{\prime}\nu^{\prime}\mu^{\prime}}$ yields%
\begin{equation}
\det\nolimits_{q}\Lambda=-\frac{1}{2}q^{8}[[2]]_{q^{2}}^{-2}[[3]]_{q^{2}}%
^{-1}\,\varepsilon_{\mu\nu\rho\sigma}\,\Lambda^{\sigma}{}_{\sigma^{\prime}%
}\Lambda^{\rho}{}_{\rho^{\prime}}\Lambda^{\nu}{}_{\nu^{\prime}}\Lambda^{\mu}%
{}_{\mu^{\prime}}\,\varepsilon^{\sigma^{\prime}\rho^{\prime}\nu^{\prime}%
\mu^{\prime}}.
\end{equation}
Using (\ref{DefqDet}) we find%
\begin{align}
\beta_{L}(\bar{\Phi}\,\gamma^{5}\Psi)  &  =\frac{q^{5}}{[[2]]_{q^{2}%
}[[3]]_{q^{2}}}\,\varepsilon_{\mu\nu\rho\sigma}\,\beta_{L}(\bar{\Phi}%
\,\gamma^{\sigma}\gamma^{\rho}\gamma^{\nu}\gamma^{\mu}\Psi)\nonumber\\
&  =\frac{q^{5}}{[[2]]_{q^{2}}[[3]]_{q^{2}}}\,\varepsilon_{\mu\nu\rho\sigma
}\,\Lambda^{\sigma}{}_{\sigma^{\prime}}\Lambda^{\rho}{}_{\rho^{\prime}}%
\Lambda^{\nu}{}_{\nu^{\prime}}\Lambda^{\mu}{}_{\mu^{\prime}}\otimes(\bar{\Phi
}\,\gamma^{\sigma^{\prime}}\gamma^{\rho^{\prime}}\gamma^{\nu^{\prime}}%
\gamma^{\mu^{\prime}}\Psi)\nonumber\\
&  =\frac{q^{5}}{[[2]]_{q^{2}}[[3]]_{q^{2}}}\,(\det\nolimits_{q}%
\Lambda)\,\varepsilon_{\mu^{\prime}\nu^{\prime}\rho^{\prime}\sigma^{\prime}%
}\otimes(\bar{\Phi}\,\gamma^{\sigma^{\prime}}\gamma^{\rho^{\prime}}\gamma
^{\nu^{\prime}}\gamma^{\mu^{\prime}}\Psi)\nonumber\\
&  =\,\det\nolimits_{q}\Lambda\,\otimes\bar{\Phi}\,\gamma^{5}\Psi,
\end{align}
and, likewise,%
\[
\beta_{R}(\bar{\Phi}\,\gamma^{5}\Psi)=\bar{\Phi}\,\gamma^{5}\Psi\otimes
\det\nolimits_{q}\Lambda^{\text{T}}\,
\]
In this manner, we conclude that $\bar{\Phi}\,\gamma^{5}\Psi$ transforms as a
pseudo\-scalar. With this observation it is not difficult to verify that
$\bar{\Phi}\,\gamma^{\mu}\gamma^{5}\Psi$ behaves as a pseudo\-vector, i.e.%
\begin{align}
\beta_{L}(\bar{\Phi}\,\gamma^{\mu}\gamma^{5}\Psi)  &  =(\det\nolimits_{q}%
\Lambda)\Lambda^{\mu}{}_{\nu}\otimes\bar{\Phi}\,\gamma^{\nu}\gamma^{5}%
\Psi,\nonumber\\
\beta_{R}(\bar{\Phi}\,\gamma^{\mu}\gamma^{5}\Psi)  &  =\bar{\Phi}\,\gamma
^{\nu}\gamma^{5}\Psi\otimes(\det\nolimits_{q}\Lambda^{\text{T}})(\Lambda
^{\text{T}})^{\mu}{}_{\nu}.
\end{align}

\section{Conclusion\label{Conclusion}}

Let us end with some comments on what we have done so far. In part I and II of
the present article we gave a systematic treatment of $q$-deformed spinor
calculus. In part I attention was concentrated on Weyl spinors, while part II
dealt with the foundations of the Dirac formalism. One of the highlights of
part I was the presentation of the Weyl spinor Fierz identities, which play a
very important role in supersymmetry, especially in the derivation of
transformation properties of superfields. In part II we could finally discuss
Dirac spinors and their bilinear covariants along the same line of reasonings
as in the undeformed case.

Furthermore, we could derive a rather complete list of $q$-analogs of trace
relations and rearrangement formulae concerning Pauli, Dirac, and spin
matrices. Our results show striking similarities to their undeformed
counterparts. The reason for this observation lies in the fact that
$q$-deformation does not change the representation theoretic content of the
theory. On these grounds, our list of relations could prove very useful for
calculations in quantum field theory.

However, it should become clear that $q$-deformation leads to a more
sophisticated structure. There are often several $q$-deformed variants to one
and the same undeformed object. Thus, combining $q$-deformed objects in the
right way often requires a lot of testing and experience accompanied by a deep
understanding of the subject.\vspace*{0.2in}

\noindent\textbf{Acknowledgements}\newline First of all we are very grateful
to Eberhard Zeidler for his invitation to the MPI Leipzig, his special
interest in our work, and his financial support. Furthermore we would like to
thank Fabian Bachmaier and Ina Stein for their support. Finally, we thank
Dieter L\"{u}st for kind hospitality.

\appendix

\section{Some explicit formulae for the four-di\-men\-sio\-nal Euclidean
space\label{AppA}}

In Sec. \ref{gammaeu4kapSec1} we gave two basis of the Clifford algebra of
four-di\-men\-sio\-nal $q$-deformed Euclidean space. In terms of the
generators of the Clifford algebra the elements of the two bases become%
\begin{align}
\gamma^{\lbrack1}\gamma^{i]_{q}}=  &  \;\gamma^{1}\gamma^{i},\qquad
i=2,3,\nonumber\\
\gamma^{\lbrack1}\gamma^{4]_{q}}=  &  \;\gamma^{1}\gamma^{4}-q^{-1}\lambda
_{+}^{-1}\mbox{1 \kern-.59em {\rm l}},\nonumber\\
\gamma^{\lbrack2}\gamma^{3]_{q}}=  &  \;\gamma^{2}\gamma^{3}-\lambda_{+}%
^{-1}\mbox{1 \kern-.59em {\rm l}},\nonumber\\
\gamma^{\lbrack i}\gamma^{4]_{q}}=  &  \;\gamma^{2}\gamma^{4},\qquad i=2,3,
\end{align}%
\begin{align}
\gamma^{\lbrack1}\gamma^{2}\gamma^{3]_{q}}=  &  \;\gamma^{1}\gamma^{2}%
\gamma^{3}-\frac{1}{2}q^{-1}\,\gamma^{1},\nonumber\\
\gamma^{\lbrack1}\gamma^{i}\gamma^{4]_{q}}=  &  \;\gamma^{1}\gamma^{i}%
\gamma^{4}+\frac{1}{2}q^{-1}\,\gamma^{i},\qquad i=2,3,\nonumber\\
\gamma^{\lbrack2}\gamma^{3}\gamma^{4]_{q}}=  &  \;\gamma^{2}\gamma^{3}%
\gamma^{4}-\frac{1}{2}q^{-1}\,\gamma^{4},
\end{align}%
\begin{align}
\gamma^{5}\gamma^{1}=  &  -\gamma^{1}+2q\,\gamma^{1}\gamma^{2}\gamma
^{3},\nonumber\\
\gamma^{5}\gamma^{2}=  &  -\gamma^{2}-2q\,\gamma^{1}\gamma^{2}\gamma
^{3},\nonumber\\
\gamma^{5}\gamma^{3}=  &  \;\gamma^{3}+2q\,\gamma^{1}\gamma^{2}\gamma
^{3},\nonumber\\
\gamma^{5}\gamma^{4}=  &  \;\gamma^{4}-2q\,\gamma^{1}\gamma^{2}\gamma^{3},
\end{align}%
\begin{align}
\gamma^{5}=  &  -2\gamma^{1}\gamma^{4}-2q\,\gamma^{2}\gamma^{3}+2q^{2}%
\lambda_{+}\,\gamma^{1}\gamma^{2}\gamma^{3}\gamma^{4}%
+\mbox{1 \kern-.59em {\rm l}}\nonumber\\
\gamma^{\lbrack1}\gamma^{2}\gamma^{3}\gamma^{4]_{q}}=  &  \;4[[2]]_{q^{2}}%
^{4}[[3]]_{q^{2}}^{2}\,\gamma^{1}\gamma^{2}\gamma^{3}\gamma^{4}-4[[2]]_{q^{2}%
}^{3}[[3]]_{q^{2}}^{2}(\gamma^{2}\gamma^{3}+q^{-1}\gamma^{1}\gamma
^{4})\nonumber\\
&  +2q^{-1}[[2]]_{q^{2}}^{3}[[3]]_{q^{2}}^{2}\mbox{1 \kern-.59em {\rm l}}.
\label{gammaexpeu4}%
\end{align}
The matrices representing the elements of the two bases are%
\begin{gather*}
\Gamma^{12}=\left(
\begin{array}
[c]{cccc}%
0 & 0 & 0 & 0\\
0 & 0 & 0 & 0\\
0 & 0 & 0 & q^{-1}\\
0 & 0 & 0 & 0
\end{array}
\right)  ,\quad\Gamma^{13}=\left(
\begin{array}
[c]{cccc}%
0 & q^{-1} & 0 & 0\\
0 & 0 & 0 & 0\\
0 & 0 & 0 & 0\\
0 & 0 & 0 & 0
\end{array}
\right)  ,\\[0.03in]
\Gamma^{14}=\lambda_{+}^{-1}\left(
\begin{array}
[c]{cccc}%
q & 0 & 0 & 0\\
0 & -q^{-1} & 0 & 0\\
0 & 0 & q & 0\\
0 & 0 & 0 & -q^{-1}%
\end{array}
\right)  ,\quad\Gamma^{23}=\lambda_{+}^{-1}\left(
\begin{array}
[c]{cccc}%
-1 & 0 & 0 & 0\\
0 & q^{-2} & 0 & 0\\
0 & 0 & q^{2} & 0\\
0 & 0 & 0 & -1
\end{array}
\right)  ,\\[0.03in]
\Gamma^{24}=\left(
\begin{array}
[c]{cccc}%
0 & 0 & 0 & 0\\
1 & 0 & 0 & 0\\
0 & 0 & 0 & 0\\
0 & 0 & 0 & 0
\end{array}
\right)  ,\quad\Gamma^{34}=\left(
\begin{array}
[c]{cccc}%
0 & 0 & 0 & 0\\
0 & 0 & 0 & 0\\
0 & 0 & 0 & 0\\
0 & 0 & 1 & 0
\end{array}
\right)  ,\\[0.03in]
\Gamma^{123}=\frac{1}{2}\left(
\begin{array}
[c]{cccc}%
0 & 0 & 0 & 0\\
0 & 0 & -q^{-1/2} & 0\\
0 & 0 & 0 & 0\\
-q^{-5/2} & 0 & 0 & 0
\end{array}
\right)  ,\quad\Gamma^{124}=-\frac{1}{2}\left(
\begin{array}
[c]{cccc}%
0 & 0 & 0 & 0\\
0 & 0 & 0 & q^{-3/2}\\
q^{-1} & 0 & 0 & 0\\
0 & 0 & 0 & 0
\end{array}
\right)  ,\\[0.03in]
\Gamma^{134}=-\frac{1}{2}\left(
\begin{array}
[c]{cccc}%
0 & 0 & q^{-1/2} & 0\\
0 & 0 & 0 & 0\\
0 & 0 & 0 & 0\\
0 & q^{-3/2} & 0 & 0
\end{array}
\right)  ,\quad\Gamma^{234}=\frac{1}{2}q^{-1/2}\left(
\begin{array}
[c]{cccc}%
0 & 0 & 0 & 0\\
0 & 0 & -1 & 0\\
0 & 0 & 0 & 0\\
1 & 0 & 0 & 0
\end{array}
\right)  ,\\[0.03in]
\Gamma^{51}=q^{-1/2}\left(
\begin{array}
[c]{cccc}%
0 & 0 & 0 & 1\\
0 & 0 & 0 & 0\\
0 & -1 & 0 & 0\\
0 & 0 & 0 & 0
\end{array}
\right)  ,\quad\Gamma^{52}=\left(
\begin{array}
[c]{cccc}%
0 & 0 & 0 & 0\\
0 & 0 & 0 & q^{-1/2}\\
q^{1/2} & 0 & 0 & 0\\
0 & 0 & 0 & 0
\end{array}
\right)  ,\\[0.03in]
\Gamma^{54}=q^{1/2}\left(
\begin{array}
[c]{cccc}%
0 & 0 & 0 & 0\\
0 & 0 & 1 & 0\\
0 & 0 & 0 & 0\\
-1 & 0 & 0 & 0
\end{array}
\right)  ,\quad\Gamma^{53}=-\left(
\begin{array}
[c]{cccc}%
0 & 0 & q^{1/2} & 0\\
0 & 0 & 0 & 0\\
0 & 0 & 0 & 0\\
0 & -q^{-1/2} & 0 & 0
\end{array}
\right)  ,\\[0.03in]
\Gamma^{1234}=-2q^{-1}[[2]]_{q^{2}}^{3}[[3]]_{q^{2}}^{2}\;\gamma^{5}.
\end{gather*}
We would like to write down the $\Sigma$-matrices.\ Clearly, we have
$\Sigma^{\mu\nu}=\Gamma^{\mu\nu}$ for $(\mu\nu)\in{\{}(12)$, $(13)$, $(14)$,
$(23)$, $(24)$, $(34){\}}$ [cf.\ (\ref{prgammaeu4}) and (\ref{defsigmaeu4})].
Moreover, it holds $\Sigma^{\mu\mu}=0$ for all $\mu=1,...,4.$ The remaining
$\Sigma$-matrices are%
\begin{gather}
\Sigma^{21}=\left(
\begin{array}
[c]{cccc}%
0 & 0 & 0 & 0\\
0 & 0 & 0 & 0\\
0 & 0 & 0 & -1\\
0 & 0 & 0 & 0
\end{array}
\right)  ,\quad\Sigma^{31}=\left(
\begin{array}
[c]{cccc}%
0 & -1 & 0 & 0\\
0 & 0 & 0 & 0\\
0 & 0 & 0 & 0\\
0 & 0 & 0 & 0
\end{array}
\right)  ,\nonumber\\[0.03in]
\Sigma^{32}=\lambda_{+}^{-1}\left(
\begin{array}
[c]{cccc}%
q^{2} & 0 & 0 & 0\\
0 & -1 & 0 & 0\\
0 & 0 & -1 & 0\\
0 & 0 & 0 & q^{-2}%
\end{array}
\right)  ,\quad\Sigma^{41}=\lambda_{+}^{-1}\left(
\begin{array}
[c]{cccc}%
-q & 0 & 0 & 0\\
0 & q^{-1} & 0 & 0\\
0 & 0 & -q & 0\\
0 & 0 & 0 & q^{-1}%
\end{array}
\right)  ,\nonumber\\[0.03in]
\Sigma^{42}=\left(
\begin{array}
[c]{cccc}%
0 & 0 & 0 & 0\\
-q & 0 & 0 & 0\\
0 & 0 & 0 & 0\\
0 & 0 & 0 & 0
\end{array}
\right)  ,\quad\Sigma^{43}=\left(
\begin{array}
[c]{cccc}%
0 & 0 & 0 & 0\\
0 & 0 & 0 & 0\\
0 & 0 & 0 & 0\\
0 & 0 & -q & 0
\end{array}
\right)  .
\end{gather}

\section{Some explicit expressions for Min\-kow\-ski space\label{AppB}}

In terms of the generators of the Clifford algebra the elements of the two
bases in (\ref{gammabasis}) and (\ref{basisgamma5mink}) become%
\begin{align}
\gamma^{\lbrack+}\gamma^{3]_{q}}=  &  \;\gamma^{+}\gamma^{3},\nonumber\\
\gamma^{\lbrack+}\gamma^{0]_{q}}=  &  \;\gamma^{+}\gamma^{0},\nonumber\\
\gamma^{\lbrack+}\gamma^{-]_{q}}=  &  \;\gamma^{+}\gamma^{-}-q\lambda_{+}%
^{-1}\mbox{1 \kern-.59em {\rm l}},\nonumber\\
\gamma^{\lbrack3}\gamma^{0]_{q}}=  &  \;\gamma^{3}\gamma^{0},\nonumber\\
\gamma^{\lbrack3}\gamma^{-]_{q}}=  &  \;\gamma^{3}\gamma^{-},\nonumber\\
\gamma^{\lbrack0}\gamma^{-]_{q}}=  &  \;\gamma^{0}\gamma^{-},
\label{gammaexpmink1N}%
\end{align}%
\begin{align}
\gamma^{\lbrack+}\gamma^{3}\gamma^{0]_{q}}=  &  \;\gamma^{+}\gamma^{3}%
\gamma^{0}+\frac{q}{2}\lambda\lambda_{+}^{-1}\,\gamma^{+},\nonumber\\
\gamma^{\lbrack+}\gamma^{3}\gamma^{-]_{q}}=  &  \;\gamma^{+}\gamma^{3}%
\gamma^{-}+q\lambda_{+}^{-1}\,\gamma^{3}+\frac{q^{2}}{2}\lambda\lambda
_{+}^{-1}\,\gamma^{0},\nonumber\\
\gamma^{\lbrack+}\gamma^{0}\gamma^{-]_{q}}=  &  \;\gamma^{+}\gamma^{0}%
\gamma^{-}+\frac{1}{2}\lambda\lambda_{+}^{-1}\,\gamma^{3}+\frac{1}{2}%
(q^{4}+1)q^{-1}\lambda_{+}^{-1}\,\gamma^{0},\nonumber\\
\gamma^{\lbrack3}\gamma^{0}\gamma^{-]_{q}}=  &  \;\gamma^{3}\gamma^{0}%
\gamma^{-}+\frac{q}{2}\lambda\lambda_{+}^{-1}\,\gamma^{-1},
\end{align}%
\begin{align}
\gamma^{5}\gamma^{+}=  &  -2q^{-1}\,\gamma^{+}\gamma^{3}\gamma^{0}%
-\lambda\lambda_{+}^{-1}\,\gamma^{+},\nonumber\\
\gamma^{5}\gamma^{3}=  &  \;2\gamma^{+}\gamma^{0}\gamma^{-}-2q^{-1}%
\lambda\,\gamma^{+}\gamma^{3}\gamma^{-}\nonumber\\
&  +2q\lambda_{+}^{-1}\,\gamma^{0}-\lambda\lambda_{+}^{-1}\,\gamma
^{3},\nonumber\\
\gamma^{5}\gamma^{0}=  &  \;2q^{-2}\,\gamma^{+}\gamma^{3}\gamma^{-}%
+2q^{-1}\lambda_{+}^{-1}\,\gamma^{3}+\lambda\lambda_{+}^{-1}\,\gamma
^{0},\nonumber\\
\gamma^{5}\gamma^{-}=  &  \;2q^{-1}\,\gamma^{3}\gamma^{0}\gamma^{-}%
+\lambda\lambda_{+}^{-1}\,\gamma^{-},
\end{align}%
\begin{align}
\gamma^{5}=  &  -2q^{-1}\,\gamma^{+}\gamma^{-}+2q^{-1}\,\gamma^{3}\gamma
^{0}-2q^{-2}\lambda_{+}\,\gamma^{+}\gamma^{3}\gamma^{0}\gamma^{-}%
+\lambda\lambda_{+}^{-1}\mbox{1 \kern-.59em {\rm l}}\nonumber\\
\gamma^{\lbrack+}\gamma^{3}\gamma^{0}\gamma^{-]_{q}}=  &  -4q^{-14}%
[[2]]_{q}^{3}[[3]]_{q}^{2}\,\gamma^{3}\gamma^{0}+4q^{-14}\lambda\lbrack
\lbrack2]]_{q}^{3}[[3]]_{q}^{2}\,\gamma^{+}\gamma^{-}\nonumber\\
&  +4q^{-16}[[2]]_{q}^{4}[[3]]_{q}^{2}\,\gamma^{+}\gamma^{3}\gamma^{0}%
\gamma^{-}-2q^{-14}\lambda\lbrack\lbrack2]]_{q}^{2}[[3]]_{q}^{2}%
\,\mbox{1 \kern-.59em {\rm l}}. \label{gammaexpmink2}%
\end{align}
Similar expressions exist for the two bases (\ref{BasGammInvMin}) and
(\ref{InvBasGam5}). Most of them follow from the expressions in
(\ref{gammaexpmink1N})-(\ref{gammaexpmink2}) by applying the substitutions
$\gamma^{\mu}\rightarrow\gamma_{\mu}^{-1}$, $\mu\in\{+$, $3$, $0$, $-\}$, with
the exception of the following ones:
\begin{align}
\gamma_{\lbrack+}^{-1}\gamma_{3}^{-1}\gamma_{0]_{q}}^{-1}=  &  \;\gamma
_{+}^{-1}\gamma_{3}^{-1}\gamma_{0}^{-1}-\frac{q}{2}\lambda\lambda_{+}%
^{-1}\,\gamma_{+}^{-1},\nonumber\\
\gamma_{\lbrack+}^{-1}\gamma_{3}^{-1}\gamma_{-]_{q}}^{-1}=  &  \;\gamma
_{+}^{-1}\gamma_{3}^{-1}\gamma_{-}^{-1}+q\lambda_{+}^{-1}\,\gamma_{3}%
^{-1}-\frac{q^{2}}{2}\lambda\lambda_{+}^{-1}\,\gamma_{0}^{-1},\nonumber\\
\gamma_{\lbrack+}^{-1}\gamma_{0}^{-1}\gamma_{-]_{q}}^{-1}=  &  \;\gamma
_{+}^{-1}\gamma_{0}^{-1}\gamma_{-}^{-1}-\frac{1}{2}\lambda\lambda_{+}%
^{-1}\,\gamma_{3}^{-1}+\frac{1}{2}(q^{4}+1)q^{-1}\lambda_{+}^{-1}\,\gamma
_{0}^{-1},\nonumber\\
\gamma_{\lbrack3}^{-1}\gamma_{0}^{-1}\gamma_{-]_{q}}^{-1}=  &  \;\gamma
_{3}^{-1}\gamma_{0}^{-1}\gamma_{-}^{-1}-\frac{q}{2}\lambda\lambda_{+}%
^{-1}\,\gamma_{-}^{-1},
\end{align}%
\begin{align}
\gamma_{5}^{-1}\gamma_{+}^{-1}=  &  \;2q^{-1}\,\gamma_{+}^{-1}\gamma_{3}%
^{-1}\gamma_{0}^{-1}-\lambda\lambda_{+}^{-1}\,\gamma_{+}^{-1},\nonumber\\
\gamma_{5}^{-1}\gamma_{3}^{-1}=  &  \;2\gamma_{+}^{-1}\gamma_{0}^{-1}%
\gamma_{-}^{-1}-2q^{-1}\lambda\,\gamma_{+}^{-1}\gamma_{3}^{-1}\gamma_{-}%
^{-1}\nonumber\\
&  -2q\lambda_{+}^{-1}\,\gamma_{0}^{-1}-\lambda\lambda_{+}^{-1}\,\gamma
_{3}^{-1},\nonumber\\
\gamma_{5}^{-1}\gamma_{0}^{-1}=  &  -2q^{-2}\,\gamma_{+}^{-1}\gamma_{3}%
^{-1}\gamma_{-}^{-1}-2q^{-1}\lambda_{+}^{-1}\,\gamma_{3}^{-1}+\lambda
\lambda_{+}^{-1}\,\gamma_{0}^{-1},\nonumber\\
\gamma_{5}^{-1}\gamma_{-}^{-1}=  &  -2q^{-1}\,\gamma_{3}^{-1}\gamma_{0}%
^{-1}\gamma_{-}^{-1}+\lambda\lambda_{+}^{-1}\,\gamma_{-}^{-1},
\end{align}%
\begin{align}
\gamma_{5}^{-1}=  &  -2q^{-1}\,\gamma_{+}^{-1}\gamma_{-}^{-1}-2q^{-1}%
\,\gamma_{3}^{-1}\gamma_{0}^{-1}\nonumber\\
&  +2q^{-2}\lambda_{+}\,\gamma_{+}^{-1}\gamma_{3}^{-1}\gamma_{0}^{-1}%
\gamma_{-}^{-1}+\lambda\lambda_{+}^{-1}\mbox{1 \kern-.59em {\rm l}}\nonumber\\
\gamma_{\lbrack+}^{-1}\gamma_{3}^{-1}\gamma_{0}^{-1}\gamma_{-]_{q}}^{-1}=  &
\;4q^{-14}[[2]]_{q}^{3}[[3]]_{q}^{2}\,\gamma_{3}^{-1}\gamma_{0}^{-1}%
+4q^{-14}\lambda\lbrack\lbrack2]]_{q}^{3}[[3]]_{q}^{2}\,\gamma_{+}^{-1}%
\gamma_{-}^{-1}\nonumber\\
&  -4q^{-16}[[2]]_{q}^{4}[[3]]_{q}^{2}\,\gamma_{+}^{-1}\gamma_{3}^{-1}%
\gamma_{0}^{-1}\gamma_{-}^{-1}\nonumber\\
&  -2q^{-14}\lambda\lbrack\lbrack2]]_{q}^{2}[[3]]_{q}^{2}%
\;\mbox{1 \kern-.59em {\rm l}}.
\end{align}
We give the matrices representing the elements of the two bases in
(\ref{gammabasis}) and (\ref{basisgamma5mink}):%
\begin{gather}
{\Gamma}^{+3}{=-\lambda}_{+}^{-1/2}\left(
\begin{array}
[c]{cccc}%
{\scriptstyle0} & {\scriptstyle0} & {\scriptstyle0} & {\scriptstyle0}\\
{\scriptstyle q}^{3/2} & {\scriptstyle0} & {\scriptstyle0} & {\scriptstyle0}\\
{\scriptstyle0} & {\scriptstyle0} & {\scriptstyle0} & {\scriptstyle0}\\
{\scriptstyle0} & {\scriptstyle0} & {\scriptstyle q}^{3/2} & {\scriptstyle0}%
\end{array}
\right)  ,\quad\Gamma^{+0}=\lambda_{+}^{-1/2}\left(
\begin{array}
[c]{cccc}%
{\scriptstyle0} & {\scriptstyle0} & {\scriptstyle0} & {\scriptstyle0}\\
{\scriptstyle-q}^{3/2} & {\scriptstyle0} & {\scriptstyle0} & {\scriptstyle0}\\
{\scriptstyle0} & {\scriptstyle0} & {\scriptstyle0} & {\scriptstyle0}\\
{\scriptstyle0} & {\scriptstyle0} & {\scriptstyle q}^{-1/2} & {\scriptstyle0}%
\end{array}
\right)  ,\nonumber\\[0.03in]
\Gamma^{+-}=\lambda_{+}^{-1}\left(
\begin{array}
[c]{cccc}%
{\scriptstyle-q} & {\scriptstyle0} & {\scriptstyle0} & {\scriptstyle0}\\
{\scriptstyle0} & {\scriptstyle1} & {\scriptstyle0} & {\scriptstyle0}\\
{\scriptstyle0} & {\scriptstyle0} & {\scriptstyle-q} & {\scriptstyle0}\\
{\scriptstyle0} & {\scriptstyle0} & {\scriptstyle0} & {\scriptstyle q}^{-1}%
\end{array}
\right)  ,\quad\Gamma^{30}=\lambda_{+}^{-1}\left(
\begin{array}
[c]{cccc}%
{\scriptstyle-q}^{2} & {\scriptstyle0} & {\scriptstyle0} & {\scriptstyle0}\\
{\scriptstyle0} & {\scriptstyle1} & {\scriptstyle0} & {\scriptstyle0}\\
{\scriptstyle0} & {\scriptstyle0} & {\scriptstyle1} & {\scriptstyle0}\\
{\scriptstyle0} & {\scriptstyle0} & {\scriptstyle0} & {\scriptstyle-q}^{-2}%
\end{array}
\right)  ,\nonumber\\[0.03in]
\Gamma^{3-}=\lambda_{+}^{-1/2}\left(
\begin{array}
[c]{cccc}%
{\scriptstyle0} & {\scriptstyle q}^{1/2} & {\scriptstyle0} & {\scriptstyle0}\\
{\scriptstyle0} & {\scriptstyle0} & {\scriptstyle0} & {\scriptstyle0}\\
{\scriptstyle0} & {\scriptstyle0} & {\scriptstyle0} & {\scriptstyle q}^{1/2}\\
{\scriptstyle0} & {\scriptstyle0} & {\scriptstyle0} & {\scriptstyle0}%
\end{array}
\right)  ,\quad\Gamma^{0-}=\lambda_{+}^{-1/2}\left(
\begin{array}
[c]{cccc}%
{\scriptstyle0} & {\scriptstyle-q}^{-3/2} & {\scriptstyle0} & {\scriptstyle0}%
\\
{\scriptstyle0} & {\scriptstyle0} & {\scriptstyle0} & {\scriptstyle0}\\
{\scriptstyle0} & {\scriptstyle0} & {\scriptstyle0} & {\scriptstyle q}^{1/2}\\
{\scriptstyle0} & {\scriptstyle0} & {\scriptstyle0} & {\scriptstyle0}%
\end{array}
\right)  ,\nonumber\\[0.03in]
\Gamma^{+30}=\frac{1}{2}\left(
\begin{array}
[c]{cccc}%
{\scriptstyle0} & {\scriptstyle0} & {\scriptstyle0} & {\scriptstyle0}\\
{\scriptstyle0} & {\scriptstyle0} & {\scriptstyle-q}^{5/2} & {\scriptstyle0}\\
{\scriptstyle0} & {\scriptstyle0} & {\scriptstyle0} & {\scriptstyle0}\\
{\scriptstyle-q}^{1/2} & {\scriptstyle0} & {\scriptstyle0} & {\scriptstyle0}%
\end{array}
\right)  ,\quad\Gamma^{+3-}=\frac{1}{2}q^{2}\lambda_{+}^{-1/2}\left(
\begin{array}
[c]{cccc}%
{0} & {0} & {-1} & {0}\\
{0} & {0} & {0} & {-1}\\
{1} & {0} & {0} & {0}\\
{0} & {1} & {0} & {0}%
\end{array}
\right)  ,\nonumber\\[0.03in]
\Gamma^{+0-}=\frac{1}{2}\lambda_{+}^{-1/2}\left(
\begin{array}
[c]{cccc}%
{\scriptstyle0} & {\scriptstyle0} & {\scriptstyle1} & {\scriptstyle0}\\
{\scriptstyle0} & {\scriptstyle0} & {\scriptstyle0} & {\scriptstyle-q}^{2}\\
{\scriptstyle q}^{2} & {\scriptstyle0} & {\scriptstyle0} & {\scriptstyle0}\\
{\scriptstyle0} & {\scriptstyle q\lambda-q}^{-2} & {\scriptstyle0} &
{\scriptstyle0}%
\end{array}
\right)  ,\quad\Gamma^{30-}=\frac{1}{2}\left(
\begin{array}
[c]{cccc}%
{\scriptstyle0} & {\scriptstyle0} & {\scriptstyle0} & {\scriptstyle-q}^{3/2}\\
{\scriptstyle0} & {\scriptstyle0} & {\scriptstyle0} & {\scriptstyle0}\\
{\scriptstyle0} & {\scriptstyle-q}^{-1/2} & {\scriptstyle0} & {\scriptstyle0}%
\\
{\scriptstyle0} & {\scriptstyle0} & {\scriptstyle0} & {\scriptstyle0}%
\end{array}
\right)  ,\nonumber\\[0.03in]
\Gamma^{5+}=\left(
\begin{array}
[c]{cccc}%
{\scriptstyle0} & {\scriptstyle0} & {\scriptstyle0} & {\scriptstyle0}\\
{\scriptstyle0} & {\scriptstyle0} & {\scriptstyle q}^{3/2} & {\scriptstyle0}\\
{\scriptstyle0} & {\scriptstyle0} & {\scriptstyle0} & {\scriptstyle0}\\
{\scriptstyle q}^{-1/2} & {\scriptstyle0} & {\scriptstyle0} & {\scriptstyle0}%
\end{array}
\right)  ,\quad\Gamma^{53}=\lambda_{+}^{-1/2}\left(
\begin{array}
[c]{cccc}%
{\scriptstyle0} & {\scriptstyle0} & {\scriptstyle q}^{2} & {\scriptstyle0}\\
{\scriptstyle0} & {\scriptstyle0} & {\scriptstyle0} & {\scriptstyle-1}\\
{\scriptstyle1} & {\scriptstyle0} & {\scriptstyle0} & {\scriptstyle0}\\
{\scriptstyle0} & {\scriptstyle-q}^{-2} & {\scriptstyle0} & {\scriptstyle0}%
\end{array}
\right)  ,\nonumber\\[0.03in]
\Gamma^{50}=\lambda_{+}^{-1/2}\left(
\begin{array}
[c]{cccc}%
{\scriptstyle0} & {\scriptstyle0} & {\scriptstyle-1} & {\scriptstyle0}\\
{\scriptstyle0} & {\scriptstyle0} & {\scriptstyle0} & {\scriptstyle-1}\\
{\scriptstyle1} & {\scriptstyle0} & {\scriptstyle0} & {\scriptstyle0}\\
{\scriptstyle0} & {\scriptstyle1} & {\scriptstyle0} & {\scriptstyle0}%
\end{array}
\right)  ,\quad\Gamma^{5-}=\left(
\begin{array}
[c]{cccc}%
{\scriptstyle0} & {\scriptstyle0} & {\scriptstyle0} & {\scriptstyle-q}^{1/2}\\
{\scriptstyle0} & {\scriptstyle0} & {\scriptstyle0} & {\scriptstyle0}\\
{\scriptstyle0} & {\scriptstyle-q}^{-3/2} & {\scriptstyle0} & {\scriptstyle0}%
\\
{\scriptstyle0} & {\scriptstyle0} & {\scriptstyle0} & {\scriptstyle0}%
\end{array}
\right)  ,\nonumber\\[0.03in]
\Gamma^{+30-}=-2q^{-13}[[2]]_{q^{2}}^{3}[[3]]_{q^{2}}^{2}\;\gamma^{5}.
\end{gather}
The matrix representations of the elements of the bases (\ref{BasGammInvMin})
and (\ref{InvBasGam5}) are
\begin{gather}
\Gamma_{+3}^{-1}=\lambda_{+}^{-1/2}\left(
\begin{array}
[c]{cccc}%
{\scriptstyle0} & {\scriptstyle q}^{1/2} & {\scriptstyle0} & {\scriptstyle0}\\
{\scriptstyle0} & {\scriptstyle0} & {\scriptstyle0} & {\scriptstyle0}\\
{\scriptstyle0} & {\scriptstyle0} & {\scriptstyle0} & {\scriptstyle q}^{1/2}\\
{\scriptstyle0} & {\scriptstyle0} & {\scriptstyle0} & {\scriptstyle0}%
\end{array}
\right)  ,\quad\Gamma_{+0}^{-1}=\lambda_{+}^{-1/2}\left(
\begin{array}
[c]{cccc}%
{\scriptstyle0} & {\scriptstyle q}^{-3/2} & {\scriptstyle0} & {\scriptstyle0}%
\\
{\scriptstyle0} & {\scriptstyle0} & {\scriptstyle0} & {\scriptstyle0}\\
{\scriptstyle0} & {\scriptstyle0} & {\scriptstyle0} & {\scriptstyle-q}^{1/2}\\
{\scriptstyle0} & {\scriptstyle0} & {\scriptstyle0} & {\scriptstyle0}%
\end{array}
\right)  ,\nonumber\\[0.03in]
\Gamma_{+-}^{-1}=\lambda_{+}^{-1/2}\left(
\begin{array}
[c]{cccc}%
{\scriptstyle q}^{-1} & {\scriptstyle0} & {\scriptstyle0} & {\scriptstyle0}\\
{\scriptstyle0} & {\scriptstyle-q} & {\scriptstyle0} & {\scriptstyle0}\\
{\scriptstyle0} & {\scriptstyle0} & {\scriptstyle q}^{-1} & {\scriptstyle0}\\
{\scriptstyle0} & {\scriptstyle0} & {\scriptstyle0} & {\scriptstyle q}%
\end{array}
\right)  ,\quad\Gamma_{30}^{-1}=\lambda_{+}^{-1}\left(
\begin{array}
[c]{cccc}%
{\scriptstyle q}^{-2} & {\scriptstyle0} & {\scriptstyle0} & {\scriptstyle0}\\
{\scriptstyle0} & {\scriptstyle-q}^{-1} & {\scriptstyle0} & {\scriptstyle0}\\
{\scriptstyle0} & {\scriptstyle0} & {\scriptstyle-q}^{-1} & {\scriptstyle0}\\
{\scriptstyle0} & {\scriptstyle0} & {\scriptstyle0} & {\scriptstyle q}^{2}%
\end{array}
\right)  ,\nonumber\\[0.03in]
\Gamma_{3-}^{-1}=-\lambda_{+}^{-1/2}\left(
\begin{array}
[c]{cccc}%
{\scriptstyle0} & {\scriptstyle0} & {\scriptstyle0} & {\scriptstyle0}\\
{\scriptstyle q}^{3/2} & {\scriptstyle0} & {\scriptstyle0} & {\scriptstyle0}\\
{\scriptstyle0} & {\scriptstyle0} & {\scriptstyle q}^{3/2} & {\scriptstyle0}\\
{\scriptstyle0} & {\scriptstyle0} & {\scriptstyle0} & {\scriptstyle0}%
\end{array}
\right)  ,\quad\Gamma_{0-}^{-1}=\lambda_{+}^{-1/2}\left(
\begin{array}
[c]{cccc}%
{\scriptstyle0} & {\scriptstyle0} & {\scriptstyle0} & {\scriptstyle0}\\
{\scriptstyle q}^{3/2} & {\scriptstyle0} & {\scriptstyle0} & {\scriptstyle0}\\
{\scriptstyle0} & {\scriptstyle0} & {\scriptstyle0} & {\scriptstyle0}\\
{\scriptstyle0} & {\scriptstyle0} & {\scriptstyle-q}^{-1/2} & {\scriptstyle0}%
\end{array}
\right)  ,\nonumber\\[0.03in]
\Gamma_{+30}^{-1}=\frac{1}{2}\left(
\begin{array}
[c]{cccc}%
{\scriptstyle0} & {\scriptstyle0} & {\scriptstyle0} & {\scriptstyle q}%
^{-1/2}\\
{\scriptstyle0} & {\scriptstyle0} & {\scriptstyle0} & {\scriptstyle0}\\
{\scriptstyle0} & {\scriptstyle q}^{3/2} & {\scriptstyle0} & {\scriptstyle0}\\
{\scriptstyle0} & {\scriptstyle0} & {\scriptstyle0} & {\scriptstyle0}%
\end{array}
\right)  ,\quad\Gamma_{+3-}^{-1}=\frac{1}{2}q^{2}\lambda_{+}^{-1/2}\left(
\begin{array}
[c]{cccc}%
{\scriptstyle0} & {\scriptstyle0} & {\scriptstyle-1} & {\scriptstyle0}\\
{\scriptstyle0} & {\scriptstyle0} & {\scriptstyle0} & {\scriptstyle-1}\\
{\scriptstyle1} & {\scriptstyle0} & {\scriptstyle0} & {\scriptstyle0}\\
{\scriptstyle0} & {\scriptstyle1} & {\scriptstyle0} & {\scriptstyle0}%
\end{array}
\right)  ,\nonumber\\[0.03in]
\Gamma_{+0-}^{-1}=\frac{1}{2}\lambda_{+}^{-1/2}\left(
\begin{array}
[c]{cccc}%
{\scriptstyle0} & {\scriptstyle0} & {\scriptstyle q\lambda-q}^{-2} &
{\scriptstyle0}\\
{\scriptstyle0} & {\scriptstyle0} & {\scriptstyle0} & {\scriptstyle q}^{2}\\
{\scriptstyle-q}^{2} & {\scriptstyle0} & {\scriptstyle0} & {\scriptstyle0}\\
{\scriptstyle0} & {\scriptstyle1} & {\scriptstyle0} & {\scriptstyle0}%
\end{array}
\right)  ,\quad\Gamma_{3-0}^{-1}=\frac{1}{2}\left(
\begin{array}
[c]{cccc}%
{\scriptstyle0} & {\scriptstyle0} & {\scriptstyle0} & {\scriptstyle0}\\
{\scriptstyle0} & {\scriptstyle0} & {\scriptstyle q}^{1/2} & {\scriptstyle0}\\
{\scriptstyle0} & {\scriptstyle0} & {\scriptstyle0} & {\scriptstyle0}\\
{\scriptstyle q}^{5/2} & {\scriptstyle0} & {\scriptstyle0} & {\scriptstyle0}%
\end{array}
\right)  ,\nonumber\\[0.03in]
\Gamma_{5+}^{-1}=\left(
\begin{array}
[c]{cccc}%
{\scriptstyle0} & {\scriptstyle0} & {\scriptstyle0} & {\scriptstyle q}%
^{-3/2}\\
{\scriptstyle0} & {\scriptstyle0} & {\scriptstyle0} & {\scriptstyle0}\\
{\scriptstyle0} & {\scriptstyle q}^{1/2} & {\scriptstyle0} & {\scriptstyle0}\\
{\scriptstyle0} & {\scriptstyle0} & {\scriptstyle0} & {\scriptstyle0}%
\end{array}
\right)  ,\quad\Gamma_{53}^{-1}=\lambda_{+}^{-1/2}\left(
\begin{array}
[c]{cccc}%
{\scriptstyle0} & {\scriptstyle0} & {\scriptstyle q}^{-2} & {\scriptstyle0}\\
{\scriptstyle0} & {\scriptstyle0} & {\scriptstyle0} & {\scriptstyle-1}\\
{\scriptstyle1} & {\scriptstyle0} & {\scriptstyle0} & {\scriptstyle0}\\
{\scriptstyle0} & {\scriptstyle-q}^{2} & {\scriptstyle0} & {\scriptstyle0}%
\end{array}
\right)  ,\nonumber\\[0.03in]
\Gamma_{50}^{-1}=\lambda_{+}^{-1/2}\left(
\begin{array}
[c]{cccc}%
{\scriptstyle0} & {\scriptstyle0} & {\scriptstyle1} & {\scriptstyle0}\\
{\scriptstyle0} & {\scriptstyle0} & {\scriptstyle0} & {\scriptstyle1}\\
{\scriptstyle-1} & {\scriptstyle0} & {\scriptstyle0} & {\scriptstyle0}\\
{\scriptstyle0} & {\scriptstyle-1} & {\scriptstyle0} & {\scriptstyle0}%
\end{array}
\right)  ,\quad\Gamma_{5-}^{-1}=\left(
\begin{array}
[c]{cccc}%
{\scriptstyle0} & {\scriptstyle0} & {\scriptstyle0} & {\scriptstyle0}\\
{\scriptstyle0} & {\scriptstyle0} & {\scriptstyle-q}^{-1/2} & {\scriptstyle0}%
\\
{\scriptstyle0} & {\scriptstyle0} & {\scriptstyle0} & {\scriptstyle0}\\
{\scriptstyle-q}^{3/2} & {\scriptstyle0} & {\scriptstyle0} & {\scriptstyle0}%
\end{array}
\right)  ,\nonumber\\[0.03in]
\Gamma_{5}^{-1}=2q^{-13}[[2]]_{q^{2}}^{3}{}[[3]]_{q^{2}}^{2}\;\gamma_{5}^{-1}.
\end{gather}
The matrices $\Sigma^{\mu\nu}$ that cannot be read off from the matrices for
$\Gamma^{A}$ are
\begin{gather}
\Sigma^{3+}=q^{-1/2}\lambda_{+}^{-1/2}\left(
\begin{array}
[c]{cccc}%
0 & 0 & 0 & 0\\
1 & 0 & 0 & 0\\
0 & 0 & 0 & 0\\
0 & 0 & 1 & 0
\end{array}
\right)  ,\quad\Sigma^{0+}=\lambda_{+}^{-1/2}\left(
\begin{array}
[c]{cccc}%
0 & 0 & 0 & 0\\
q^{-1/2} & 0 & 0 & 0\\
0 & 0 & 0 & 0\\
0 & 0 & -q^{3/2} & 0
\end{array}
\right)  ,\nonumber\\[0.03in]
\Sigma^{-+}=\lambda_{+}^{-1}\left(
\begin{array}
[c]{cccc}%
q & 0 & 0 & 0\\
0 & -q^{-1} & 0 & 0\\
0 & 0 & q & 0\\
0 & 0 & 0 & -q^{-1}%
\end{array}
\right)  ,\quad\Sigma^{33}=\lambda\lambda_{+}^{-1}\left(
\begin{array}
[c]{cccc}%
-q^{1} & 0 & 0 & 0\\
0 & 1 & 0 & 0\\
0 & 0 & 1 & 0\\
0 & 0 & 0 & -q^{-3}%
\end{array}
\right)  ,\nonumber\\[0.03in]
\Sigma^{03}=\lambda_{+}^{-1}\left(
\begin{array}
[c]{cccc}%
q^{-1} & 0 & 0 & 0\\
0 & -q^{-2} & 0 & 0\\
0 & 0 & -q^{2} & 0\\
0 & 0 & 0 & 1
\end{array}
\right)  ,\quad\Sigma^{-3}=-q^{-3/2}\lambda_{+}^{-1/2}\left(
\begin{array}
[c]{cccc}%
0 & 1 & 0 & 0\\
0 & 0 & 0 & 0\\
0 & 0 & 0 & 1\\
0 & 0 & 0 & 0
\end{array}
\right)  ,\nonumber\\[0.03in]
\Sigma^{-0}=\lambda_{+}^{-1/2}\left(
\begin{array}
[c]{cccc}%
0 & q^{1/2} & 0 & 0\\
0 & 0 & 0 & 0\\
0 & 0 & -q^{-3/2} & 0\\
0 & 0 & 0 & 0
\end{array}
\right)  .
\end{gather}
The matrices $\Sigma_{\mu\nu}^{-1}$ that cannot be read off from the matrices
for $\Gamma_{A}^{-1}$ are%
\begin{gather}
\Sigma_{3+}^{-1}=-q^{-3/2}\lambda_{+}^{-1/2}\left(
\begin{array}
[c]{cccc}%
0 & 1 & 0 & 0\\
0 & 0 & 0 & 0\\
0 & 0 & 0 & 1\\
0 & 0 & 0 & 0
\end{array}
\right)  ,\quad\Sigma_{0+}^{-1}=\lambda_{+}^{-1/2}\left(
\begin{array}
[c]{cccc}%
0 & -q^{1/2} & 0 & 0\\
0 & 0 & 0 & 0\\
0 & 0 & 0 & q^{-3/2}\\
0 & 0 & 0 & 0
\end{array}
\right)  ,\nonumber\\
\Sigma_{-+}^{-1}=\lambda_{+}^{-1}\left(
\begin{array}
[c]{cccc}%
-q^{-1} & 0 & 0 & 0\\
0 & q & 0 & 0\\
0 & 0 & -q^{-1} & 0\\
0 & 0 & 0 & q
\end{array}
\right)  ,\quad\Sigma_{33}^{-1}=\lambda\lambda_{+}^{-1/2}\left(
\begin{array}
[c]{cccc}%
q^{-2} & 0 & 0 & 0\\
0 & -1 & 0 & 0\\
0 & 0 & q^{-2} & 0\\
0 & 0 & 0 & 1
\end{array}
\right)  ,\nonumber\\
\Sigma_{03}^{-1}=\lambda_{+}^{-1}\left(
\begin{array}
[c]{cccc}%
-1 & 0 & 0 & 0\\
0 & q^{2} & 0 & 0\\
0 & 0 & q^{-2} & 0\\
0 & 0 & 0 & -q
\end{array}
\right)  ,\quad\Sigma_{-3}^{-1}=q^{-1/2}\lambda_{+}^{-1/2}\left(
\begin{array}
[c]{cccc}%
0 & 0 & 0 & 0\\
1 & 0 & 0 & 0\\
0 & 0 & 0 & 0\\
0 & 0 & 1 & 0
\end{array}
\right)  ,\nonumber\\
\Sigma_{-0}^{-1}=\lambda_{+}^{-1/2}\left(
\begin{array}
[c]{cccc}%
0 & 0 & 0 & 0\\
-q^{-1/2} & 0 & 0 & 0\\
0 & 0 & 0 & 0\\
0 & 0 & q^{3/2} & 0
\end{array}
\right)  .
\end{gather}

\section{$q$-Deformed quantum spaces\label{AppA3}}

The aim of this appendix is the following. For the quantum spaces under
consideration we list the defining commutation relations. In addition to this,
we write down the non-vanishing elements of their quantum metric and
$q$-deformed epsilon tensor.

In the case of three-dimensional $q$-deformed Euclidean space the commutation
relations between its coordinates $X^{A},$ $A\in\{+,3,-\},$ read
\begin{align}
X^{3}X^{\pm}  &  =q^{\pm2}X^{\pm}X^{3},\\
X^{-}X^{+}  &  =X^{+}X^{-}+\lambda X^{3}X^{3}.\nonumber
\end{align}
The non-vanishing elements of the quantum metric are
\begin{equation}
g^{+-}=-q,\quad g^{33}=1,\quad g^{-+}=-q^{-1}. \label{metriceu3app}%
\end{equation}
As usual, covariant coordinates are introduced by
\begin{equation}
X_{A}\equiv g_{AB}X^{B},
\end{equation}
with $g_{AB}$ being the inverse of $g^{AB}$. The non-vanishing components of
the three-dimensional $q$-deformed epsilon tensor take the form%
\begin{align}
\varepsilon^{-3+}  &  =-q^{-4}, & \varepsilon^{3-+}  &  =q^{-2},\nonumber\\
\varepsilon^{-+3}  &  =q^{-2}, & \varepsilon^{+-3}  &  =-q^{-2},\nonumber\\
\varepsilon^{3+-}  &  =-q^{-2}, & \varepsilon^{+3-}  &  =1.\nonumber\\
\varepsilon^{333}  &  =-q^{-2}\lambda. &  &  \label{epseu3app}%
\end{align}

Next we come to four-dimensional $q$-deformed Euclidean space. For its
coordinates $X^{i},$ $i=1,\ldots,4,$ we have the relations
\begin{align}
X^{1}X^{j}  &  =qX^{j}X^{1},\nonumber\\
X^{j}X^{4}  &  =qX^{4}X^{j},\quad j=1,2,\nonumber\\
X^{2}X^{3}  &  =X^{3}X^{2},\nonumber\\
X^{4}X^{1}  &  =X^{1}X^{4}+\lambda X^{2}X^{3}. \label{koordeu4app}%
\end{align}
The metric has the non-vanishing components
\begin{equation}
g^{14}=q^{-1},\quad g^{23}=g^{32}=1,\quad g^{41}=q. \label{metriceu4app}%
\end{equation}
Its inverse denoted by $g_{ij}$ can again be used to introduce covariant
coordinates, i.e.
\begin{equation}
X_{i}\equiv g_{ij}X^{j}.
\end{equation}
The non-vanishing components of the epsilon tensor of four-dimensional
$q$-deformed Euclidean become (see also Refs. \cite{Maj-Eps, Fio93, Mey96,
qliealg})%
\begin{align}
\varepsilon^{1234}  &  =1, & \varepsilon^{1432}  &  =-q^{2}, & \varepsilon
^{2413}  &  =-q^{2},\nonumber\\
\varepsilon^{2134}  &  =-q, & \varepsilon^{4132}  &  =q^{2}, & \varepsilon
^{4213}  &  =q^{3},\nonumber\\
\varepsilon^{1324}  &  =-1, & \varepsilon^{3412}  &  =q^{2}, & \varepsilon
^{2341}  &  =-q^{2},\nonumber\\
\varepsilon^{3124}  &  =q, & \varepsilon^{4312}  &  =-q^{3}, & \varepsilon
_{3241}  &  =q^{2},\nonumber\\
\varepsilon^{2314}  &  =q^{2}, & \varepsilon^{1243}  &  =-q, & \varepsilon
^{2431}  &  =q^{3},\nonumber\\
\varepsilon^{3214}  &  =-q^{2}, & \varepsilon^{2143}  &  =q^{2}, &
\varepsilon^{4231}  &  =-q^{4},\nonumber\\
\varepsilon^{1342}  &  =q, & \varepsilon^{1423}  &  =q^{2}, & \varepsilon
^{3421}  &  =-q^{3},\nonumber\\
\varepsilon^{3142}  &  =-q^{2}, & \varepsilon^{4123}  &  =-q^{2}, &
\varepsilon^{4321}  &  =q^{4}, \label{epseu4app}%
\end{align}
together with the non-classical components%
\begin{equation}
\varepsilon^{3232}=-\varepsilon^{2323}=-q^{2}\lambda.
\end{equation}

Now, we come to $q$-deformed Min\-kow\-ski space \cite{SWZ91,OSWZ92,Maj91}.
Its coordinates are subjected to the relations
\begin{gather}
X^{\mu}X^{0}=X^{0}X^{\mu},\quad\mu\in{\{}0,+,-,3{\},}\nonumber\\
X^{3}X^{\pm}-q^{\pm2}X^{\pm}X^{3}=-q\lambda X^{0}X^{\pm},\nonumber\\
X^{-}X^{+}-X^{+}X^{-}=\lambda(X^{3}X^{3}-X^{0}X^{3}), \label{koordminkapp}%
\end{gather}
and its metric is given by%
\begin{equation}
\eta^{00}=-1,\quad\eta^{33}=1,\quad\eta^{+-}=-q,\quad\eta^{-+}=-q^{-1}.
\label{metricminkapp}%
\end{equation}
As usual, the metric can be used to raise and lower indices. The non-vanishing
components of the $q$-deformed epsilon tensor read%
\begin{align}
\varepsilon^{+30-}  &  =1, & \varepsilon^{+-03}  &  =-q^{-2}, & \varepsilon
^{3-+0}  &  =q^{-2},\nonumber\\
\varepsilon^{3+0-}  &  =-q^{-2}, & \varepsilon^{-+03}  &  =q^{-2}, &
\varepsilon^{-3+0}  &  =q^{-4},\nonumber\\
\varepsilon^{+03-}  &  =-1, & \varepsilon^{0-+3}  &  =q^{-2}, & \varepsilon
^{30-+}  &  =-q^{-2},\nonumber\\
\varepsilon^{0+3-}  &  =1, & \varepsilon^{-0+3}  &  =-q^{-2}, & \varepsilon
^{03-+}  &  =q^{-2},\nonumber\\
\varepsilon^{30+-}  &  =q^{-2}, & \varepsilon^{+3-0}  &  =-1, & \varepsilon
^{3-0+}  &  =q^{-2},\nonumber\\
\varepsilon^{03+-}  &  =-q^{-2}, & \varepsilon^{3+-0}  &  =q^{-2}, &
\varepsilon^{-30+}  &  =-q^{-4},\nonumber\\
\varepsilon^{+0-3}  &  =q^{-2}, & \varepsilon^{+-30}  &  =q^{-2}, &
\varepsilon^{0-3+}  &  =-q^{-4},\nonumber\\
\varepsilon^{0+-3}  &  =-q^{-2}, & \varepsilon^{-+30}  &  =-q^{-2}, &
\varepsilon^{-03+}  &  =q^{-4}, \label{epsminkapp}%
\end{align}
and%
\begin{align}
\varepsilon^{0-0+}  &  =q^{-3}\lambda, & \varepsilon^{-0+0}  &  =-q^{-3}%
\lambda,\\
\varepsilon^{0333}  &  =-q^{-2}\lambda, & \varepsilon^{3330}  &
=q^{-2}\lambda,\nonumber\\
\varepsilon^{3033}  &  =+q^{-2}\lambda, & \varepsilon^{3030}  &
=-q^{-2}\lambda,\nonumber\\
\varepsilon^{3303}  &  =-q^{-2}\lambda, & \varepsilon^{+0-0}  &
=-q^{-1}\lambda,\nonumber\\
\varepsilon^{0303}  &  =q^{-2}\lambda, & \varepsilon^{0+0-}  &  =q^{-1}%
\lambda.\nonumber
\end{align}
Lowering the indices of the epsilon tensor is achieved by the quantum metric:%
\begin{equation}
\varepsilon_{\mu\nu\rho\sigma}=\eta_{\mu\mu^{\prime}}\eta_{\nu\nu^{\prime}%
}\eta_{\rho\rho^{\prime}}\eta_{\sigma\sigma^{\prime}}\,\varepsilon
^{\mu^{\prime}\nu^{\prime}\rho^{\prime}\sigma^{\prime}}.
\end{equation}


\begin{thebibliography}{99}                                                                                               %
\bibitem {qspinor1}A. Schmidt and H. Wachter, \textit{Spinor calculus for
$q$-deformed quantum spaces I}, preprint, [hep-th/0705.1640].

\bibitem {Ku83}P. P. Kulish and N. Yu. Reshetikin, \textit{Quantum linear
problem for the Sine-Gordon equation and higher representations},\textit{ }J.
Sov. Math. \textbf{23} (1983) 2345.

\bibitem {Wor87}S. L. Woronowicz, \textit{Compact matrix pseudogroups}%
,\textit{ }Commun. Math. Phys. \textbf{111} (1987) 613.

\bibitem {Dri85}V. G. Drinfeld, \textit{Hopf algebras and the quantum
Yang-Baxter equation},\textit{ }Sov. Math. Dokl. \textbf{32} (1985) 254.

\bibitem {Jim85}M. Jimbo, \textit{A q-analogue of U(g) and the Yang-Baxter
equation}, Lett. Math. Phys. \textbf{10}\ (1985) 63.

\bibitem {Drin86}V. G. Drinfeld, \textit{Quantum groups}, in A. M. Gleason,
ed., Proceedings of the International Congress of Mathematicians, Amer. Math.
Soc., 798 (1986).

\bibitem {Man88}Yu. I. Manin, \textit{Quantum Groups and Non-Commutative
Geometry}, Centre de Recherche Math\'{e}matiques, Montreal (1988).

\bibitem {RFT90}N. Yu. Reshetikhin, L. A. Takhtadzhyan and L. D. Faddeev,
\textit{Quantization of Lie Groups and Lie Algebras}, Leningrad Math. J.
\textbf{1} (1990) 193.

\bibitem {Tak90}M. Takeuchi, \textit{Matrix Bialgebras and Quantum
Groups},\textit{ }Israel J. Math. \textbf{72} (1990) 232.

\bibitem {CSSW90}U. Carow-Watamura, M. Schlieker, M. Scholl and S. Watamura,
\textit{Tensor Representations of the Quantum Group }$SL_{q}(2)$ \textit{and
Quantum Minkowski Space},\textit{ }Z. Phys. C\textbf{ 48} (1990) 159.

\bibitem {PW90}P. Podle\'{s} and S. L. Woronowicz, \textit{Quantum Deformation
of Lorentz Group}, Commun. Math. Phys. \textbf{130} (1990) 381.

\bibitem {SWZ91}W. B. Schmidke, J. Wess and B. Zumino, \textit{A q-deformed
Lorentz Algebra in Minkowski phase space},\textit{ }Z. Phys. C\textbf{ 52
}(1991) 471.

\bibitem {Maj91}S. Majid, \textit{Examples of braided groups and braided
matrices}, J. Math. Phys. \textbf{32} (1991) 3246.

\bibitem {OSWZ92}O. Ogievetsky, W. B. Schmidke, J. Wess and B. Zumino,
\textit{q-Deformed Poincar\'{e} Algebra}, Commun. Math. Phys. \textbf{150}
(1992) 495.

\bibitem {LWW97}A. Lorek, W. Weich and J. Wess, \textit{Non-commutative
Euclidean and Minkowski Structures},\textit{ }Z. Phys. C \textbf{76} (1997)
375, [q-alg/9702025].

\bibitem {FLW96}M. Fichtenm\"{u}ller, A. Lorek and J. Wess, \textit{q-deformed
Phase Space and its Lattice Structure, }Z. Phys. C\textbf{ 71} (1996) 533, [hep-th/9511106].

\bibitem {CW98}B. L. Cerchiai and J. Wess, \textit{q-Deformed Minkowski Space
based on a q-Lorentz Algebra, }Eur. Phys. J. C\textbf{ 5} (1998) 553, [math.qa/9801104].

\bibitem {GKP96}H. Grosse, C. Klim\v{c}ik and P. Pre\v{s}najder,
\textit{Towards finite quantum field theory in non-commutative geometry}%
,\textit{ }Int. J. Theor. Phys. \textbf{35} (1996) 231, [hep-th/9505175].

\bibitem {MajReg}S. Majid, \textit{On the q-regularisation},\textit{ }Int. J.
Mod. Phys. A\textbf{ 5} (1990) 4689.

\bibitem {Oec99}R. Oeckl, \textit{Braided Quantum Field Theory},\textit{
}Commun. Math. Phys. \textbf{217} (2001) 451.

\bibitem {Lu92}J. Lukierski, A. Nowicki and H. Ruegg, \textit{New Quantum
Poincar\'{e} Algebra and }$\kappa$\textit{-deformed Field Theory}, Phys. Lett.
B\textbf{ 293} (1992) 344.

\bibitem {Cas93}L. Castellani, \textit{Differential Calculus on }$ISO_{q}%
(N)$\textit{, Quantum Poincar\'{e} Algebra and q-Gravity}, preprint, [hep-th/9312179].

\bibitem {Dob94}V. K. Dobrev, \textit{New q-Minkowski space-time and q-Maxwell
equations hierarchy from q-conformal invariance}, Phys. Lett. B \textbf{341}
(1994) 133.

\bibitem {DFR95}S. Doplicher, K. Fredenhagen and J. E. Roberts, \textit{The
Quantum Structure of Space-Time at the Planck Scale and Quantum fields},
Commun. Math. Phys. \textbf{172} (1995) 187.

\bibitem {ChDe95}M. Chaichian and A. P. Demichev, \textit{Quantum Poincar\'{e}
group without dilatation and twisted classical algebra}, J. Math. Phys.
\textbf{36} (1995)\textbf{ }398.

\bibitem {ChKu04}M. Chaichian, P. P. Kulish, K. Nishijima and A. Tureanu,
\textit{On a Lorentz-Invariant Interpretation of Noncommutative Space-Time and
its Implications on Noncommutative QFT},\textit{ }Phys. Lett. B \textbf{604}
(2004) 98, [hep-th/0408062].

\bibitem {Koch04}F. Koch and E. Tsouchnika, \textit{Construction of }$\theta
$-\textit{Poincar\'{e} Algebras and their invariants on }$\mathcal{M}_{\theta
}$, Nucl. Phys. B \textbf{717} (2005) 387,\textbf{ }[hep-th/0409012].

\bibitem {Heis}W. Heisenberg, \textit{\"{U}ber die in der Theorie der
Elementarteilchen auftretende universelle L\"{a}nge},\textit{ }Ann. Phys.
\textbf{32} (1938) 20.

\bibitem {Sny47}H. S. Snyder, \textit{Quantized Space-Time},\textit{ }Phys.
Rev. \textbf{71 }(1947) 38.

\bibitem {Yan47}C. H. Yang, \textit{On Quantized Space-Time}, Phys. Rev.
\textbf{72 }(1947) 847.

\bibitem {Fli48}H. T. Flint, \textit{The Quantization of Space and Time},
Phys. Rev. \textbf{74} (1948) 209.

\bibitem {Hill55}E. L. Hill, \textit{Relativistic Theory of Discrete Momentum
Space and Discrete Space-Time}, Phys. Rev. \textbf{100} (1955) 1780.

\bibitem {Das60}A. Das, \textit{Cellular space-time and quantum field theory},
Nuovo Cimento \textbf{18} (1960) 483.

\bibitem {Gol63}Yu. A. Gol'fand, \textit{Quantum field theory in constant
curvature p-space}, Sov. Phys. JETP \textbf{16} (1963) 184.

\bibitem {Hopf}H. Hopf, \textit{\"{U}ber die Topologie der
Gruppenmannigfaltigkeiten und ihre Verallgemeinerungen}, Ann. Math. \textbf{42
}(1941) 22.

\bibitem {Maj95}S. Majid, \textit{Foundations of Quantum Group Theory}%
,\textit{ }University Press, Cambridge (1995).

\bibitem {ChDe96}M. Chaichian and A. P. Demichev, \textit{Introduction to
Quantum Groups}, World Scientific, Singapore \textit{(}1996).

\bibitem {Klimyk}A. Klimyk and K. Schm\"{u}dgen, \textit{Quantum Groups and
their Representations,} Springer Verlag, Berlin (1997).

\bibitem {Schlieck1}M. Schliecker, \textit{Quantendeformation der
Lorentzgruppe und $SO_{q}(N)$ kovariante Differentialrechnung auf
Quantenebenen, }PhD thesis, Ludwig-Maximilians-Universit\"{a}t M\"{u}nchen,
Fakult\"{a}t f\"{u}r Physik (1990).

\bibitem {Schlieck2}M. Schliecker and M. Scholl, \textit{Spinor calculus for
quantum groups},\textit{ }Z. Phys. C \textbf{47} (1990) 625.

\bibitem {Schlieck3}U. Carow-Watamura, M. Schliecker, M. Scholl and S.
Watamura, \textit{Tensor representations of the quantum group $SL_{q}%
(2,\mathbb{C})$ and quantum minkowski space},\textit{ }Z. Phys. C \textbf{48}
(1990) 159.

\bibitem {WSSW90}U. Carow-Watamura, M. Schliecker, M. Scholl and S. Watamura,
\textit{A Quantum Lorentz Group},\textit{ }Int. J. Mod. Phys. A \textbf{6}
(1990) 3081.

\bibitem {Mey96}U. Meyer, \textit{Wave equations on q-deformed Minkowski
space}, Commun. Math. Phys. \textbf{174 }(1996) 457.

\bibitem {Peskin}M. E. Peskin and D. V. Schroeder, \textit{An Introduction to
Quantum Field Theory}, Westview Press, (1995).

\bibitem {Landau}L. D. Landau and M. E. Lifschitz, \textit{Lehrbuch der
theoretischen Physik, Vol. }4\textit{ Quantenelektrodynamik},\textit{
}Akademie Verlag, Berlin (1991).

\bibitem {Wessbagger}J. Wess and J. Bagger, \textit{Supersymmetry and
Supergravity}, Second Edition, Princeton University Press, Princeton (1991).

\bibitem {bailin}D. Bailin and A. Love, \textit{Supersymmetric Gauge Field
Theory and String Theory}, Graduate Student Series in Physics, Institute of
Physics Publishing, Bristol (1996).

\bibitem {Core}V. I. Borodulin, R. N. Rogalyov and S. R. Slabospitsky,
\textit{CORE COmpendium of RElations},\textit{ }version\textit{ }2.1,\textit{
}[hep-ph/9507456].

\bibitem {Grimm}P. Bin{\'{e}}truy, G. Girardi and R. Grimm,
\textit{Supergravity couplings: a geometric formulation},\textit{ }Phys. Rept.
\textbf{343} (2001) 255, [hep-th/0005225].

\bibitem {Wol}S. Wolfram, \textit{The Mathematica Book},\textit{ }fourth ed.,
University Press, Cambridge (1999).

\bibitem {Good55}R. H. Good, \textit{Properties of the Dirac Matrices}, Rev.
Mod. Phys. \textbf{27} (1955) 187.

\bibitem {MSW04}D. Mikulovic, A. Schmidt and H. Wachter, \textit{Grassmann
variables on quantum spaces},\textit{\ }Eur. Phys. J. C \textbf{45} (2006)
529, [hep-th/0407273].

\bibitem {BW01}C. Bauer and H. Wachter, \textit{Operator Representations on
Quantum Spaces},\textit{ }Eur. Phys. J. C\textbf{31 }(2003) 261, math-ph/0201023.

\bibitem {Song92}X. C. Song, \textit{Covariant differential calculus on
quantum minkowski space and q-analog of Dirac equation}, Z. Phys. C
\textbf{55} (1992) 417.

\bibitem {AKR96}J. A. de Azc\'{a}rraga, P. P. Kulish and F. Rodenas,
\textit{Quantum groups and deformed special relativity}, Fortschr. Phys.
\textbf{44} (1996) 1.

\bibitem {Pod97}P. Podle\'{s}, \textit{The Dirac operator and gamma matrices
for quantum Minkowski spaces}, J. Math. Phys. \textbf{38} (1997) 4474.

\bibitem {Blo03}C. Blohmann, \textit{Free q-deformed relativistic wave
equations by representation theory}, Eur. Phys. J. C\textbf{ 30 }(2003) 435, [hep-th/0111172].

\bibitem {Wess00}J. Wess, \textit{q-deformed Heisenberg Algebras}, in H.
Gausterer, H. Grosse and L. Pittner, eds., Proceedings of the 38.
Internationale Universit\"{a}tswochen f\"{u}r Kern- und Teilchenphysik, no.
543 in Lect. Notes in Phys., Springer-Verlag, Schladming (2000), [math-phy/9910013].

\bibitem {Schmidke}A. Lorek, W. B. Schmidke and J. Wess, \textit{$SU_{q}(2)$
Covariant $\hat{R}$-Matrices for Reducible Representations},\textit{ }Lett.
Math. Phys. 31 (1994) 279.

\bibitem {SW04}A. Schmidt and H. Wachter, \textit{Superanalysis on quantum
spaces}, JHEP \textbf{0601} (2006) 84,\textit{ }[hep-th/0411180].

\bibitem {FioDet}G. Fiore, \textit{Quantum groups }$SO_{q}(N),$ $Sp_{q}%
(N)$\textit{ have quantum determinants, too}, J. Phys. A \textbf{27} (1994) 3795.

\bibitem {Umeyer}U. Meyer \textit{Quantum determinants},\textit{ }hep-th/9406172.

\bibitem {Fio93}G. Fiore, \textit{The }$\mathit{SO}_{q}\mathit{(N)}%
$\textit{-symmmetric harmonic oszillator on the quantum Euclidean space
}$R_{q}^{N}$ \textit{and its Hilbert space structure}, Int. J. Mod. Phys. A
\textbf{8} (1993) 4679.

\bibitem {Maj-Eps}S. Majid, \textit{q-epsilon tensor for quantum and braided
spaces}, J. Math. Phys. \textbf{36} (1995) 1991.

\bibitem {qliealg}A. Schmidt and H. Wachter, \textit{$q$-Deformed quantum Lie
algebras},\textit{ }J. Geom. Phys. \textbf{56} (2006) 2289, [mat-ph/0500932].
\end{thebibliography}
\end{document}